\documentclass[12pt]{article}
\usepackage[latin9]{inputenc}
\usepackage{geometry}
\usepackage[active]{srcltx}
\usepackage{float}
\usepackage{amsmath}
\usepackage{amsthm}
\usepackage{amssymb}
\usepackage{graphicx,psfrag,epsf}
\usepackage{rotfloat}
\usepackage{enumerate}
\usepackage{natbib}

\usepackage{setspace}
\usepackage{esint}
\usepackage[font=normalsize,labelfont=bf]{caption}
\usepackage{helvet}
\usepackage{comment}
\usepackage{hyperref}
\usepackage{units}
\makeatletter

\providecommand{\tabularnewline}{\\}
\floatstyle{ruled}
\newfloat{algorithm}{tbp}{loa}
\providecommand{\algorithmname}{Algorithm}
\floatname{algorithm}{\protect\algorithmname}

\usepackage{latexsym}
\usepackage{amsthm}\usepackage{amsfonts}\usepackage{graphicx}\usepackage{epsfig}

\newcommand*{\patchAmsMathEnvironmentForLineno}[1]{%
      \expandafter\let\csname old#1\expandafter\endcsname\csname #1\endcsname
      \expandafter\let\csname oldend#1\expandafter\endcsname\csname end#1\endcsname
      \renewenvironment{#1}%
         {\linenomath\csname old#1\endcsname}%
         {\csname oldend#1\endcsname\endlinenomath}}%
    \newcommand*{\patchBothAmsMathEnvironmentsForLineno}[1]{%
      \patchAmsMathEnvironmentForLineno{#1}%
      \patchAmsMathEnvironmentForLineno{#1*}}%
    \AtBeginDocument{%
    \patchBothAmsMathEnvironmentsForLineno{equation}%
    \patchBothAmsMathEnvironmentsForLineno{align}%
    \patchBothAmsMathEnvironmentsForLineno{flalign}%
    \patchBothAmsMathEnvironmentsForLineno{alignat}%
    \patchBothAmsMathEnvironmentsForLineno{gather}%
    \patchBothAmsMathEnvironmentsForLineno{multline}%
    }
\usepackage{setspace}
\def\dispmuskip{\thinmuskip= 3mu plus 0mu minus 2mu \medmuskip=  4mu plus 2mu minus 2mu \thickmuskip=5mu plus 5mu minus 2mu}
\def\textmuskip{\thinmuskip= 0mu                    \medmuskip=  1mu plus 1mu minus 1mu \thickmuskip=2mu plus 3mu minus 1mu}
\def\beq{\dispmuskip\begin{equation}}    \def\eeq{\end{equation}\textmuskip}
\def\beqn{\dispmuskip\begin{displaymath}}\def\eeqn{\end{displaymath}\textmuskip}
\def\bea{\dispmuskip\begin{eqnarray}}    \def\eea{\end{eqnarray}\textmuskip}
\def\bean{\dispmuskip\begin{eqnarray*}}  \def\eean{\end{eqnarray*}\textmuskip}

\newcommand{\blind}{0}

\def\N{{\cal N}}
\def\R{{\cal R}}

\def\vech{\text{\rm vech}}
\def\vechsq{P}

\def\Eqref{Eq. \eqref}

\usepackage[mathlines,displaymath]{lineno}
\newcommand\blfootnote[1]{%
	\begingroup
	\renewcommand\thefootnote{}\footnote{#1}%
	\addtocounter{footnote}{-1}%
	\endgroup
}

\usepackage{xr}
\makeatletter
\newcommand*{\addFileDependency}[1]{% argument=file name and extension
  \typeout{(#1)}
  \@addtofilelist{#1}
  \IfFileExists{#1}{}{\typeout{No file #1.}}
}
\makeatother

\newcommand*{\myexternaldocument}[1]{
    \externaldocument{#1}
    \addFileDependency{#1.tex}
    \addFileDependency{#1.aux}
}
\myexternaldocument{reply2refereesJBES}
\myexternaldocument{DSGEsupplementR1}
\begin{document}

\title{\bf The Block-Correlated Pseudo Marginal Sampler for State Space Models}
\author{David Gunawan\textsuperscript{$\star$}, Pratiti Chatterjee\textsuperscript{$\ddagger$}, and Robert Kohn\textsuperscript{$\star\star$}}
\maketitle
\if1\blind%%
{
   %\title {\bf Flexible Density Tempering Approaches for State Space Models with an Application to Factor Stochastic Volatility Models}
     \title{\bf The Block-Correlated Pseudo Marginal Sampler for State Space Models}		
\author{}
\maketitle
} \fi

%NOTE: alternative title \lq Correlated block pseudo marginal estimation of state space models\rq{} 
%\footnote{Corresponding Author: Robert Kohn, School of Economics, UNSW Business School, University of New South Wales, Email: {r.kohn@unsw.edu.au}}}

%Fourth, it uses delayed acceptance to increase the efficiency  of
%the sampler.

\begin{abstract}
Particle Marginal Metropolis-Hastings (PMMH) is a general approach to
Bayesian inference when the likelihood is intractable, but can be estimated unbiasedly.
Our article develops an efficient PMMH method that 
scales up better to  higher dimensional state vectors than previous approaches. The improvement is achieved by the following innovations. First, 
the trimmed mean of the unbiased likelihood estimates of the multiple particle filters is used. 
Second, a novel block version of PMMH that works with multiple particle filters is proposed. 
Third, the article develops an efficient auxiliary disturbance particle filter, which
is necessary when the bootstrap disturbance filter is inefficient, but the state transition density cannot be expressed in closed form. Fourth, a novel sorting algorithm, which is as effective as previous approaches but significantly faster than them, is developed to preserve the correlation between the logs of the likelihood estimates at the current and proposed parameter values.
The performance of the sampler  is investigated empirically by applying it to non-linear Dynamic Stochastic General Equilibrium models with relatively high state dimensions and with intractable  state transition densities and to multivariate stochastic volatility in the mean models.
Although our focus is on applying the  method to state
space models, the approach will be useful in a wide range of applications
 such as large panel data models and stochastic differential
equation models with mixed effects.
\end{abstract}

%\section*{RK comment}
%\begin{itemize}
%    \item \lq  to
%carry out Bayesian inference  \rq{}  to  \lq to
%Bayesian inference\rq 
%\item \lq complex and high-dimensional state space models
%having the following features\rq{} to \lq complex  state space models and has
% the following features.\rq 
 
% I dropped the high dimensional here, because we don't do really high dimensional, we mention higher dimensional States; third, it should be high dimensional states as high dimensional can also mean lots of parameters.   
% \item \lq (trimmed) means\rq{} to  \lq trimmed means\rq{} 
% \item \lq Second, it combines
%block and correlated PMMH sampling\rq{} to \lq Second, it uses a block version of PMMH\rq{} 
%\item   \lq  These first two  \rq{} to \lq These two \rq{} 
%\item \lq and multivariate stochastic volatility model\rq{} to \lq and to multivariate stochastic volatility in the mean models\rq{} 
%\end{itemize}
%\section*{RK end comment}

%\begin{center}

%\section*{RK comment}
%\begin{itemize}
%    \item \lq Keywords: Block PMMH;
%correlated PMMH; delayed acceptance; dynamic stochastic general equilibrium (DSGE) model;
%trimmed mean likelihood estimate\rq 
%\item 
%get rid of the centering for the keywords
%\end{itemize}
%\section*{RK end comment}

\blfootnote{
	\textsuperscript{$\ddagger$}
	Level 4, West Lobby, School of Economics, University of New South Wales Business School -- building E-12, Kensington Campus, UNSW Sydney -- 2052, \textit{Email:} {pratiti.chatterjee@unsw.edu.au}, \textit{Phone Number:} {(+61) 293852150}. Website: {http://www.pratitichatterjee.com}\\
	\textsuperscript{$\star$} 39C. 164, School of Mathematics and Applied Statistics (SMAS), University of Wollongong, Wollongong, 2522; Australian Center of Excellence for Mathematical and Statistical Frontiers (ACEMS); National Institute for Applied Statistics Research Australia (NIASRA); \textit{Email}: dgunawan@uow.edu.au. \textit{Phone Number:} {(+61) 424379015}. \\
	\textsuperscript{$\star\star$} 	Level 4, West Lobby, School of Economics, University of New South Wales Business School -- Building E-12, Kensington Campus, UNSW Sydney -- 2052, and ACEMS \textit{Email:} {r.kohn@unsw.edu.au}, \textit{Phone Number:} {(+61) 424802159}.}

\section{Introduction \label{sec:Introduction-1}}
Particle marginal Metropolis-Hastings (PMMH) \citep{Andrieu:2009}  is a Bayesian inference method for  statistical models  having an intractable  likelihood, when the likelihood
can be estimated unbiasedly. Our article  develops  a PMMH sampling method for efficiently estimating the parameters of complex state space models. The method scales better than current approaches 
 in the number of observations and latent states and can handle  state transition densities that cannot be expressed in closed form; e.g.,
many dynamic stochastic general equilibrium (DSGE) models, which are a popular class of macroeconomic time series state space models, do not have closed form transition densities.

A key issue in efficiently estimating statistical models using a PMMH approach is that the variance of the log of
the estimated likelihood grows  with the number of observations and the dimension of the latent states \citep{Deligiannidis2018}. \citet{Pitt:2012} show that to obtain a
balance between computational time and the mixing of the Markov chain Monte Carlo
(MCMC) chain, the
number of particles used in the particle filter should be such that the variance
of the log of the estimated likelihood is in the range 1 to 3, depending on the efficiency of the proposal for $\theta$.  \citet{Pitt:2012} also show that the efficiency
of PMMH schemes deteriorates exponentially as that variance increases; we further note that 
in many complex statistical applications, it is computationally very expensive ensuring that the variance of the log of the estimated likelihood is within the required range.

\citet{Deligiannidis2018}  propose a more
efficient PMMH scheme, called the correlated pseudo-marginal (CPM) method,
which correlates the random numbers in the (single) particle filters used to
construct the estimated
likelihoods at the current and proposed values of the parameters. This
dependence induces a positive correlation between
the estimated likelihoods and reduces the variance of the difference
in the logs of the estimated likelihoods which appear in the Metropolis-Hastings
(MH) acceptance ratio. They show that the CPM scales up with
the number of observations compared to the standard pseudo marginal
method of \citet{Andrieu:2010} when the state dimension is small.

\citet{Tran2016} propose an alternative correlated pseudo marginal approach to that of 
\citeauthor{Deligiannidis2018}, calling it 
the block pseudo marginal (BPM) method;  the BPM divides the random
numbers used in constructing likelihood estimators into blocks and then updates the parameters jointly with one randomly chosen
block of the random numbers in each MCMC iteration; this
induces a positive correlation between the numerator and denominator
of the MH acceptance ratio, similarly to the CPM. They show that for large samples the
correlation of the logs of the estimated likelihoods at the current
and proposed values is close to $1-1/S$, where $S$ is the number
of blocks. However, they do not apply the BPM method for estimating time series state space models using the particle filter.

Our article introduces a new PMMH sampler, referred to as the multiple PMMH algorithm (MPM), which extends the CPM method of Deligiannidis (2018) and the BPM method of Tran (2016). The MPM sampler is innovative and addresses various issues in the following ways: (a)~The likelihood estimator is a trimmed mean of unbiased likelihood estimators from multiple independent particle filters (PFs); these PFs can be run in parallel.  This approach reduces the variance of the likelihood estimator compared to using a single particle filter. The algorithm is exact when there is no trimming and approximate otherwise, but our empirical results suggest that the approximation is very close to being exact. (b)~The unknown parameters, and only the random numbers used in one of the PFs, are updated jointly. This is a novel block version of PMMH that works with multiple particle filters. See section~\ref{subsec:Multiple-Particle-Filter} for details. (c)~An auxiliary disturbance PF (ADPF) algorithm is proposed to estimate the likelihood efficiently for state space models such as DSGE models, where the bootstrap filter is very inefficient and the state transition density does not have a closed form so that an auxiliary PF cannot be constructed; see section \ref{subsec:Disturbance-Particle-Filter} for details. (d)~A novel sorting algorithm, which is as effective as previous approaches but is significantly faster than them, is proposed to maintain the correlation between the logs of the likelihood estimates at the current and proposed values. The standard resampling step in the particle filter introduces discontinuities that break down the correlation between the logs of the likelihood estimates at the current and proposed values, even when the current and proposed parameters are close. Section \ref{subsec:Linear-Gaussian-State Space Model} shows that the MPM sampler with the proposed sorting algorithm can maintain the correlation between the logs of the estimated likelihoods for relatively high dimensional state space models. Note that the proposed PMMH sampler works with any particle filter algorithm, including the tempered particle filter of \citet{Herbst2019}, the standard bootstrap filter of \citet{Gordon:1993}, the disturbance particle filter of \citet{Murray2013b}, and the proposed auxiliary disturbance filter. When the number of state dimensions is greater than the number of disturbance dimensions, a disturbance particle filter is more efficient than a state-based particle filter. Additionally, the disturbance particle filter is useful for state space models with intractable state transition density.

%\section*{RK comment}
%\begin{itemize}
%    \item \lq call the mixed PMMH algorithm (MPM)\rq{} to \lq call the multiple PMMH algorithm (MPM)\rq{} 
    
%    David, I am unsure what mixed stands for in the current context. Multiple to me would be using multiple particle filters. 
%    \item 
%    To do or discuss. For a trimmed mean, we can work out the variance of the log of the trimmed mean. One way is by the bootstrap. 
    
%    We can also assume that the correlation between numerator and denominator is $1-1/S$. Hence, assuming log of the trimmed mean is normal, we can do a bias correction. David, check this argument out. 
%\end{itemize}
%\section*{RK end comment}

%\end{document}

%(d) Section \ref{subsec:Dynamic-Stochastic-GeneralModel} develops a delayed-acceptance version \citep{Christen2005} of the
%MPM algorithm  to speed up the computation for
%DSGE models. The motivation for the delayed acceptance
%algorithm is to minimize the expensive computation of the  likelihood or its estimate when it appears likely that the proposal will be rejected.

%We also compare the performance of the auxiliary disturbance particle filter (ADPF) to the tempered particle
%filter (TPF) proposed by \citet{Herbst2019}; however, we note that \citet{Herbst2019} only apply the TPF to linear (first order) DSGE models. 

We illustrate  the MPM sampler empirically, and compare its performance to the CPM method,
using a standard linear Gaussian
state space model, a multivariate stochastic volatility in the mean model and two non-linear Dynamic Stochastic General Equilibrium (DSGE) models,
using simulated and real data.
Our work in estimating non-linear DSGE models is also related to \citet{FVVRR2007} and \citet{Hall2014} who use standard PMMH methods.
We note that the MPM sampler will  also be useful for other complex statistical models, such as  panel data models and stochastic differential equation mixed
effects models.

The rest of the article is organised as follows. Section \ref{sec:statespacemodel}
introduces  the state space model and gives some examples. Section
\ref{subsec:Multiple-Particle-Filter} discusses the MPM sampler. Section
\ref{sec:Examples} presents results from both simulated and real
data.   Section \ref{sec:Conclusions} concludes with a discussion of our major results and findings.
The paper  has an online supplement
containing some further technical and empirical results.

%Sections \ref{subsec:Linear-Gaussian-State Space Model} to \ref{subsec:SecondOrderSmallScale} show that: (1) the ADPF is much more efficient than the standard bootstrap particle filter (BPF) and the tempered particle filter (TPF); (2) the MPM, which runs multiple independent particle filters and only updates the random numbers used in one particle filter, maintains the correlation between the logs of the estimated likelihoods much better than both CPM and BPM. The BPM method of \citet{Tran2016}, which runs a single particle filter and updates a block of random numbers for all time periods in the particle filter algorithm, is unsuitable for time-series state space model;   (3) the delayed acceptance version of the MPM sampler is much more efficient than the standard MPM sampler.

\section{Bayesian Inference for State Space Models \label{sec:statespacemodel}}

%\subsection{Notation\label{SS:general SS models}}
\subsection*{Notation}
We use the colon notation for collections of variables, i.e., $a_{t}^{r:s}:=\left(a_{t}^{r},...,a_{t}^{s}\right)$
and for $t\leq u$, $a_{t:u}^{r:s}:=\left(a_{t}^{r:s},...,a_{u}^{r:s}\right)$.
Suppose $\left\{ \left(Z_{t},Y_{t}\right), t \geq 0 \right\} $ is a stochastic process,
with parameter $\theta$, where the $Y_{t}$ are the observations and
the $Z_{t}$ are  the latent state  vectors; random variables are denoted by capital letters and their realizations by lower case letters.  We consider the state space model
with $p\left(z_{0}|\theta\right)$ the density of $Z_{0},$
$p\left(z_{t}|z_{t-1},\theta\right)$  the density
 of $Z_{t}$ given $Z_{0:t-1}=z_{0:t-1}$
for $t\geq1$,  and $p\left(y_{t}|z_{t},\theta\right)$
is  the density of $Y_{t}$ given $Z_{0:t}=z_{0:t}$,
$Y_{1:t-1}=y_{1:t-1}$.

%=y_t
%=z_t
%\bibliographystyle{apalike}
%\bibliography{references_v1}
%\end{document}
%p\left(\theta\right)
%\section{Bayesian Inference}

%f
% a higher-order approximation is used,  such as  \Eqref{equation:dsge4} in Section~\ref{DescriptionOfPrunedDSGEModel},
%for the evolution of the state and other

\subsection{Bayesian Inference \label{sec:Bayesian-Inference}}

The objective of Bayesian inference is to obtain the posterior distribution
of the model parameters $\theta$ and the latent states $z_{0:T}$, given
the observations $y_{1:T}$ and a prior distribution $p\left(\theta\right)$;
i.e.,
\begin{eqnarray}
p\left(\theta,z_{0:T}|y_{1:T}\right) & = & p\left(y_{1:T}|\theta,z_{0:T}\right)p\left(z_{0:T}|\theta\right)
p\left(\theta\right)/p\left(y_{1:T}\right),\label{eq:posteriorwithstates}
\end{eqnarray}
where
\begin{equation*}
p\left(y_{1:T}\right)=\int\int p\left(y_{1:T}|\theta,z_{0:T}\right)p\left(z_{0:T}|\theta\right)p\left(\theta\right)dz_{0:T}d\theta
\end{equation*}
is the marginal likelihood. The likelihood
\[
p\left(y_{1:T}|\theta\right)=\int p\left(y_{1:T}|\theta,z_{0:T}\right)p\left(z_{0:T}|\theta\right)dz_{0:T}
\]
can be calculated exactly using the Kalman filter for linear Gaussian
state space models (LGSS) and for  linear DSGE models so that posterior samples can
be obtained using an MCMC sampling
scheme. However, the likelihood can be estimated unbiasedly, but not computed exactly, for non-linear and non-Gaussian
state space models and the non-linear DSGE models described in 
section~\ref{subsec:Dynamic-Stochastic-GeneralModel}.

%the likelihood cannot be computed exactly for non-linear and non-Gaussian
%state space models and the non-linear DSGE models described in Section \ref{subsec:Dynamic-Stochastic-GeneralModel}. In these cases, the likelihood can only be estimated

%\footnote{I suggest "The likelihood can be estimated unbiasedly, but not computed exactly, for non-linear and non-Gaussian
%state space models and the non-linear DSGE models described in 
%section~\ref{subsec:Dynamic-Stochastic-GeneralModel}}.

%\Eqref{eq:posteriorwithstates}

The bootstrap particle filter \citep{Gordon:1993} provides an unbiased
estimator of the likelihood for a general state space model. \citet{Andrieu:2009}
and \citet{Andrieu:2010} show that it is possible to use this unbiased
estimator of the likelihood to carry out exact Bayesian inference for
the parameters of the general state space model. They call this MCMC approach pseudo
marginal Metropolis-Hastings (PMMH).

The non-linear (second-order) DSGE models considered in section~\ref{subsec:Dynamic-Stochastic-GeneralModel} lie within the class
of general non-linear state space models whose state transition density is difficult to work with or cannot be expressed in closed form. In such cases, it is useful to express
the model in terms of the density of its latent disturbance variables as it can be expressed in closed form. The posterior in Eq. \eqref{eq:posteriorwithstates} 
becomes
\begin{equation}
\label{disturbamcefilterposterior}
p\left(\theta,\epsilon_{1:T},z_0|y_{1:T}\right)\propto
\prod_{t=1}^{T}p\left(y_{t}|\epsilon_{1:t},\theta,z_0\right)p\left(\epsilon_{t}\right)p\left(z_0|\theta\right)p\left(\theta\right),
\end{equation}
which \citet{Murray2013b} call  the disturbance state-space model.
The standard state space model can be recovered from the disturbance state space model by using
the deterministic function $F\left(z_{t-1},\epsilon_{t};\theta\right)\rightarrow z_{t}$.
This gives us a state trajectory $z_{0:T}$ from any sample $\left(\epsilon_{1:T},z_{0}\right)$.
In the disturbance state-space model the target becomes the posterior
distribution over the parameters $\theta$, the latent noise variables
$\epsilon_{1:T}$ and the initial state $z_0$, rather than $\left(\theta,z_{0:T}\right)$.
%\section*{RK comment}
%\begin{itemize}
%    \item \lq  The posterior in becomes    \rq{} to \lq The posterior in \eqref{eq:posteriorwithstates} becomes \rq 
%\end{itemize}

%\section*{RK end comment}

%where the initial state $z_0=0$ (see Section \ref{subsec:Dynamic-Stochastic-GeneralModel} for further details)

\subsection{Standard Pseudo Marginal Metropolis-Hastings \label{subsec:Standard-Pseudo-Marginal}}

This section outlines the standard PMMH scheme
which  carries out MCMC
on an expanded space using an unbiased estimate of the likelihood.
% instead of the likelihood  $p\left(y|\theta\right)$.
Our paper focuses on MCMC based on the parameters, but it is straightforward to also use it to obtain posterior inference for the unobserved states; see, e.g., \cite{Andrieu:2010}. 
Let   $u$ consist of all the random variables required to compute the unbiased likelihood estimate
$\widehat{p}_{N}\left(y|\theta,u\right)$,
with $p(u)$ the  density of $u$; let $p\left(\theta\right)$ be the prior of $\theta$.
The joint posterior density $\theta$ and $u$ is
\begin{equation*}
p\left(\theta,u|y_{1:T}\right)=\widehat{p}_{N}\left(y_{1:T}|\theta,u\right)p\left(\theta\right)p\left(u\right)/p\left(y_{1:T}\right),
\end{equation*}
so that
\[
p\left(\theta|y_{1:T}\right)=\int p\left(\theta,u|y_{1:T}\right)du=p\left(y_{1:T}|\theta\right)p\left(\theta\right)/p\left(y_{1:T}\right)
\]
is the posterior of $\theta$ and $\intop\widehat{p}_{N}\left(y_{1:T}|\theta,u\right)p\left(u\right)du=p\left(y_{1:T}|\theta\right)$
because the likelihood estimate is unbiased. We can therefore sample from the posterior
density $p\left(\theta|y_{1:T}\right)$ by sampling $\theta$ and
$u$ from $p\left(\theta,u|y_{1:T}\right)$. The subscript $N$ indicates the number of particles used to estimate the likelihood.
%\section*{RK comment}
%\begin{itemize}
%    \item \lq Our  paper focuses on  estimating the posterior density of the parameters $\theta$,
%but not of the states.\rq{} to 

%\lq Our paper focuses on MCMC based on the parameters, but it is straightforward to also obtain posterior inference for the unobserved states; see, e.g., ...\rq{} 

%need a reference here. We should make it clear that it is straightforward to obtain posterior inference for the States.

%DG: Note that generating the states using multiple particle filter has not been done before, so not clear to me who to cite. That paper \cite{Andrieu:2010} shows how to also obtain infrernces on states using PMMH. 

%I know then you are going to say that we can do it in two steps. First, generate the parameters, then generate the states given the parameters. And you can generate states from multiple particle filters.
%But then, another question that needs to be investgated is that how do we combine the multiple states generated from multiple particle filters, of course, we may be able to take the mean of median of states at each time period.

%Another way is to just remove the sentence completely.
%\end{itemize}
%\section*{RK end comment}

Let $q\left(\theta^{'}|\theta\right)$ be the proposal density for $\theta^{\prime}$
with the current value  $\theta$ and $q\left(u^{'}|u\right)$ the proposal
density for $u^{'}$ given $u$. We always assume that $q\left(u^{'}|u\right)$
satisfies the reversibility condition
\begin{equation*}
q\left(u^{'}|u\right)p\left(u\right)=q\left(u|u^{'}\right)p\left(u^{'}\right);
\end{equation*}
it  is clearly satisfied by the standard PMMH where $q\left(u^{'}|u\right)=p\left(u^{'}\right)$.
We generate a proposal $\theta^{'}$ from $q\left(\theta^{'}|\theta\right)$
and $u^{'}$ from $q\left(u^{'}|u\right)$, and accept both proposals
with probability
\begin{eqnarray}
\alpha\left(\theta,u;\theta^{'},u^{'}\right) & = & \min\left(1,\frac{\widehat{p}_{N}\left(y|\theta^{'},u^{'}\right)p\left(\theta^{'}\right)p\left(u^{'}\right)q\left(\theta|\theta^{'}\right)q\left(u|u^{'}\right)}{\widehat{p}_{N}\left(y|\theta,u\right)p\left(\theta\right)p\left(u\right)q\left(\theta^{'}|\theta\right)q\left(u^{'}|u\right)}\right)\nonumber \\
 & = & \min\left(1,\frac{\widehat{p}_{N}\left(y|\theta^{'},u^{'}\right)p\left(\theta^{'}\right)q\left(\theta|\theta^{'}\right)}{\widehat{p}_{N}\left(y|\theta,u\right)p\left(\theta\right)q\left(\theta^{'}|\theta\right)}\right).\label{eq:acceptancePMMH}
\end{eqnarray}
The expression in \Eqref{eq:acceptancePMMH} is identical
to a standard Metropolis-Hastings algorithm, except that estimates
of the likelihood at the current and proposed parameters are used.
\citet{Andrieu:2009}  show that the resulting
PMMH algorithm has the correct invariant distribution regardless of
the variance of the estimated likelihoods. However, the performance
of the PMMH approach crucially depends on the number of particles $N$ used to
estimate the likelihood. The variance
of the log of the estimated likelihood should be between 1 and 3 depending on the quality
of the proposal
for  $\theta$; see \citet{Pitt:2012} and
\citet{Sherlock2015}. In many
applications of the non-linear state space models considered in this
paper, it is computationally very expensive to ensure that the variance
of the log of the estimated likelihood is within the required range.
Section \ref{subsec:Multiple-Particle-Filter} discusses the new PMMH sampler, which we call the multiple PMMH algorithm (MPM),  that builds on and extends, the
CPM of \citet{Deligiannidis2018} and BPM
of \citet{Tran2016}.

%similarly
% to the approach of \citet{Sherlock2017}

%Suppose that  $G$ particle filters are run in parallel. Let $\widehat{p}_{N}\left(y|\theta,u_{g}\right)$
%be the unbiased estimate of the likelihood obtained from the $g$th particle filter, for
%$g=1,...,G$.

%=
%\sum_{g=1}^{G}\widehat{p}_{N}\left(y|\theta,u_{g}\right)/G

%average of the $G$ unbiased likelihood estimates and hence it is
%also unbiased \citep{Sherlock2017}

\subsection{Multiple PMMH (MPM) \label{subsec:Multiple-Particle-Filter}}

This section discusses Algorithm \ref{alg:MPM-PMMH}, which is the proposed multiple
PMMH (MPM) method that uses multiple particle filters to obtain
the estimate of the likelihood. The novel version of block-correlated PMMH that works with multiple particle filters to induce a
high correlation between successive logs of the estimated likelihoods
is also discussed. 
We now define the joint target density of $\theta$
and $\widetilde{u}=\left(u_{1},...,u_{S}\right)$ as
\begin{equation}
p\left(\theta,\widetilde{u}|y_{1:T}\right)\propto
\overline{\widehat{p}}_{N}\left(y|\theta,\widetilde{u}\right)
p\left(\theta\right)\prod_{s=1}^{S}p\left(u_{s}\right),
\end{equation}
where
%\[
$\overline{\widehat{p}}_{N}\left(y|\theta,u\right)$
%\]
is the likelihood estimates obtained from $S$ particle filters discussed below. We update the parameters $\theta$
jointly with a randomly selected block $u_{s}$ in each MCMC iteration,
with $\Pr\left(S=s\right)=1/S$ for any $s=1,...,S$. The selected block $u_{s}$ is updated using
$u_{s}^{'}=\rho_{u}u_{s}+\sqrt{1-\rho_{u}^{2}}\eta_{u}$, where $\rho_{u}$
is the non-negative correlation between the random numbers $u_{s}$
and $u_{s}^{'}$ and $\eta_{u}$ is a standard normal vector of the
same length as $u_{s}$. This is similar to component-wise MCMC
whose convergence is well established in the literature; see, e.g. \citet{johnson2013component}. Using this scheme, the acceptance probability
is
\begin{equation}
\alpha\left(\theta,\widetilde{u};\theta^{'},
\widetilde{u}^{'}\right)=\min\left(1,\frac{\overline{\widehat{p}}_{N}
\left(y|\theta^{'},\widetilde{u}^{'}=\left(u_{1},...,u_{s-1},u_{s}^{'},u_{s+1},...,u_{S}\right)
\right)p\left(\theta^{'}\right)q\left(\theta|\theta^{'}\right)}{\overline{\widehat{p}}_{N}
\left(y|\theta,\widetilde{u}=\left(u_{1},...,u_{s-1},u_{s},u_{s+1},...,u_{S}\right)\right)
p\left(\theta\right)q\left(\theta^{'}|\theta\right)}\right).
\label{MHacceptance}
\end{equation}

\begin{algorithm}[H]
\caption{The Multiple PMMH (MPM) algorithm \label{alg:MPM-PMMH}}
\begin{itemize}
\item Set the initial values of $\theta^{\left(0\right)}$ arbitrarily.
\item Sample $u_s\sim N\left(0,I\right)$ for $s=1,...,S$, and run $S$ particle filters to
 estimate the likelihood $\overline{\widehat{p}}_{N}\left(y|\theta,\widetilde{u}\right)$ as the trimmed mean of $\widehat{p}_{N}\left(y|\theta,u_{s}\right), s=1, \dots, S$; a 0\% is the mean and a 50\% trimmed mean is the median.
\item For each MCMC iteration $p$, $p=1,...,P$,
\begin{itemize}
\item Sample $\theta^{'}$ from the proposal density $q\left(\theta^{'}|\theta\right)$.
\item Choose index $s$ with probability $1/S$, sample $\eta_{u}\sim N\left(0,I\right)$, and set $u_{s}^{'}=\rho_{u} u_{s}+\sqrt{1-{\rho_{u}}^{2}}{\eta_{u}}$.
\item Run $S$ particle filters to compute
the estimate of likelihood $\overline{\widehat{p}}_{N}\left(y|\theta^{'},\widetilde{u}^{'}\right)$.
\item With the probability in Eq. \eqref{MHacceptance},
set $\overline{\widehat{p}}\left(y_{1:T}|\theta,\widetilde{u}\right)^{\left(p\right)}=\overline{\widehat{p}}\left(y_{1:T}|\theta^{'},\widetilde{u}^{'}\right)$, $\widetilde{u}^{(p)}=\widetilde{u}^{'}$, and $\theta^{\left(p\right)}=\theta^{'}$; otherwise, set
$\overline{\widehat{p}}\left(y_{1:T}|\theta,\widetilde{u}\right)^{\left(p\right)}=\overline{\widehat{p}}\left(y_{1:T}|\theta,\widetilde{u}\right)^{\left(p-1\right)}$,  $\widetilde{u}^{(p)}=\widetilde{u}^{(p-1)}$, and $\theta^{\left(p\right)}=\theta^{\left(p-1\right)}$.
\end{itemize}
\end{itemize}
\end{algorithm}

%\section*{RK comment}
%\begin{itemize}
%    \item \lq  proposed mixed
%PMMH (MPM) method    \rq{} 

%see my earlier comment on replacing \lq mixed
%PMMH\rq{} by \lq multiple PMMH\rq{} 
%\item \lq Sample $u_s\sim N\left(0,I\right)$ for $s=1,...,S$, and run $S$ particle filters to
% estimate the likelihood $\overline{\widehat{p}}_{N}\left(y|\theta,\widetilde{u}\right)=\frac{1}{S}\sum_{s=1}^{S}\widehat{p}_{N}\left(y|\theta,u_{s}\right)$.\rq{} 
 
%do you want \lq Sample $u_s\sim N\left(0,I\right)$ for $s=1,...,S$, and run $S$ particle filters to
%estimate the likelihood $\overline{\widehat{p}}_{N}\left(y|\theta,\widetilde{u}\right)$ as the trimmed mean of $\widehat{p}_{N}\left(y|\theta,u_{s}\right), s=1, \dots, S$; a 0\% is the mean and a 50\% trimmed mean is the median. \rq{} 
%\end{itemize}
%\section*{RK end comment}

%, and run ancestral tracing algorithm in Section~\ref{sec:AncestralTracing} after each particle filter algorithm to obtain the initial $G$ trajectories of $\epsilon_{g,1:T}$. The mean $\widehat{\mu}_{t}$ and the covariance matrix $\widehat{\Sigma}_{t}$ of the proposal defined in Section \ref{subsec:Disturbance-Particle-Filter} are set as the mean and the covariance matrix of these $G$ trajectories of $\epsilon_{g,1:T}$ at each time $t$.

%\subsection{Obtaining likelihood estimates from multiple particle filters\label{likelihoodmultiplefilters}}

%in Section \ref{subsec:Multiple-Particle-Filter}
%. This likelihood estimate is used

We now discuss approaches to obtain a likelihood estimate from multiple particle filters which we then use in the MPM algorithm. Suppose that $S$ particle filters are run in parallel. Let $\widehat{p}_{N}\left(y|\theta,u_{s}\right)$
be the unbiased estimate of the likelihood obtained from the $s$th particle filter, for
$s=1,...,S$. The first approach takes the average of the $S$ unbiased likelihood estimates from the particle filter. The resulting likelihood estimate 
\begin{equation}
\overline{\widehat{p}}_{N}\left(y|\theta,u\right)=
\sum_{s=1}^{S}\widehat{p}_{N}\left(y|\theta,u_{s}\right)/S
\end{equation}
%\]
is also unbiased \citep{Sherlock2017}. The second approach is to take the  trimmed mean as the likelihood estimate in MPM. For example, the $20\%$ trimmed mean averages the middle $60\%$ of the likelihood values.   Although the trimmed mean does not give an unbiased estimate of the likelihood, we show in  section \ref{sec:Examples} that the posterior based on the trimmed means approximates the exact posterior well. 

%$(100(1-\alpha))\%$
%for $\alpha = 0.2$, 
%\section*{RK comment}
%\begin{itemize}
%\item heading of section  \lq Obtaining likelihood estimates from multiple particle filters\rq {} 

%to \lq Using multiple particle filters to estimate the likelihod\rq 
%    \item  I suggest combining the previous and current sections. 
%\end{itemize}
%\section*{RK end comment}

%Section \ref{Multidimensional Sorting} proposes a novel multidimensional sorting algorithm.

%\footnote{I suggest putting this paragraph in the next section} 

%Algorithm \ref{alg:Multidimensional-Euclidean-Sorting}
%gives the proposed fast multidimensional Euclidean sorting algorithm.

\subsection{Multidimensional Euclidean Sorting Algorithm \label{Multidimensional Sorting}}

It is important that the
logs of the likelihood estimates evaluated at the current and proposed
values of $\theta$ and $\widetilde{u}$ are highly correlated to reduce
the variance of $\log\overline{\widehat{p}}_{N}\left(y|\theta^{'},\widetilde{u}^{'}\right)-\log\overline{\widehat{p}}_{N}\left(y|\theta,\widetilde{u}\right)$ because this helps the Markov chain to mix well. However, the standard
resampling step in the particle filter introduces discontinuities which breaks down the correlation between the logs of the likelihood estimates at the current and proposed values
even when  the current parameters $\theta$ and the proposed parameters $\theta^{'}$ are close.
Sorting the particles from smallest to largest before resampling helps to preserve the correlation between the logs of the likelihood estimates at the current and proposed values
\citep{Deligiannidis2018}.
However, such simple sorting is unavailable for multidimensional state particles.

This section discusses Algorithm \ref{alg:Multidimensional-Euclidean-Sorting}, which is the fast multidimensional Euclidean sorting algorithm
used to sort the multidimensional state or disturbance
particles in the particle filter algorithm described in Algorithm~\ref{alg:The-Disturbance-Particle filter} in section \ref{Disturbance particle filter} of the online supplement. The algorithm is written in terms of state particles, but similar steps can be implemented for the disturbance particles.
The algorithm takes the particles $z_{t}^{1:N}$ and the normalised weights $\overline{w}_{t}^{1:N}$ as the inputs; it outputs the sorted particles $\widetilde{z}_{t}^{1:N}$, sorted weights
$\widetilde{\overline{w}}_{t}^{1:N}$, and sorted indices $\zeta_{1:N}$. 
Let $z_{t}^{i}$ be
$d$-dimensional particles at a time $t$ for the $i$th particle, $z_{t}^{i}=\left(z_{t,1}^{i},...,z_{t,d}^{i}\right)^{\top}$.
Let $d\left(z_{t}^{j},z_{t}^{i}\right)$ be the Euclidean distance
between two multidimensional particles. 
The first step is to calculate the means of the multidimensional particles
$\overline{z}_{t}^{i}=(\nicefrac{1}{d})\sum_{k=1}^{d}z_{t,k}^{i}$ at time $t$ for $i=1,...,N$. The first sorting index, $\zeta_1$, is the index of the multidimensional particle having the smallest value of $\overline{z}_{t}^{i}$ for $i=1,...,N$. Therefore, the first sorted particle $\widetilde{z}_{t}^{1}$ is $z_{t}^{\zeta_1}$ with its associated weight $\widetilde{\overline{w}}_{t}^{1}=\overline{w}_{t}^{\zeta_1}$.
We then calculate the Euclidean distance between the selected multidimensional particle and the set of all remaining multidimensional particles, $d\left(\widetilde{z}_{t}^{1},z_{t}^{i}\right)$, for all $i \neq \zeta_1$. The next step is to sort the particles and weights according to the Euclidean distance from smallest to largest to obtain the sorted particles $\widetilde{z}_{t}^{2:N}$, sorted weights $\widetilde{\overline{w}}_{t}^{2:N}$, and sorted indices $\zeta_{2:N}$.

Algorithm \ref{alg:Multidimensional-Euclidean-Sorting} simplifies the sorting algorithm proposed by \citet{Choppala2016}. In \citet{Choppala2016}, the first sorting index  is the index of the multidimensional particle having the smallest value along its first dimension. To select the second sorted index, $\zeta_{2}$, we need to calculate the Euclidean distance between the first selected particle and the remaining multidimensional particles. Similarly, to select the third sorted index, $\zeta_{3}$, we need to calculate the Euclidean distance between the second selected particle and the remaining multidimensional particles. This process is repeated $N$ times, which is expensive for a large number of particles and long time series. In contrast, Algorithm \ref{alg:Multidimensional-Euclidean-Sorting} only needs to calculate the Euclidean distance of the first selected particle and the remaining particles once. 
\citet{Deligiannidis2018} use the Hilbert sorting method of \citet{Skilling:2004} to order the multidimensional state particles, which requires calculating the Hilbert index for each multidimensional particle and is much more expensive than Algorithm \ref{alg:Multidimensional-Euclidean-Sorting}. Table \ref{sortingmethodcomparison} shows the CPU time (in seconds) of the three sorting methods for different numbers of particles $N$ and state dimensions $d$. The table shows that the proposed sorting algorithm is much faster than that in \citet{Choppala2016} and the Hilbert sorting method. For example, for state dimensions $d=30$ and $N=2000$ particles, Algorithm \ref{alg:Multidimensional-Euclidean-Sorting} is $290$ and $335$ times faster than the \citet{Choppala2016} and the Hilbert sorting methods, respectively

%\footnote{This is just a question and you don't need to do anything about it now. But is the quality of the sorting different for the three algorithms; and how do we measure the quality of the sorting?} . 

\begin{algorithm}[H]

\caption{Multidimensional Euclidean Sorting Algorithm \label{alg:Multidimensional-Euclidean-Sorting}}

Input: $z_{t}^{1:N}$, $\overline{w}_{t}^{1:N}$

Output: sorted particles $\widetilde{z}_{t}^{1:N}$, sorted weights
$\widetilde{\overline{w}}_{t}^{1:N}$, sorted indices $\zeta_{1:N}$
\begin{itemize}
\item Calculate the means of the multidimensional particles $\overline{z}_{t}^{i}=\frac{1}{d}\sum_{k=1}^{d}z_{t,k}^{i}$ at time $t$ for $i=1,...,N$. The first sorting index, $\zeta_1$, is the index of the multidimensional particle having the smallest value of $\overline{z}_{t}^{i}$ for $i=1,...,N$. The first selected particle $\widetilde{z}_{t}^{1}$ is $z_{t}^{\zeta_1}$ with its associated weight $\widetilde{\overline{w}}_{t}^{1}=\overline{w}_{t}^{\zeta_1}$.
\item Calculate the Euclidean distance between the selected multidimensional particle and the set of all remaining multidimensional particles, $d\left(\widetilde{z}_{t}^{1},z_{t}^{i}\right)$, for $\forall i \neq \zeta_1$. 
\item Sort the particles and weights according to the Euclidean distance from smallest to largest to obtain $\zeta_i$ for $i=2,...,N$. The sorted particles $\widetilde{z}_{t}^{i}=z_{t}^{\zeta_i}$ and $\widetilde{\overline{w}}_{t}^{i} =\overline{w}_{t}^{\zeta_i}$ for $i=2,...,N$.
\end{itemize}
\end{algorithm}

\begin{table}[H]
\caption{Comparing different sorting methods for various combinations of  
$N$ and $d$ in terms of CPU time (in seconds):
I: The fast Euclidean sorting algorithm, II: \citet{Choppala2016} sorting
method, and III: Hilbert sorting method. The entries in columns I to III are multiples of the time in column I. The column headed \lq \lq $I$ actual\rq\rq{}  is the actual time for column I; e.g. the 30.5 in column II and  row 1 means that I is 30.5 times  faster than the \citet{Choppala2016} sorting. The corresponding entry for 
I actual is 0.0002. \label{sortingmethodcomparison}}

\centering{}%
\begin{tabular}{cccccc}
\hline 
$d$ & $N$ & $I$ & $II$ & $III$ & $I$ \textrm{actual} \tabularnewline
\hline 
10 & 500 & 1.00 & 30.50 & 87.50 & 0.0002\tabularnewline
 & 1000 & 1.00 & 66.67 & 115.00 & 0.0003\tabularnewline
 & 2000 & 1.00 & 172.40 & 141.00 & 0.0005\tabularnewline
\hline 
30 & 500 & 1.00 & 39.67 & 141.00 & 0.0003\tabularnewline
 & 1000 & 1.00 & 113.00 & 214.00 & 0.0004\tabularnewline
 & 2000 & 1.00 & 289.80 & 335.00 & 0.0005\tabularnewline
\hline 
\end{tabular}
\end{table}

%provides an unbiased estimate of the likelihood and

\subsection{The Auxiliary Disturbance Particle Filter \label{subsec:Disturbance-Particle-Filter} }

This section  discusses the auxiliary disturbance particle filter we use to
obtain the estimates of the likelihood in the MPM sampler described in
section~\ref{subsec:Multiple-Particle-Filter}. It is particularly useful for state space models when the state dimension is substantially bigger than the disturbance dimension and the state transition density is intractable.
Suppose that $z_{1}=\Phi\left(z_{0},\epsilon_{1};\theta\right)$,
where $z_0$ is the initial state vector with density $p(z_0|\theta)$, and   $z_{t}=F\left(z_{t-1},\epsilon_{t};\theta\right) ( t \geq 2) $,
where $\epsilon_{t}$ is a $n_{e}\times1$ vector of normally distributed
latent noise with density $p\left(\epsilon_{t}\right). $
\citet{Murray2013b}
express the standard state-space model in terms of the latent noise
variables $\epsilon_{1:T}$, and call
\[
\epsilon_{t}\sim p\left(\epsilon_{t}\right),y_{t}|\epsilon_{1:t}\sim p\left(y_{t}|\epsilon_{1:t},z_{0};\theta\right)=p\left(y_{t}|z_{t};\theta\right),\,\, t=1,...,T,
\]
 the disturbance state-space model. We note that the conditional distribution of $y_{t}$
depends on all the latent error variables,  $\epsilon_{1:t}.$

%\subsection*{Auxiliary Disturbance Particle Filter
%(ADPF) \label{subsec:Proposals-Distribution}}
%\section*{RK comment}
%\begin{itemize}
%    \item  do we need the heading \lq Auxiliary Disturbance Particle Filter
%(ADPF)\rq{} as it is also the heading of section 2.6? I suggest dropping it. 
%\end{itemize}
%\section*{RK end comment}

%\footnote{I suggest adding "See algorithm~\ref{alg:Constructing-proposal-for ADPF}}.

We now discuss the proposal for $\epsilon_t$ used in the disturbance filter. The defensive mixture proposal density \citep{Hesterberg1995} is 
\begin{equation}
m\left(\epsilon_{t}|u_{\epsilon,t},\theta\right)=\pi p\left(\epsilon_{t}|\theta\right)+(1-\pi) q\left(\epsilon_{t}|\theta,y_{1:T}\right), \quad {\rm with }\quad 0 <  \pi \ll 1,
\label{eqnprop}
\end{equation}
where $u_{\epsilon,t}$ is the vector
random variable used to generate the particles $\epsilon_{t}$
given $\theta$.
If the observation density $p\left(y_{t}|z_{t},\theta\right)$ is bounded, then   \Eqref{eqnprop} guarantees that the weights are bounded in the disturbance particle filter algorithm defined in \Eqref{importanceweights} of section~\ref{Disturbance particle filter} of the online supplement.
In practice, we take
$q\left(\epsilon_{t}|\theta,y_{1:T}\right)
=N\left(\epsilon_{t}|\widehat{\mu}_{t},\widehat{\Sigma}_{t}\right)$ and
set $\pi=0.05$ in all the examples in section~\ref{sec:Examples}; see algorithm~\ref{alg:Constructing-proposal-for ADPF}.
If $\pi = 1$ in \Eqref{eqnprop}, then $m\left(\epsilon_{t}\right) = p\left(\epsilon_{t}\right)$
is the bootstrap disturbance particle filter. However, the empirical performance of this
filter is usually poor because the resulting likelihood estimate is too variable.

%Algorithm \ref{alg:Constructing-proposal-for ADPF} discusses how we obtain $\widehat{\mu}_{t,p}$ and the covariance matrix
%$\widehat{\Sigma}_{t,p}$, for $t=1,...,T$, at iteration $p$ of the MPM algorithm. It is computationally cheap to obtain $\widehat{\mu}_{t,p}$ and $\widehat{\Sigma}_{t,p}$ for $t=1,...,T$ because they are obtained from the output of the $S$ disturbance particle filters run at iteration $p$ and the ancestral tracing method is fast. The proposal mean $\widehat{\mu}_{t,p}$ and the covariance matrix
%$\widehat{\Sigma}_{t,p}$, for $t=1,...,T$, obtained at iteration $p$ is used to estimate the likelihood at iteration $p+1$. At iteration $p=1$, we use the bootstrap filter to initialise the mean $\widehat{\mu}_{t,1}$ and the covariance matrix $\widehat{\Sigma}_{t,1}$, for $t=1,...,T$. Algorithm \ref{alg:MPM-PMMH-ADPF} gives the MPM algorithm with ADPF. 

Algorithm \ref{alg:Constructing-proposal-for ADPF} discusses how we obtain $\widehat{\mu}_{t}$ and the covariance matrix
$\widehat{\Sigma}_{t}$, for $t=1,...,T$, at each iteration of the  MPM algorithm.  It is computationally cheap to obtain $\widehat{\mu}_{t}$ and $\widehat{\Sigma}_{t}$ for $t=1,...,T$ because they are obtained from the output of the $S$ disturbance particle filters run at each MCMC iteration  and the ancestral tracing method is fast. The proposal mean $\widehat{\mu}_{t}$ and the covariance matrix
$\widehat{\Sigma}_{t}$, for $t=1,...,T$, obtained at iteration $p$ are used to estimate the likelihood at the next iteration. At the first iteration, we use the bootstrap filter to initialise the mean $\widehat{\mu}_{t}$ and the covariance matrix $\widehat{\Sigma}_{t}$, for $t=1,...,T$. Algorithm \ref{alg:MPM-PMMH-ADPF} gives the MPM algorithm with ADPF.

%\footnote{I suggest "Algorithm \ref{alg:Constructing-proposal-for ADPF} discusses how we obtain $\widehat{\mu}_{t}$ and the covariance matrix
%$\widehat{\Sigma}_{t}$, for $t=1,...,T$, at each iteration of the  MPM algorithm.  It is computationally cheap to obtain $\widehat{\mu}_{t}$ and $\widehat{\Sigma}_{t}$ for $t=1,...,T$ because they are obtained from the output of the $S$ disturbance particle filters run at iteration  and the ancestral tracing method is fast. The proposal mean $\widehat{\mu}_{t}$ and the covariance matrix
%$\widehat{\Sigma}_{t}$, for $t=1,...,T$, obtained at each iteration $p$ are used to estimate the likelihood at the next iteration. At the first iteration, we use the bootstrap filter to initialise the mean $\widehat{\mu}_{t}$ and the covariance matrix $\widehat{\Sigma}_{t}$, for $t=1,...,T$. Algorithm \ref{alg:MPM-PMMH-ADPF} gives the MPM algorithm with ADPF." I suggest this change because the actual algorithm does not have $p$ in it and the initial mention of mean $\widehat{\mu}_{t}$ and the covariance matrix
%$\widehat{\Sigma}_{t}$ does not mention $p$ either.}  

\begin{algorithm}[H]
\caption{The Multiple PMMH (MPM) with ADPF algorithm \label{alg:MPM-PMMH-ADPF}}

%and to obtain the particles $\epsilon_{1:T}^{1:N}$, the ancestor indices $A_{1:T-1}^{1:N}$,
%and the weights $\overline{w}_{1:T}^{1:N}$.

\begin{itemize}
\item Set the initial values of $\theta^{\left(0\right)}$ arbitrarily.
%\item Sample $u_s\sim N\left(0,I\right)$ for $s=1,...,S$, run 

\item Sample $u_s\sim N\left(0,I\right)$ for $s=1,...,S$,  run $S$ (disturbance) particle filters to
estimate the likelihood $\overline{\widehat{p}}_{N}\left(y|\theta,\widetilde{u}\right)$ as the trimmed mean of $\widehat{p}_{N}\left(y|\theta,u_{s}\right), s=1, \dots, S$; a 0\% trimmed mean is the mean and a 50\% trimmed mean is the median, and run the algorithm \ref{alg:Constructing-proposal-for ADPF} to construct the (initial) proposal for the auxiliary disturbance particle filter.
 
 %The mean $\widehat{\mu}_{t}$ and the covariance matrix $\widehat{\Sigma}_{t}$ of the proposal defined in Section \ref{subsec:Disturbance-Particle-Filter} are set as the sample mean and the sample covariance matrix of these $S$ trajectories of $\epsilon_{s,1:T}$ at each time $t$.
\item For each MCMC iteration $p$, $p=1,...,P$,
\begin{itemize}
\item Sample $\theta^{'}$ from the proposal density $q\left(\theta^{'}|\theta\right)$.
\item Choose index $s$ with probability $1/S$, sample $\eta_{u}\sim N\left(0,I\right)$, and set $u_{s}^{'}=\rho_{u} u_{s}+\sqrt{1-{\rho_{u}}^{2}}{\eta_{u}}$.
\item Run $S$ (disturbance) particle filters to compute
the estimate of likelihood $\overline{\widehat{p}}_{N}\left(y|\theta^{'},\widetilde{u}^{'}\right)$
\item Run algorithm \ref{alg:Constructing-proposal-for ADPF} to construct the proposal for the auxiliary disturbance particle filter at iteration $p+1$.
\item With the probability in Eq. \eqref{MHacceptance},
set $\overline{\widehat{p}}\left(y_{1:T}|\theta,\widetilde{u}\right)^{\left(p\right)}=\overline{\widehat{p}}\left(y_{1:T}|\theta^{'},\widetilde{u}^{'}\right)$, $\widetilde{u}^{(p)}=\widetilde{u}^{'}$,  and $\theta^{\left(p\right)}=\theta^{'}$; otherwise, set
$\overline{\widehat{p}}\left(y_{1:T}|\theta,\widetilde{u}\right)^{\left(p\right)}=\overline{\widehat{p}}\left(y_{1:T}|\theta,\widetilde{u}\right)^{\left(p-1\right)}$,  $\widetilde{u}^{(p)}=\widetilde{u}^{(p-1)}$, and $\theta^{\left(p\right)}=\theta^{\left(p-1\right)}$.
%\item The mean $\widehat{\mu}_{t}$ and the covariance matrix $\widehat{\Sigma}_{t}$ are set as the sample mean and the sample covariance matrix of these $S$ trajectories of $\epsilon_{s,1:T}^{(i)}$ at each time $t$. , and $\epsilon_{1:T}^{(i)}=\epsilon_{1:T}^{'}$ , and $\epsilon_{1:T}^{(i)}=\epsilon_{1:T}^{(i-1)}$
\end{itemize}
\end{itemize}
\end{algorithm}

%The details of the MPM algorithm with ADPF are given in algorithm \ref{alg:MPM-PMMH-ADPF} in section \ref{MPM-PMMH-ADPF algorithm} of the online supplement.

%, initial or previous iteration estimates of $\widehat{\mu}_{t-1}$ and $\widehat{\Sigma}_{t-1}$

\begin{algorithm}[H]

\caption{Constructing proposal for the auxiliary disturbance particle filter
(ADPF) \label{alg:Constructing-proposal-for ADPF}}
Input: the number of particle filters $S$, the number of particles for each particle filter $N$, particles $\epsilon_{1:S,1:T}^{1:N}$, weights $\overline{w}_{1:S,1:T}^{1:N}$, and ancestor indices $a_{1:S,1:T-1}^{1:N}$.

Output: the mean $\widehat{\mu}_{t}$ and the covariance matrix $\widehat{\Sigma}_{t}$, for $t=1,...,T$.
\begin{enumerate}
%\item Run $S$ (disturbance) particle filters with estimates of $\widehat{\mu}_{t-1}$ and $\widehat{\Sigma}_{t-1}$ from iteration $t-1$ in parallel to give particles
%$\epsilon_{1:S,1:T}^{1:N}$, with weights $\overline{w}_{1:S,1:T}^{1:N}$ and
%ancestor indices $a_{1:S,1:T-1}^{1:N}$. See 
%section~\ref{Disturbance particle filter} of the online supplement for details. 
\item Given the particles $\epsilon_{1:S,1:T}^{1:N}$ with weights $\overline{w}_{1:S,1:T}^{1:N}$
and ancestor indices $a_{1:S,1:T-1}^{1:N}$ from the output of the disturbance
particle filters, the ancestral tracing algorithm of \cite{Kitagawa:1996} is used to sample from  particle approximations from the smoothing distribution $p(\epsilon_{1:T}|\theta,y_{1:T})$. This consists
of sampling one particle trajectory from each of the $S$ particle filters in parallel. For each particle filter, we first sample an index $j_s$ with the probability  
$\overline{w}_{s,T}^{j_s}$, tracing back its ancestral lineage $b_{s,1:T}^{j_s}\left(b_{s,T}^{j_s}=j_s\;\textrm{and}\;b_{s,t-1}^{j_s}=a_{s,t-1}^{b_{s,t}^{j_s}}\right)$
and choosing the particle trajectory $\epsilon_{s,1:T}^{j_s}=\left(\epsilon_{s,1}^{b_{s,1}^{j_s}},...,\epsilon_{s,T}^{b_{s,T}^{j_s}}\right)$ for $s=1,...,S$.

%$\epsilon_{s,1:T}^{b_{s,1:T}^{j}}=\left(\epsilon_{s,1}^{b_{s,1}^{j}},...,\epsilon_{s,T}^{b_{s,T}^{j}}\right)$,
%for $s=1,...,S$, from the smoothing distribution $p(\epsilon_{1:T}|\theta,y_{1:T})$,
%where $b_{s,1:T}^{j}$ is the ancestral lineage of the $s$th particle
%filter. 

%\item Given the particles $\epsilon_{s,1:T}^{1:N}$ with weights $w_{s,1:T}^{1:N}$
%and ancestor indices $a_{s,1:T-1}^{1:N}$ from the output of the disturbance
%particle filters, the ancestral tracing algorithm given in Section \ref{sec:AncestralTracing}
%of the online supplement is used to generate particle trajectories
%$\epsilon_{s,1:T}^{b_{s,1:T}^{j}}=\left(\epsilon_{s,1}^{b_{s,1}^{j}},...,\epsilon_{s,T}^{b_{s,T}^{j}}\right)$,
%for $s=1,...,S$, from the smoothing distribution $p(\epsilon_{1:T}|\theta,y_{1:T})$,
%where $b_{s,1:T}^{j}$ is the ancestral lineage of the $s$th particle
%filter. 
\item The mean $\widehat{\mu}_{t}$ and the covariance matrix $\widehat{\Sigma}_{t}$
are 
\[
\widehat{\mu}_{t}:=\frac{1}{S}\sum_{s=1}^{S}\epsilon_{s,t}^{b_{s,t}^{j_s}}, \quad 
\text{and} \quad 
\widehat{\Sigma}_{t}:=\frac{1}{S-1}\sum_{s=1}^{S}\left(\epsilon_{s,t}^{b_{s,t}^{j_s}}-\widehat{\mu}_{t}\right)\left(\epsilon_{s,t}^{b_{s,t}^{j_s}}-\widehat{\mu}_{t}\right)^{\top}.
\]
\end{enumerate}
\end{algorithm}

%\section*{RK comment}
%\begin{itemize}
%    \item  An aside only. We could use robust versions of $\wh \mu_t$ and $\wh \Sigma_t$. I think in general in all code we may want to use robust versions. 
%\end{itemize}
%\section*{RK end comment}

\section{Examples\label{sec:Examples}}
Section \ref{SS: preliminaries} discusses the inefficiency measures used
to compare the performance of different particle filters or PMMH samplers used in our article. Section \ref{subsec:Linear-Gaussian-State Space Model} investigates empirically the performance of the proposed methods for estimating a high-dimensional linear Gaussian state space model. Section \ref{SVinmeanexample} discusses a multivariate stochastic volatility in mean model with GARCH diffusion processes. Sections \ref{subsec:SecondOrderSmallScale} and \ref{mediumscaleDSGEmodelexample} apply the proposed samplers to estimate  non-linear small scale DSGE and medium scale DSGE models, respectively.

%linear small-scale DSGE example used by \citet{Herbst2019}. Section \ref{subsec:Non-linear-RBC} discusses a non-linear Real Business Cycle (RBC) using simulated datasets. Section \ref{subsec:SecondOrderSmallScale} applies the MPM sampler to estimate a non-linear small scale DSGE model.

\subsection{Definitions of Inefficiency\label{SS: preliminaries}}
%We define  the time normalised variance (TNV) of a particle filter method
%\begin{equation*}
%\textrm{TNV}_{PF}:=\widehat{\textrm{V}}
%\left(\log\widehat{p}\left(y|\theta\right)\right)\times\textrm{CT},
%\end{equation*}
%as the measure of inefficiency  of the method
%that takes computing time into account;
%$\textrm{CT}$ is the computing time to obtain a single log of the estimated likelihood  in seconds,
%and $\widehat{\textrm{V}}\left(\log\widehat{p}\left(y|\theta\right)\right)$
%is the estimated variance of the log of the likelihood estimate.
%The relative time normalised variance (RTNV) of a particle filter method is defined as $\textrm{RTNV}_{PF}:= \textrm{TNV}_{PF}/ \textrm{TNV}_{ADPF}$.

We use the inefficiency factor (IF) (also called the integrated autocorrelation time)
\[
\textrm{IF}_{\psi}:=1+2\sum_{j=1}^{\infty}\rho_{\psi}\left(j\right),
\]
to measure the inefficiency of a PMMH sampler at estimating the posterior expectation
of a univariate function $\psi\left(\theta\right)$ of $\theta$;
here, $\rho_{\psi}\left(j\right)$ is the $j$th autocorrelation of
the iterates $\psi\left(\theta\right)$ in the MCMC chain after it
has converged to its stationary distribution. We estimate the $\textrm{IF}_{\psi}$
using the CODA R package of \citet{Plummer2006}. A low value of the $\textrm{IF}_{\psi}$
estimate suggests that the Markov chain mixes well. Our measure of the inefficiency
of a PMMH sampler that takes computing time (CT) into account for a given parameter $\theta$ based on $\textrm{IF}_\psi$ is
the time normalised inefficiency factor (TNIF) defined as
$
\textrm{TNIF}_\psi := \textrm{IF}_\psi\times\textrm{CT}.
$
For a given sampler, let $\textrm{IF}_{\psi,\textrm{MAX}}$ and $\textrm{IF}_{\psi,\textrm{MEAN}}$
be the maximum and mean of the  IF values over all the
parameters in the model. The relative time normalized inefficiency factor (RTNIF) is a measure of the TNIF relative to the benchmark method, where the benchmark method depends on the example.

%\section*{RK comment}
%You also need to define RTINF as you use it later on. I think it means relative TINF
%\section*{RK end comment}

\subsection{Linear Gaussian State Space Model \label{subsec:Linear-Gaussian-State Space Model}}

%This section examines empirically the ability of the following methods to maintain the correlation between successive  values of the log of the estimated likelihood for a linear Gaussian state space model: (1) the block PMMH (BPM) of \citet{Tran2016}, (2) the correlated PMMH (CPM) of \citet{Deligiannidis2018},
%and (3) the mixed PMMH (MPM).

%This section reports the results of studies using data simulated from the linear Gaussian state space model. We consider the model  discussed in \citet{Guarniero2017}
%and \cite{Deligiannidis2018},  where $\left\{ X_{t};t\geq1\right\} $
%and $\left\{ Y_{t};t\geq1\right\} $ are $\R^{d}$ valued with
%\begin{eqnarray*}
%Y_{t} & = & X_{t}+W_{t},\\
%X_{t+1} & = & A_{\theta}X_{t}+V_{t+1},
%\end{eqnarray*}
%with $X_{1}\sim\N\left(0_{d},I_{d}\right)$, $V_{t}\sim N\left(0_{d},I_{d}\right)$,
%$W_{t}\sim N\left(0_{d},I_{d}\right)$, and $A_{\theta}^{i,j}=\theta^{|i-j|+1}$ for $i<j$;
%the true value of $\theta$ is $0.4$.  This study investigates (1) the efficiency of different approaches for obtaining likelihood estimates from $S$ particle filters, and (2) the performance of different PMMH samplers for estimating a single parameter $\theta$, regardless of the dimensions of the states. Although it is possible to use the more efficient fully adapted particle filter \citep{Pitt:2012}, we use the bootstrap
%filter for all methods to show that the proposed methods are useful for models where it is difficult to use better particle filter algorithms. 

This section investigates (1) the efficiency of different approaches for obtaining likelihood estimates from $S$ particle filters, and (2) the performance of different PMMH samplers for estimating a single parameter $\theta$, regardless of the dimension of the states. We consider the linear Gaussian state space model  discussed in \citet{Guarniero2017}
and \cite{Deligiannidis2018},  where $\left\{ X_{t};t\geq1\right\} $
and $\left\{ Y_{t};t\geq1\right\} $ are $\R^{d}$ valued with
\begin{eqnarray*}
Y_{t} & = & X_{t}+W_{t},\\
X_{t+1} & = & A_{\theta}X_{t}+V_{t+1},
\end{eqnarray*}
with $X_{1}\sim\N\left(0_{d},I_{d}\right)$, $V_{t}\sim N\left(0_{d},I_{d}\right)$,
$W_{t}\sim N\left(0_{d},I_{d}\right)$, and $A_{\theta}^{i,j}=\theta^{|i-j|+1}$ for $i<j$;
the true value of $\theta$ is $0.4$. Although it is possible to use the more efficient fully adapted particle filter \citep{Pitt:2012}, we use the bootstrap
filter for all methods to show that the proposed methods are useful for models where it is difficult to use better particle filter algorithms.

The first study compares  the   
0\%, 5\%, 10\%, 25\%, and 50\% trimmed means of the
likelihood estimates obtained from $S$ particle filters. 
The simulated data is generated from the model above
with $T=200$ and $T=300$ time periods and $d=1,5$ and $10$ dimensions.

\begin{table}[H]
\caption{Comparing the variance of the log of the estimated likelihood for five different estimators of the likelihood: I: Averaging the likelihood (0\% trimmed mean), II: Averaging
the likelihood (5\% trimmed mean), III: Averaging the likelihood (10\%
trimmed mean), IV: Averaging the likelihood (25\% trimmed mean), and
V: Averaging the likelihood (50\% trimmed mean), for $d=10$ dimension
linear Gaussian state space model with $T=300$. The variance of the
log of the estimated likelihood of a single particle filter is reported
in column $VI$. The results are based on $1000$ independent
runs. The entries in columns headed I to V are relative to the variance in column V. The entries in column "V actual" are the actual variances for estimator V.  For example, the entry on row 4, column I is 70.45, which means the variance of the log of the likelihood estimate is 70.45 times the corresponding variance in column V. The entries in column VI are relative to column V with $S=1000$ particles. \label{table10dimT300}}

\centering{}%
\begin{tabular}{ccccccccc}
\hline 
$N$ & $S$ & $I$ & $II$ & $III$ & $IV$ & $V$ & $VI$ & $V$ actual\tabularnewline
\hline 
100 & 1 &  &  &  &  &  & $563.65$\tabularnewline
 & 20 & $3.26$ & $2.02$ & $1.48$ & $1.11$ & $1.00$ & & $29.30$\tabularnewline
 & $100$ & $11.47$ & $2.60$ & $1.80$ & $1.22$ & $1.00$ & & $5.73$\tabularnewline
 & 1000 & $70.45$ & $2.32$ & $1.58$ & $1.01$ & $1.00$ & & $0.66$\tabularnewline
250 & 1 &  &  &  &  &  & $629.83$\tabularnewline
 & 20 & $3.60$ & $2.03$ & $1.67$ & $1.12$ & $1.00$ & & $15.75$\tabularnewline
 & 100 & $11.90$ & $2.48$ & $1.56$ & $1.12$ & $1.00$ & & $3.20$\tabularnewline
 & 1000 & $76.55$ & $2.24$ & $1.62$ & $1.12$ & $1.00$ & & $0.32$\tabularnewline
1000 & 1 &  &  &  &  &  & $620.74$\tabularnewline
 & 20 & $3.69$ & $1.94$ & $1.51$ & $0.99$ & $1.00$ & & $6.14$\tabularnewline
 & 100 & $13.75$ & $2.50$ & $1.79$ & $1.12$ & $1.00$ & & $1.16$\tabularnewline
 & 1000 & $82.18$ & $2.26$ & $1.59$ & $1.08$ & $1.00$ & & $0.13$\tabularnewline
\hline 
\end{tabular}
\end{table}

%\footnote{I suggest "holding the current parameter at $\theta=0.4$ and the proposed parameter at $\theta^{'}=0.4,0.399,0.385$"}

Table~\ref{table10dimT300} shows the variance of the log of the estimated likelihood obtained by using the five estimators of the likelihood for the $d=10$ dimensional linear Gaussian state space model with $T=300$ time periods. The table shows that: (a) there is no substantial reduction in the variance of the log of the estimated likelihood for the mean (0\% trimmed mean). For example, the variance decreases by only a factor of
$2.31$ times when $S$ increases from $20$ to $1000$ when $N=250$. (2)~The 5\%, 10\%, 25\%, and 50\% trimmed means of the individual likelihood estimates decrease the variance substantially as  $S$ and/or $N$ increase. The  25\% and 50\% trimmed mean estimates  have the smallest variance for all cases. Similar results are obtained for $d=5$ with $T=200$ and $T=300$ and $d=10$ with $T=200$; see  tables \ref{table5dimT200}, \ref{table5dimT300}, and \ref{table10dimT200} in section \ref{LGSSexample} of the online supplement.   

We now examine empirically the ability of the proposed MPM samplers to maintain the correlation between successive values of the log of the estimated likelihood for the 10 dimensional linear Gaussian state space model with $T=300$. We ran the different MPM approaches for $1000$ iterations
holding the current parameter at $\theta=0.4$ and the proposed parameter at $\theta^{'}=0.4,0.399,0.385$. At each
iteration we generated $\widetilde{u}$, where $\widetilde{u}=\left(u_{1},...,u_{S}\right)$
and $\widetilde{u}^{'}$ and obtained
$\log\overline{\widehat{p}}_{N}\left(y|\theta,\widetilde{u}^{\left(s\right)}\right)$
and $\log\overline{\widehat{p}}_{N}\left(y|\theta^{'},\widetilde{u}^{\left(s\right)'}\right)$
for the MPM approaches and computed their sample correlations.

Figure~\ref{corr_LGSS_sim} in Section \ref{LGSSexample} of the online supplement
reports the correlation estimates of the log of estimated likelihood obtained using different MPM approaches. The figure
show that: (1) when the current and proposed values of the parameters are equal to $0.4$, all MPM methods		maintain a high correlation between the logs of the estimated likelihoods. The estimated
correlations between logs of the estimated likelihoods when the number
of particle filters $S$ is 100 are about 0.99; (2) when the current parameter $\theta$ and the proposed parameter $\theta^{'}$ are (slightly) different, the MPM methods can still maintain some of the correlations between the log of the estimated likelihoods. The MPM methods with 25\% and 50\% trimmed means of the likelihood perform the best to maintain the correlation between the logs of estimated likelihoods. 

We now compare the efficiency of different sampling schemes for estimating the parameter $\theta$ of the linear Gaussian state space model. In all examples, we run the samplers for $25000$ iterations, with the initial $5000$ iterations  discarded as warm up. The particle
filter and the parameter samplers are implemented in Matlab running on $20$ CPU cores, a high performance computer cluster. The optimal number of CPU cores required for the MPM algorithm is equal
to the number of particle filters $S$, which means the properties of the sampler
can be easily tuned to provide maximum parallel efficiency on a large range of
hardware. The adaptive random walk proposal of \cite{Roberts:2009} is used for all samplers. 

\begin{figure}[H]
\caption{Left: Trace plots of the parameter $\theta$ estimated using (1) MPM (blocking)
with 5\% trimmed mean and (2) MPM (no blocking) 
with 5\% trimmed mean. Middle: Trace plots
of the parameter $\theta$ estimated using (1) MPM (blocking)
with 10\% trimmed mean and (2) MPM (no blocking) 
with 10\% trimmed mean. Right: Trace plots of the parameter
$\theta$ estimated using (1) MPM (blocking)
with 25\% trimmed mean and (2) MPM (no blocking) 
with 25\% trimmed mean.  
Linear Gaussian space model with $d=10$ dimensions, $T=300$ time periods, the number of particle filters $S = 100$, and the number of particles in each particle filter $N = 250$.  \label{traceplotsdim10T300standard}}

\centering{}\includegraphics[width=15cm,height=6cm]{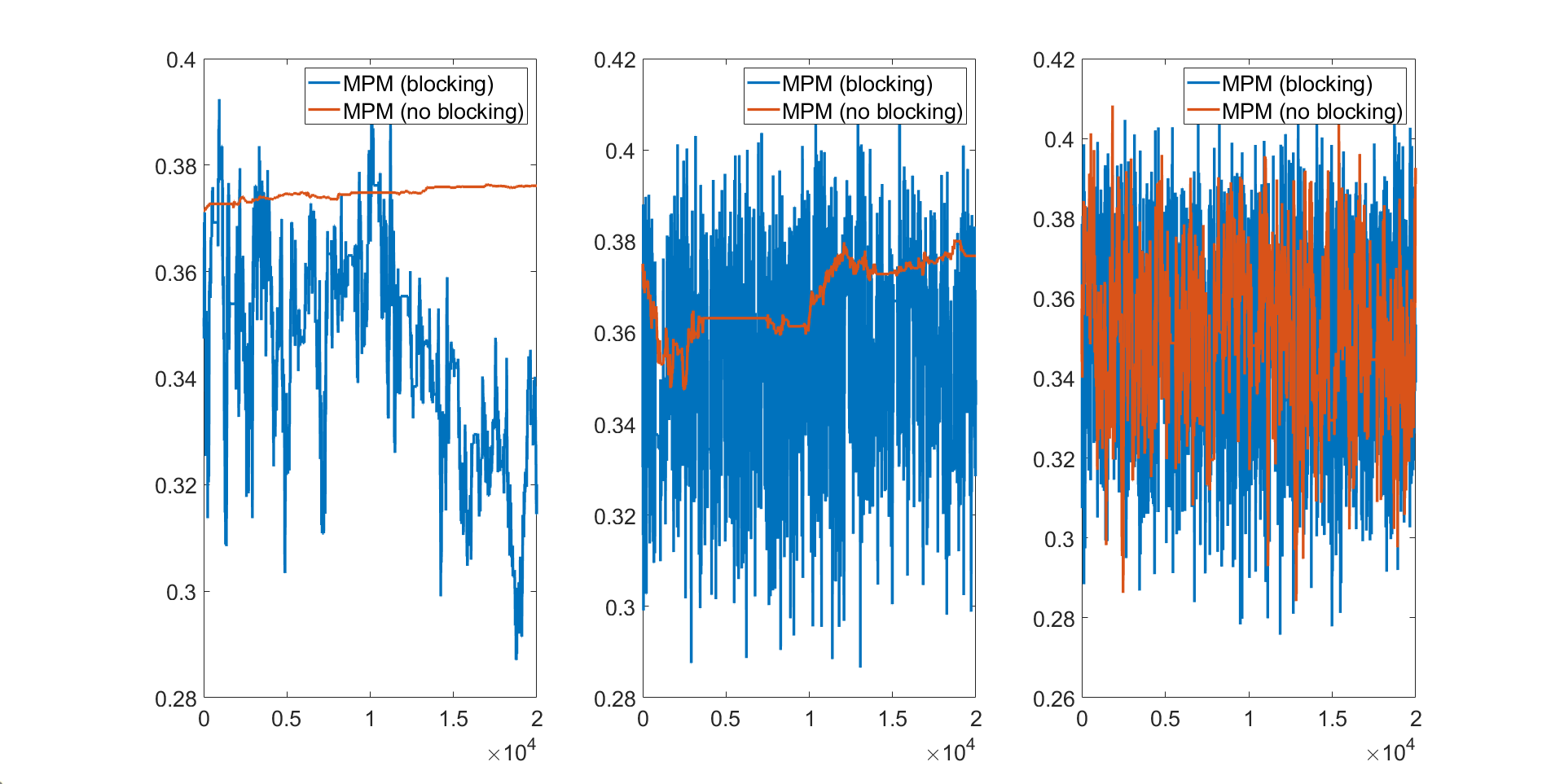}
\end{figure}

\begin{figure}[H]
\caption{Trace plots of the parameter $\theta$ of the linear Gaussian state space model with $d=10$ dimensions and $T=200$ time periods estimated using different MPM samplers with different number of particle filters $S$ and number of particles $N$ in each particle filter:
(I) MPM with mean likelihood estimate, (II) MPM with 5\% trimmed mean estimate, (III) MPM with 10\% trimmed
mean estimate, (IV) MPM with 25\% trimmed mean estimate.   \label{traceplotsdim10T200}}

\centering{}\includegraphics[width=15cm,height=8cm]{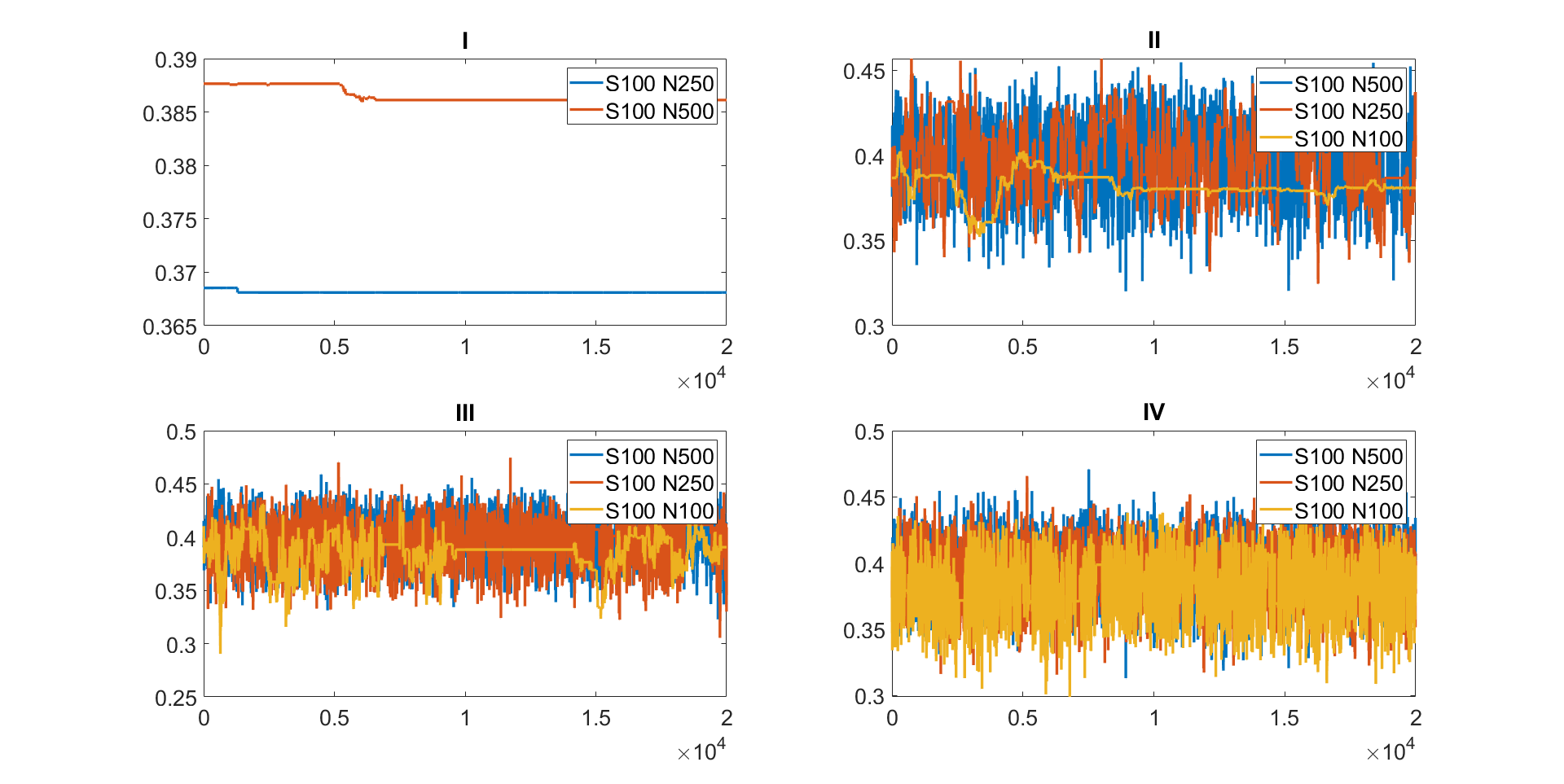}
\end{figure}

Figure~\ref{traceplotsdim10T200} shows the trace plots of the parameter $\theta$ estimated using the MPM algorithm with the different trimmed means for estimating the 10 dimensional linear Gaussian state space model with $T=200$. The figure shows that: (1) the MCMC chain gets stuck for the 5\% trimmed mean approach with $S=100$ and $N=250$ and $N=500$ and for the 10\%  trimmed mean approach with $S=100$ and $N=100$; 
(2)~the 10\% trimmed mean approach requires at least $S=100$ and $N=250$ to make the MCMC chain mix well; (3)~the 25\% trimmed mean approach is the most efficient sampler as its MCMC chain does not get stuck even with $S=100$ and $N=100$. 
Figure~\ref{traceplotsdim10T300standard} compares the MPM algorithms with and without blocking with $S=100$ and $N=250$ for estimating the 10 dimensional linear Gaussian state space model with $T=300$. 
The figure shows that the MPM (no blocking) with 5\% and 10\% trimmed means get stuck and the MPM (no blocking) with 25\% trimmed means mixes poorly. The $\widehat{\textrm{IF}}$ of the MPM (blocking) with 25\% trimmed mean is $9.624$ times smaller than the MPM (no blocking) with 25\% trimmed mean. 
All three panels show that the MPM (blocking) algorithms using trimmed means perform better than the MPM (no blocking). This suggests the usefulness of the MPM (blocking) sampling scheme.  

\begin{figure}[H]
\caption{ 
Linear Gaussian state space model with $d=10, T=300$. Kernel density estimates of the posterior density of $\theta$ estimated using: 
Top Left: 
 (1)~Metropolis-Hastings  with exact likelihood; 
(2)~MPM with 5\% trimmed mean estimate, $S=100,
N=500$; (3)~MPM with 5\% trimmed mean,
$S=100, N=1000$. Top Right: (1) Metropolis-Hastings with exact likelihood; 
 (2) MPM with the 10\%
trimmed mean estimate, $S=100, N=500$; (3) MPM with 10\% trimmed mean estimate, $S=100$, $N=1000$. Bottom Left:  (1) Metropolis-Hastings using the exact likelihood; (2) MPM with 25\% trimmed mean estimate of the likelihood, $S=100$, $N=500$; (3) MPM with
25\% trimmed mean estimate of the likelihood, $S=100$, $N=1000$.
Bottom Right:  (1) Metropolis-Hastings algorithm with exact likelihood; (2) MPM with 50\% trimmed
mean estimate of the likelihood, $S=101$, $N=500$; (3) MPM with 50\%
trimmed mean estimate of the likelihood, $S=101$, $N=1000$.   \label{kerneldensitydim10T300}}

\centering{}\includegraphics[width=15cm,height=8cm]{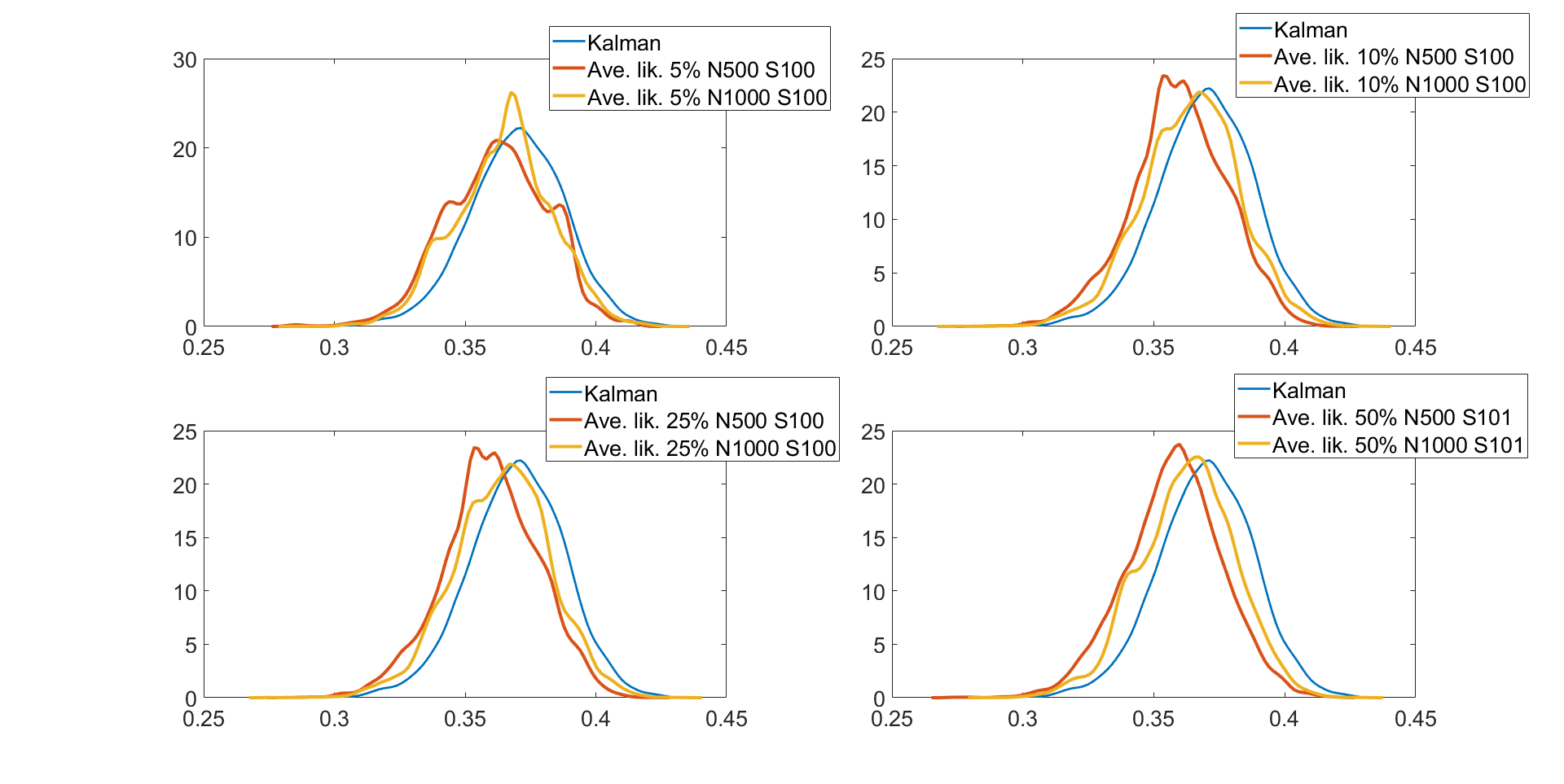}
\end{figure}

Table \ref{tabledim10T300mcmc} shows the $\widehat{\textrm{IF}}$, $\widehat{\textrm{TNIF}}$, and $\widehat{\textrm{RTNIF}}$ values for the parameter $\theta$ in the linear Gaussian state space model with $d=10$ dimensions and $T=300$ time periods estimated using the following five different MCMC samplers: (1) the correlated PMMH of \cite{Deligiannidis2018}, (2) the MPM  with 5\% trimmed mean, (3) the MPM with 10\% trimmed mean, (4) the MPM with 25\% trimmed mean, and (5) the MPM with 50\% trimmed mean. In this paper, instead of using the Hilbert sorting method, we implement the correlated PMMH using the fast multidimensional Euclidean sorting method in section \ref{Multidimensional Sorting}.
The computing time reported in the table is the time to run a single particle filter for the CPM and $S$ particle filters for the MPM approach.
The table shows the following points. (1) The correlated PMMH requires more than $50000$ particles to improve the mixing of the MCMC chain for the parameter $\theta$. (2) The CPU time for running a single particle filter with $N=50000$ particles is $4.09$ times higher than running multiple particle filters with $S=100$ and $N=500$. The MPM method can be much faster than the CPM method if it
is run using high-performance computing with a large number of cores. (3) The
MPM allows us to use much smaller number of particles for each independent
PF and these multiple PFs can be run independently. (4) The $\widehat{\textrm{IF}}$ values for the parameter $\theta$ estimated using the correlated PMMH with $N=50000$ particles is $11.05$, $41.93$, $46.12$, and $56.37$ times larger than the 5\%, 10\%, 25\%, and 50\% trimmed means approaches with $S=100$ and $N=500$ particles, respectively. (5) In terms of $\widehat{\textrm{RTNIF}}$, the 5\%, 10\%, 25\%, and 50\% trimmed means with $S=100$ and $N=500$ are $45.27$, $171.52$, $188.67$, and $230.60$ times smaller than the correlated PMMH with $N=50000$ particles. (6) The best sampler for this example is the 50\% trimmed mean approach with $S=100$ and $N=250$ particles. Table \ref{tabledim10T200} in section \ref{LGSSexample} of the online supplement gives similar results for the 10 dimensional linear Gaussian state space model with $T=200$. Figure \ref{kerneldensitydim10T300} shows the kernel density estimates of the parameter $\theta$ estimated using Metropolis-Hastings algorithm with the (exact) Kalman filter method and the MPM algorithm  with 5\%, 10\%, 25\%, and 50\% trimmed means of the likelihood. The figure shows the approximate posteriors obtained by various approaches using trimmed means are very close to the true posterior. Similar results are obtained for the case $d=10$ dimensions and $T=200$ time periods given in Figure \ref{kerneldensitydim10T200} in Section \ref{LGSSexample} of the online supplement. Figure \ref{kerneldensitydim10T200} also shows that the approximate posterior obtained using the MPM with trimmed mean of the likelihood is very close to the exact posterior obtained using the correlated PMMH.

%\section*{RK comment}
%\begin{itemize}
%    \item  note that RTNIF is not defined yet. you should define it either here or, better, in section 3.1 
%    \item \lq following five different MCMC samplers: (1) the correlated PMMH of \cite{Deligiannidis2018}, (2) the 5\% trimmed mean, (3) the 10\% trimmed mean, (4) the 25\% trimmed mean, and (5) the median (50\% trimmed mean).\rq{} 
    
%    you need to say that estimators (2)--(5) use blocking to induce correlation. 
    
%    MPM means we use multiple PMMH, so no need to say blocking or not.
%\end{itemize}
%\section*{RK end comment}

In summary, the example suggests that: (1) for a high dimensional state space model, there is no substantial reduction in the variance of the log of the estimated likelihood for the method which uses the average of the likelihood from $S$ particle filters when  $S$ and/or $N$ increases. Methods that use trimmed means of the likelihood reduce the variance substantially when $S$ and/or $N$ increases; (2) the 25\% and 50\% trimmed means approaches give the smallest variance of the log of the estimated likelihood for all cases; (3) the MPM method with 50\% trimmed mean is  best at maintaining the correlation between the logs of the estimated likelihoods in successive iterates; (4) the MPM (blocking) method is more efficient than the MPM (no blocking) for the same values of $S$ and $N$; (5) the approximate approaches that use the trimmed means of the likelihood gives accurate approximations to the true posterior; (6) the best sampler for estimating the 10 dimensional linear Gaussian state space model with $T=300$ is the 50\% trimmed mean approach with $S=100$ and $N=250$ particles; (7) the
MPM allows us to use much smaller number of particles for each independent
PF and these multiple PFs can be run independently.
These approaches can be made much faster if they are run using a high-performance computer cluster with a large number of cores.

%\section*{RK comment}
%\begin{itemize}
%    \item  caption of Table 1 needs rewriting: \lq Comparing different methods for calculating the variance of the log
%of the estimated likelihood: I: Averaging the likelihood, II: Averaging
%the likelihood (5\% trimmed mean), III: Averaging the likelihood (10\%
%trimmed mean), IV: Averaging the likelihood (25\% trimmed mean), and
%V: Averaging the likelihood (50\% trimmed mean), for $d=10$ dimension
%linear Gaussian state space model with $T=300$. The variance of the
%log of estimated likelihood of a single particle filter is reported
%under the column ``Single''. The results are based on $1000$ independent
%runs.\rq{} 

%to 

%\lq Comparing the variance of the log of the estimated likelihood for five different methods of estimating the likelihood: ...\rq{} 
%\item \lq The variance of the
%log of estimated likelihood of a single particle filter is reported
%under the column ``Single''.\rq{}

%there is no column called "Single"; you call it VII
%\item 
%I think you should follow Table 1 with a discussion of what it contains, followed by a discussion of similar results in the supplement. 

%Yes, I have discussed table 1 in the text. Please have a read.
%\end{itemize}
%\section*{RK end comment}

%The RTNIF is the TNIF relative to the MPM with 50\% trimmed mean with $N=250$ and $S=100$. 

\begin{table}[H]
\caption{Comparing the performance of different PMMH samplers with different number of particle filters $S$ and different number of particles $N$ in each particle filter for estimating the linear Gaussian state space model using a simulate dataset with $T=300$
and $d=10$ dimensions. Sampler I: Correlated PMMH of \citet{Deligiannidis2018}. Sampler II: MPM with 5\% trimmed
mean. Sampler III: MPM with 10\% trimmed mean.
Sampler IV: MPM with 25\% trimmed mean. Sampler
V: MPM with 50\% trimmed mean. Time denotes the time taken in seconds for
one iteration of the method. 
The $\widehat{\textrm{IF}}$, CT, and $\widehat{\textrm{RTNIF}}$ entries in columns headed I to V are relative to the entries in column V (MPM with 50\% trimmed mean with $N=250$ and $S=100$). The entries in column "V actual" are the actual $\widehat{\textrm{IF}}$, CT, and $\widehat{\textrm{RTNIF}}$ values for estimator V.  For example, the entry on row 5, column I is 415.08, which means the $\widehat{\textrm{RTNIF}}$ is 415.08 times the corresponding $\widehat{\textrm{RTNIF}}$ in column V (MPM with 50\% trimmed mean with $N=250$ and $S=100$). 
\label{tabledim10T300mcmc}}.

\centering{}%
\begin{tabular}{cccccccccccccccc}
\hline 
 & {\footnotesize{}I} & \multicolumn{2}{c}{{\footnotesize{}II}} & \multicolumn{2}{c}{{\footnotesize{}III}} & \multicolumn{2}{c}{{\footnotesize{}IV}} & \multicolumn{2}{c}{{\footnotesize{}V}}  & \multicolumn{2}{c}{{\footnotesize{}V actual}}\tabularnewline
\hline 
{\footnotesize{}N} & {\footnotesize{}50000} & {\footnotesize{}250} & {\footnotesize{}500} & {\footnotesize{}250} & {\footnotesize{}500} & {\footnotesize{}250} & {\footnotesize{}500}  & {\footnotesize{}250} & {\footnotesize{}500} & {\footnotesize{}250} & {\footnotesize{}500} \tabularnewline
{\footnotesize{}S} & {\footnotesize{}1} & {\footnotesize{}100} & {\footnotesize{}100} & {\footnotesize{}100} & {\footnotesize{}100} & {\footnotesize{}100} & {\footnotesize{}100} & {\footnotesize{}100} & {\footnotesize{}100} & {\footnotesize{}100} & {\footnotesize{}100} \tabularnewline
{\footnotesize{}$\widehat{\textrm{IF}}$} & {\footnotesize{}50.73} & {\footnotesize{}88.09} & {\footnotesize{}4.59} & {\footnotesize{}3.03} & {\footnotesize{}1.21} & {\footnotesize{}1.40} & {\footnotesize{}1.10} & {\footnotesize{}1.00} & {\footnotesize{}0.90} & {\footnotesize{}9.98} & {\footnotesize{}9.00}\tabularnewline
{\footnotesize{}CT} & {\footnotesize{}8.18} & {\footnotesize{}1.00} & {\footnotesize{}2.00} & {\footnotesize{}1.00} & {\footnotesize{}2.00} & {\footnotesize{}1.00} & {\footnotesize{}2.00} & {\footnotesize{}1.00} & {\footnotesize{}2.00} & {\footnotesize{}0.99} & {\footnotesize{}1.98}\tabularnewline
{\footnotesize{}$\widehat{\textrm{RTNIF}}$} & {\footnotesize{}415.08} & {\footnotesize{}88.10} & {\footnotesize{}9.17} &  {\footnotesize{}3.03} & {\footnotesize{}2.42} &  {\footnotesize{}1.40} & {\footnotesize{}2.20} & {\footnotesize{}1.00} & {\footnotesize{}1.80} & {\footnotesize{}9.88} & {\footnotesize{}17.82}\tabularnewline
\hline 
\end{tabular}
\end{table}

%\section*{RK comment}
%Figure 1 caption  
%\begin{itemize}
%    \item  
%\lq Left: Trace plots of the parameter $\theta$ estimated using (1) MPM
%with averaging likelihood (5\% trimmed mean) and (2) standard PMMH
%with averaging likelihood (5\% trimmed mean).\rq 

%to 

%\lq Left: Trace plots of the parameter $\theta$ estimated using (1) MPM (blocking)
%with 5\% trimmed mean and (2) standard PMMH (no blocking) 
%with 5\% trimmed mean.\rq{}

%I think we need to emphasise that standard PMMH means no blocking.
%\item what is a reader meant to see in right panel? \item write in caption for centre \& right panels what i did for left panel.
%\item 
%you labelled the left, centre and right panels as I, II and III. Do we need these labels? 
%\item 
%Explain in text (not caption) the interpretation of Figure 1, all the figures and tables have been explained in the text.  Do you want to put the figures and tables close to where they are discussed in the text. I usually put all the figures and tables at the end of each section. 
%\end{itemize}
%\section*{RK end comment}

%\section*{RK comment}
%what is $N$? It is in the figures.
%\section*{RK end comment}

%\begin{figure}[H]
%\caption{Left: Trace plots of the parameter $\theta$ estimated using (1) MPM
%with 5\% trimmed mean and (2) standard PMMH
%with 5\% trimmed mean. Middle: Trace plots
%of the parameter $\theta$ estimated using (1) MPM with 10\% trimmed mean and (2) standard PMMH with 10\% trimmed mean. Right: Trace plots of the parameter
%$\theta$ estimated using (1) MPM with 25\%
%trimmed mean and (2) standard PMMH with 25\%
%trimmed mean. Linear Gaussian state
%space model with $T=300, S = 100, N = 250$. \label{traceplotsdim10T300standard}}

%\centering{}\includegraphics[width=15cm,height=6cm]{trace_plots_LGSS_dim10T300_MPMstandardPMMH}
%\end{figure}
%\section*{RK comment}
%Figure 3 caption  
%\begin{itemize}
%    \item  
%see some comments on caption of Fig 1 
%\item Fig 3 right panel  shows that MPM %with 25\% trimmed mean  is better than standard PMMH with ...; but it is hard to see. Need to explain more in text. 
%\end{itemize}
%\section*{RK end comment}

%Kernel density estimates of the parameter $\theta$
%estimated using:

\subsection{Multivariate Stochastic Volatility in Mean Model\label{SVinmeanexample}}

This section investigates the performance of the proposed MPM samplers
for estimating large multivariate stochastic volatility in mean models \citep[see, e.g.,][]{carriero2018measuring,cross2021macroeconomic}, where each of the log volatility processes follows a GARCH (Generalized Autoregressive Conditional Heteroskedasticity) diffusion process \citep{shephard2004likelihood}.
The GARCH diffusion model does not have a closed form state transition
density, making it challenging to estimate using the Gibbs type MCMC sampler used in \cite{cross2021macroeconomic}. It is well-known that the Gibbs sampler is inefficient for generating parameters for a diffusion process, especially for the variance parameter $\tau^2$ \citep{stramer2011bayesian}. Conversely, MPM samplers can efficiently estimate models with non-closed form state transition densities, making them well-suited for the GARCH diffusion model. This section demonstrates that MPM samplers can estimate this model efficiently and overcome the limitations of the Gibbs sampler.

%In this section, the authors investigate the performance of the proposed MPM (Marginal Particle Metropolis) samplers for estimating large multivariate stochastic volatility in mean models, where each of the log volatility processes follows a GARCH (Generalized Autoregressive Conditional Heteroskedasticity) diffusion process. The GARCH diffusion model does not have a closed form state transition density, which makes it challenging to estimate using the Gibbs type MCMC (Markov Chain Monte Carlo) sampler used in previous works such as \cite{cross2021macroeconomic}.

%It is well-known that the Gibbs sampler can be inefficient for generating parameters for a diffusion process, especially for the variance parameter $\tau^2$. In contrast, MPM samplers can efficiently estimate models with non-closed form state transition densities, making them well-suited for the GARCH diffusion model. The authors demonstrate that MPM samplers can estimate this model efficiently and overcome the limitations of the Gibbs sampler.

Suppose that $P_{t}$ is a $d\times1$ vector of daily stock prices
and define $y_{t}=\log P_{t}-\log P_{t-1}$ as the log-return of the
stocks. Let $h_{i,t}$ be the log-volatility process of the $i$th
stock at time $t$. We also define, $h_{\cdotp,t}=\left(h_{1,t},...,h_{d,t}\right)^{\top}$
and $h_{i,\cdotp}=\left(h_{i,1},...,h_{i,T}\right)^{\top}$. The model
for $y_{t}$ is 
\begin{equation}
y_{t}=Ah_{\cdot,t}+\epsilon_{t},\;\epsilon_{t}\sim N\left(0,\Sigma_{t}\right),
\end{equation}
where $A$ is a $d\times d$ matrix that captures the effects of
log-volatilities on the levels of the variables. The time-varying
error covariance matrix $\Sigma_{t}$ depends on the unobserved latent variables $h_{\cdotp,t}$
such that 
\begin{equation}
\Sigma_{t}:=\textrm{diag}\left(\exp\left(h_{1,t}\right),...,\exp\left(h_{d,t}\right)\right).
\end{equation}
Each log-volatility $h_{i,t}$ is assumed to follow a continuous time
GARCH diffusion process \citep{shephard2004likelihood} satisfying
\begin{equation}
\label{garchtransition}
dh_{i,t}=\left\{ \alpha\left(\mu-\exp\left(h_{i,t}\right)\right)\exp\left(-h_{i,t}\right)-\frac{\tau^{2}}{2}\right\} dt+\tau dW_{i,t},
\end{equation}
for $i=1,...,d$, where the $W_{i,t}$ are independent Wiener processes.
 The following Euler scheme approximates the evolution of the log-volatilities ${h}_{i,t}$ in \eqref{garchtransition} by placing $M-1$ evenly spaced points between times $t$ and $t+1$.
We denote the intermediate volatility components by $h_{i,t,1},...,h_{i,t,M-1}$, and it is convenient to set $h_{i,t,0}=h_{i,t}$ and $h_{i,t,M}=h_{i,t+1}$. The equation for the Euler evolution, starting at $h_{i,t,0}$ is (see, for example, \cite{stramer2011bayesian}, pg. 234)
\begin{equation}
h_{i,t,j+1}|h_{i,t,j}\sim N\left(h_{i,t,j}+\left\{ \alpha\left(\mu-\exp\left(h_{i,t,j}\right)\right)\exp\left(-h_{i,t,j}\right)-\frac{\tau^{2}}{2}\right\} \delta,\tau^{2}\delta\right),
\label{eq:GARCH Euler transitiondensity}
\end{equation}
for $j=0,...,M-1$, where $\delta = 1/M$. 
The initial state of $h_{i,t}$ is assumed normally distributed $N(0,1)$ for $i=1,...,d$.
For this example, we assume for simplicity that the parameters $\mu$, $\alpha$, and $\tau^2$ are the same across all stocks and $A^{i,j}=\psi^{|i-j|+1}$ for $i<j$. This gives the same number of parameters regardless of the dimension of the stock returns so that we can focus on  comparing different PMMH samplers.

We first report on a simulation study for the above model. We simulated ten independent datasets from the model described above with $d=20$ dimensions and $T=100$ observations.
The true parameters are $\alpha=2$, $\mu=1.81$, $\tau^{2}=0.38$,
and $\psi=0.01$. The priors are $\alpha\sim G\left(1,1\right)$,
$\tau^{2}\sim G\left(0.5,0.5\right)$, $\mu\sim G\left(1,1\right)$,
and $\psi\sim N\left(0,1\right)$. These prior densities are non-informative
and cover almost all possible values in practice. We use $M=3$ latent
points for the Euler approximations of the state transition densities. 
For each dataset, we run the MPM sampler with the  25\% trimmed mean approach for $25000$ iterations;  the initial $5000$ iterations are discarded as warm up. The particle
filter and the parameter samplers are implemented in Matlab running on 20 CPU cores, a high performance computer cluster. The adaptive random walk proposal of \cite{Roberts:2009} is used for all samplers. 

\begin{figure}[H]
\caption{Kernel density estimates of the parameters of the multivariate stochastic
volatility in mean model estimated using MPM with a 25\% trimmed mean estimate of the likelihood based on a simulated dataset (data 1) with $d=20$
 and $T=100$.\label{diffusiondata1}}

\centering{}\includegraphics[width=15cm,height=8cm]{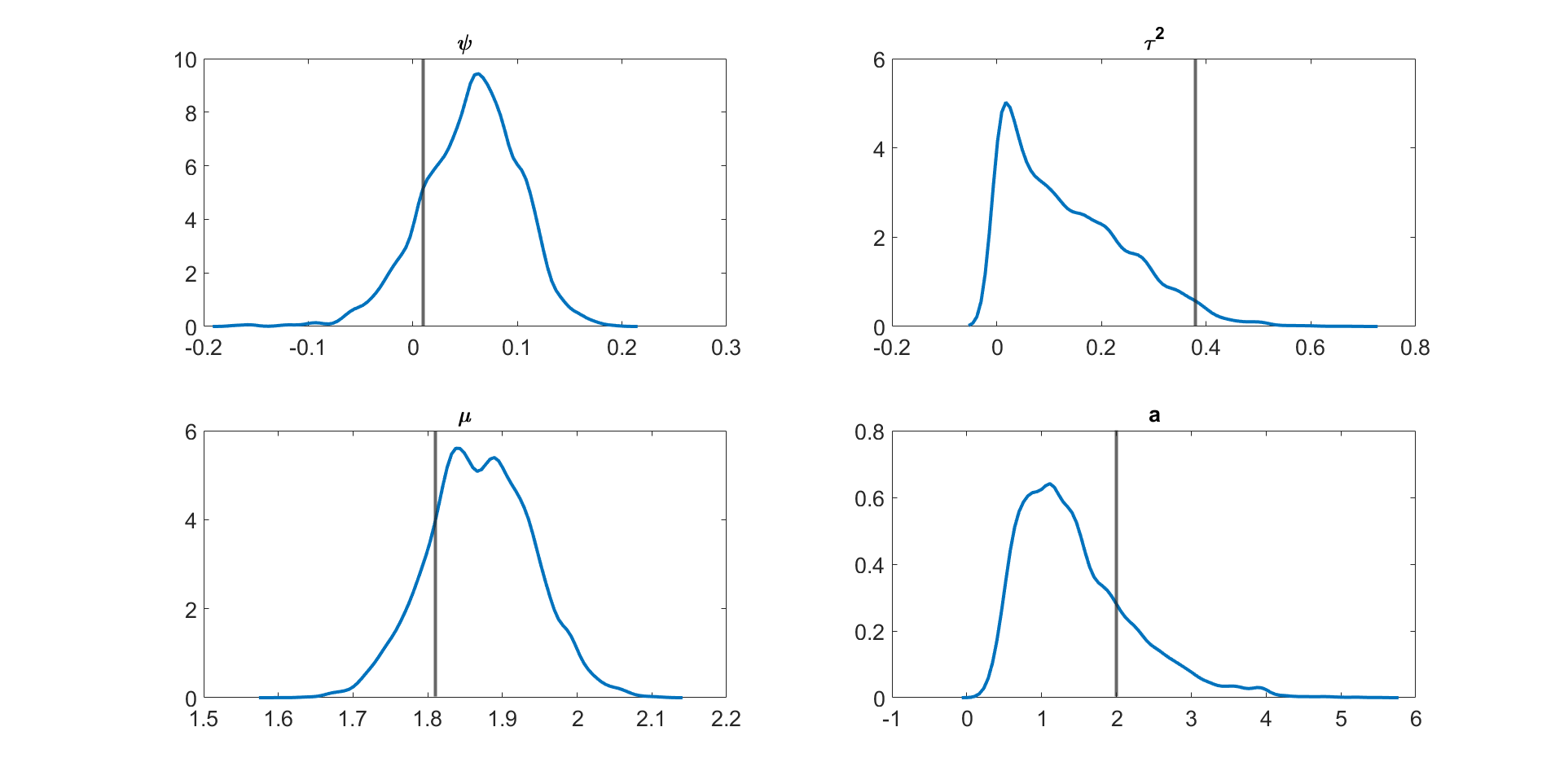}
\end{figure}

\begin{figure}[H]
\caption{The inefficiency factors ($\widehat{\textrm{IF}}$) of the parameters of the multivariate
stochastic volatility in mean model estimated using  MPM 
with a 25\% trimmed mean estimate of the likelihood for 10 simulated datasets
with $d=20$  and $T=100$.\label{boxplotdiffusionprocess}}

\centering{}\includegraphics[width=15cm,height=8cm]{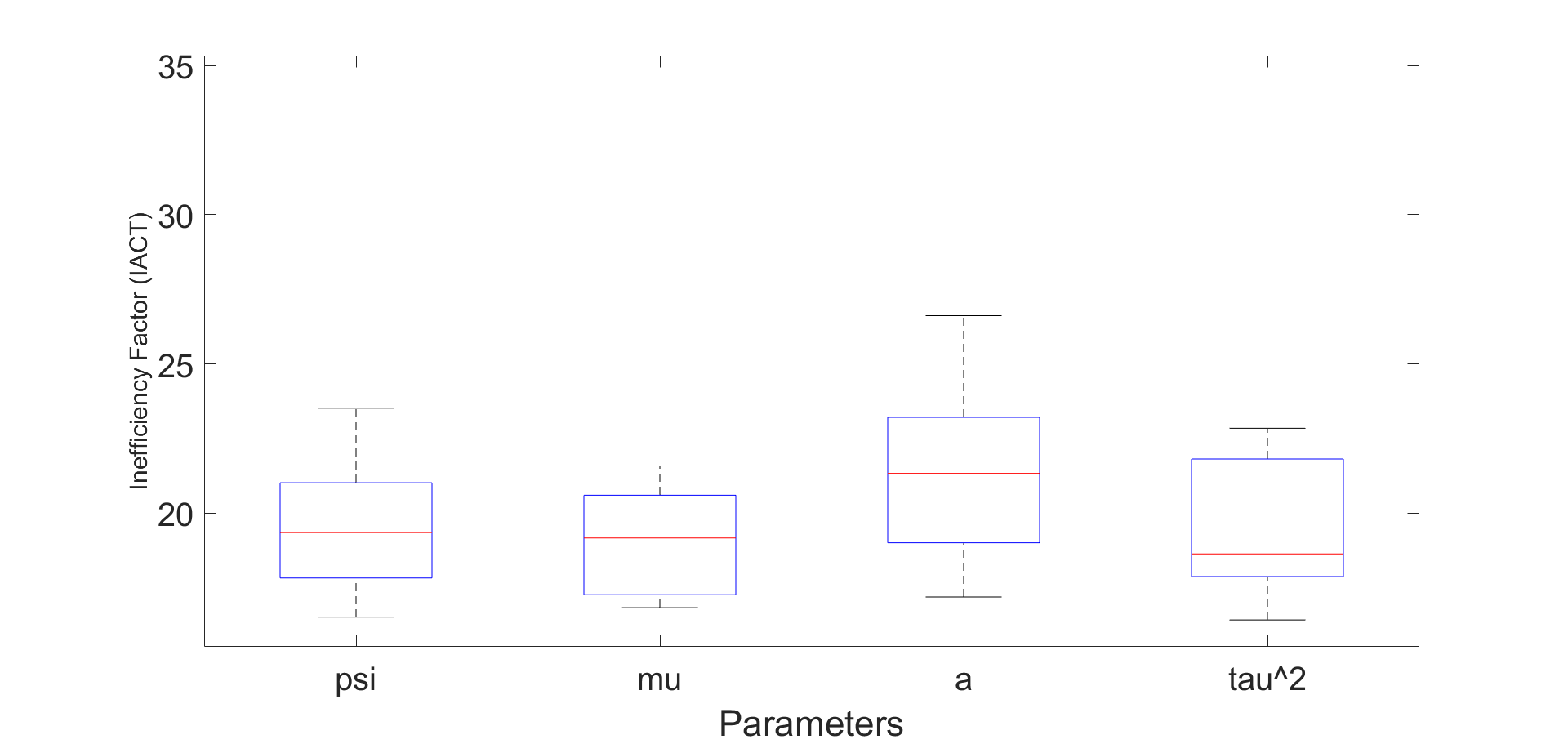}
\end{figure}

Figure~\ref{diffusiondata1} and 
Figures~\ref{diffusiondata2} to \ref{diffusiondata10} in section~\ref{SVexample} of the online supplement show the kernel density estimates of the parameters of the multivariate stochastic volatility in mean model described above estimated by the MPM sampler ($S=100$ and $N=250$) using a 25\% trimmed mean; the vertical lines show the true parameter values. The figures show that the model parameters are accurately estimated. Figure \ref{boxplotdiffusionprocess} shows the inefficiency factors ($\widehat{\textrm{IF}}$) of the parameters  over 10 simulated datasets with $d=20$ dimensions and $T=100$ time periods. The results suggest that the MPM sampler with only $N=250$ particles and $S=100$ particle filters is efficient because the $\widehat{\textrm{IF}}$ values of the parameters are reasonably small. 

%\section*{RK comment}
%\begin{enumerate}
%   \item \lq are always within the posterior density estimates\rq{} 
   
%   this is clumsy. what do you mean to say? 
%\end{enumerate}
%\section*{RK end comment}

We now apply the methods to a sample of daily US industry stock returns data obtained
from the Kenneth French website\footnote{http://mba.tuck.dartmouth.edu/pages/faculty/ken.french/datalibrary.html}, using a sample from January 4th, 2021 to the 26th
of May, 2021, a total of 100 observations. 
%\section*{RK comment}
%\begin{enumerate}
%   \item why such a small sample? Is it because macro data is so small ?
   
%   It is a typical number of observations used for high dimensional model. And actually, no one ever estimates a 30 dimensional diffusion model as I did here, in particular with bootstrap filter. 
%   Note that this is a diffusion model, so
%   when you run a particle filter, you need to generate the intermediate latent variables between time period as well and that takes a lot of time. 
%\end{enumerate}
%\section*{RK end comment}

Table \ref{20dimSVmodel} shows the $\widehat{\textrm{IF}}$, $\widehat{\textrm{TNIF}}$, and $\widehat{\textrm{RTNIF}}$ for the parameters of the 20 dimensional multivariate stochastic volatility in mean model estimated using (1) the correlated PMMH, (2) the MPM sampler with the 25\% trimmed mean of the likelihood estimate, and (3) the MPM sampler with the 50\% trimmed mean of the likelihood  estimate. The table shows that: (1) the MCMC chain obtained using the correlated PMMH with $N=100000$ still gets stuck; (2) the CPU time for the correlated PMMH with $N=100000$ particles is $26$ times slower than the MPM algorithm with $S=250$ and $N=100$; (3) the MPM sampler with $N=250$ and $S=100$ is less efficient than the MPM sampler with $N=100$ and $S=250$. The MPM sampler with $N=100$ and $S=250$ can be made more efficient by running it on a high-performance computer cluster with many cores; (4) the best sampler in terms of $\widehat{\textrm{TNIF}}_{\textrm{MEAN}}$  and  $\widehat{\textrm{TNIF}}_{\textrm{MAX}}$ for estimating the 20 dimensional multivariate stochastic volatility in mean model is the MPM sampler with a 50\% trimmed mean with $S=250$ and $N=100$.

Table \ref{30dimSVmodel} shows the $\widehat{\textrm{IF}}$, $\widehat{\textrm{TNIF}}$, and $\widehat{\textrm{RTNIF}}$ for the parameters of the $d=30$ dimensional multivariate stochastic volatility in the mean model estimated using: (1) 
the MPM sampler with the 25\% trimmed mean of the likelihood, and (2) the MPM sampler with 50\% trimmed mean of the likelihood. We do not consider the correlated PMMH as it is very expensive to run for this high dimensional model. The table shows that (1) increasing the number of particle filters from $S=100$ to $S=200$ reduces the $\widehat{\textrm{IF}}$, but increases the computational time substantially. However, the computational time can be reduced by using a high performance computer cluster with a larger number of cores; (2) the MPM sampler with $N=500$ and $S=100$ is less efficient than the MPM sampler with $N=250$ and $S=200$. We recommend that if there is access to large computing resources, then the number of particle filters $S$ should be increased while keeping the number of particles $N$ in each particle filter reasonably small.

\begin{table}[H]
\caption{Comparing the performance of different PMMH samplers with different numbers $S$
of particle filters and different number $N$ of particles in each particle filter for estimating a multivariate stochastic volatility in mean model with GARCH diffusion processes
using a real dataset with $T=100$ and $d=20$. The model parameters are $\psi$, $\mu$, $a$, and $\tau^2$. Sampler
I: Correlated PMMH of \cite{Deligiannidis2018}. Sampler II: MPM
with 25\% trimmed mean. Sampler III: MPM with
50\% trimmed mean. Time denotes the time taken in seconds for
one iteration of the method. The RTNIF is the TNIF relative to the MPM with 25\% trimmed mean with $S=100$ and $N=250$.\label{20dimSVmodel} }  

\centering{}%
\begin{tabular}{cccccc}
\hline 
Param. & I & \multicolumn{2}{c}{II} & \multicolumn{2}{c}{III}\tabularnewline
\hline 
N & 100000 & 250 & 100 & 250 & 100\tabularnewline
S & 1 & 100 & 250 & 100 & 250\tabularnewline
$\psi$ & NA & $97.621$ & $67.459$ & $83.523$ & $62.680$\tabularnewline
$\mu$ & NA & $65.678$ & $42.299$ & $52.293$ & $39.390$\tabularnewline
$a$ & NA & $20.037$ & $10.409$ & $12.857$ & $9.490$\tabularnewline
$\tau^{2}$ & NA & $56.617$ & $32.900$ & $43.705$ & $37.928$\tabularnewline
$\widehat{\textrm{IF}_{\psi,\textrm{MAX}}}$ & NA & $97.621$ & $67.459$ & $83.523$ & $62.680$\tabularnewline
$\widehat{\textrm{TNIF}}_{\textrm{MAX}}$ & NA & $173.765$ & $124.799$ & $148.671$ & $115.958$\tabularnewline
$\widehat{\textrm{RTNIF}}_{\textrm{MAX}}$ & NA & $1$ & $0.718$ & $0.856$ & $0.667$\tabularnewline
$\widehat{\textrm{IF}_{\psi,\textrm{MEAN}}}$ & NA & $59.988$ & $38.267$ & $48.095$ & $37.372$\tabularnewline
$\widehat{\textrm{TNIF}}_{\textrm{MEAN}}$ & NA & $106.779$ & $70.794$ & $85.609$ & $69.138$\tabularnewline
$\widehat{\textrm{RTNIF}}_{\textrm{MEAN}}$ & NA & $1$ & $0.663$ & $0.802$ & $0.647$\tabularnewline
\hline 
Time & 47.65  & 1.78 & 1.85 & 1.78 & 1.85\tabularnewline
\hline 
\end{tabular}
\end{table}
%\section*{RK comment}
%Table 3 caption 
%\begin{enumerate}
%   \item define entries for $\psi, \mu$ etc 
%\end{enumerate}
%\section*{RK end comment}

\begin{table}[H]
\caption{Comparing the performance of different PMMH samplers with different number
of particle filters S and different number of particles N in each particle filter for estimating multivariate stochastic volatility in mean model with GARCH diffusion processes
using a real dataset with $T=100$ and $d=30$. The model parameters are $\psi$, $\mu$, $a$, and $\tau^2$. Sampler I:~MPM
with 25\% trimmed mean. Sampler II:~MPM with
50\% trimmed mean. Time is the time  in seconds for
one iteration of the method. The RTNIF is the TNIF relative to the MPM with 25\% trimmed mean with $S=100$ and $N=250$. \label{30dimSVmodel}}

\centering{}%
\begin{tabular}{cccccc}
\hline 
Param. & \multicolumn{3}{c}{I} & \multicolumn{2}{c}{II}\tabularnewline
\hline 
N & 250 & 250 & 500 & 250 & 250\tabularnewline
S & 100 & 200 & 100 & 100 & 200\tabularnewline
$\psi$ & $183.559$ & $124.258$ & $ 227.685 $ & $151.745$ & $80.073$\tabularnewline
$\mu$ & $74.070$ & $49.250$ & $ 384.767 $ & $74.485$ & $45.554$\tabularnewline
$a$ & $46.159$ & $45.662$ & $ 126.968 $ & $32.721$ & $14.753$\tabularnewline
$\tau^{2}$ & $102.234$ & $70.255$ & $ 218.866 $ & $78.171$ & $42.075$\tabularnewline
$\widehat{\textrm{IF}_{\psi,\textrm{MAX}}}$ & $183.559$ & $124.258$ & $ 384.767 $ & $151.745$ & $80.073$\tabularnewline
$\widehat{\textrm{TNIF}}_{\textrm{MAX}}$ & $523.143$ & $877.262$ & $ 3351.321 $ & $432.473$ & $565.315$\tabularnewline
$\widehat{\textrm{RTNIF}}_{\textrm{MAX}}$ & $1$ & $1.677$ & $6.406 $ & $0.827$ & $1.081$\tabularnewline
$\widehat{\textrm{IF}_{\psi,\textrm{MEAN}}}$ & $101.505$ & $72.356$ & $239.571 $ & $84.280$ & $45.614$\tabularnewline
$\widehat{\textrm{TNIF}}_{\textrm{MEAN}}$ & $289.289$ & $510.833$ & $ 2077.953 $ & $240.198$ & $322.035$\tabularnewline
$\widehat{\textrm{RTNIF}}_{\textrm{MEAN}}$ & $1$ & $1.766$ & $ 7.183 $ & $0.830$ & $1.113$\tabularnewline
\hline 
Time & 2.85 & 7.06 & 8.71 & 2.85 & 7.06\tabularnewline
\hline 
\end{tabular}
\end{table}

%\section*{RK comment}
%Table 4 caption 
%\begin{enumerate}
%   \item define entries for $\psi, \mu$ etc 
%\end{enumerate}
%\section*{RK end comment}

%\section*{RK comment}
%Table 9: 
%\begin{enumerate}
%    \item Explain Table 9 in text. 
%    \item What are you trying to show in the table. 
%    \item Which datasets are used for the table 
%\end{enumerate}
%\section*{RK end comment}

%\section*{RK comment}
%Figures 9 to  16 
%\begin{enumerate}
%    \item What are you trying to how in the figures
%    \item I think one figure will do here. The rest can be in the supplement. . 
%    \item Which datasets are used for the table 
%\end{enumerate}

%\section*{RK end comment}
%\bibliographystyle{apalike}
%\bibliography{references_v1}
%\end{document}

\subsection{DSGE models \label{subsec:Dynamic-Stochastic-GeneralModel}}

In this section, we evaluate the effectiveness of the MPM samplers for estimating two non-linear DSGE models: the small scale and medium scale DSGE models as detailed in sections~\ref{Description of SmallScale Model} and \ref{Description of MediumScale Model} of the online supplement. Due to the large dimension of the states compared to the dimension of the disturbances in these state space models, we employ the disturbance filter method introduced in section~\ref{subsec:Disturbance-Particle-Filter}.

%This section investigates the performance of the MPM samplers for estimating two non-linear DSGE models: the small scale DSGE and the medium scale DSGE models described in sections~\ref{Description of SmallScale Model} and \ref{Description of MediumScale Model} of the online supplement. These DSGE models are examples of state space models where the dimension of the states is much larger than the dimension of the disturbances, thus motivating the use of the disturbance filter in section \ref{subsec:Disturbance-Particle-Filter}\footnote{maybe better "In this section, we evaluate the effectiveness of the MPM samplers for estimating two non-linear DSGE models: the small scale and medium scale DSGE models as detailed in sections~\ref{Description of SmallScale Model} and \ref{Description of MediumScale Model} of the online supplement. Due to the large dimension of the states compared to the dimension of the disturbances in these state space models, we employ the disturbance filter method introduced in section~\ref{subsec:Disturbance-Particle-Filter}."}.  

The state transition densities of the two non-linear DSGE models are intractable. 
Section~\ref{Supp: overview of DSGE models} of the online supplement describes 
the  state space representation of these models, which are solved using Dynare.

%It  consists of a consumption Euler equation, a new Keynesian Philip curve, a monetary policy rule, fiscal policy rule,  three exogenous shock processes, and eight endogenous latent variables.

\subsubsection{Nonlinear Small Scale DSGE Model \label{subsec:SecondOrderSmallScale}}
This section reports on  how well a number of PMMH samplers of interest estimate the parameters of the
nonlinear (second order) small scale DSGE model. The small scale DSGE model follows the setting in \citet{Herbst2019} with nominal rigidities through price adjustment costs following \citet{Rotemberg}, and three structural shocks. There are $4$ state variables, $4$ control variables ($8$ variables combining the state and control variables) and $3$ exogenous shocks in the model. We estimate this model using $3$ observables, which means that the dimensions of $y_t$, the state vector and the disturbance vector are $3$, $8$, and $3$, respectively.
 Section~\ref{Description of SmallScale Model} of the online supplement describes 
 the model. 

In this section, we use the delayed acceptance  version of the MPM algorithm, calling it  \lq \lq delayed acceptance multiple PMMH\rq\rq{}  (DA-MPM), to speed up the computation. The delayed acceptance sampler  \citep{Christen2005}  avoids computing the expensive likelihood estimate
if it is likely  that the proposed draw will ultimately
be rejected. The first accept-reject stage uses a cheap
(or deterministic) approximation to the likelihood instead of the expensive likelihood estimate in the MH acceptance ratio.
The particle filter is then used to estimate the likelihood
only for a proposal that is
accepted in the first stage;
a second accept-reject stage  ensures that detailed
balance is satisfied with respect to the true posterior. We use the likelihood obtained from the  central difference Kalman filter (CDKF) proposed by \citet{Norgaard2000} in the first accept-reject stage of the delayed acceptance scheme. Section ~\ref{subsec:Delayed-Acceptance-Multiple} of the online supplement gives further details. The dimension of states is usually much larger than the dimension of the disturbances for the DSGE models. Two different sorting methods are also considered. The first sorts the state particles and the second sorts the disturbance particles using the sorting algorithm in 
section~\ref{Multidimensional Sorting}. 

The PMMH samplers considered are: (1) The MPM (0\% trimmed mean) with the auxiliary disturbance particle filter (ADPF), disturbance sorting, and the correlation between random numbers used in estimating the log of estimated likelihoods set to $\rho_u=0.9$; (2) the MPM (10\% trimmed mean) with the ADPF, disturbance sorting, and the $\rho_u=0.9$; (3) the MPM (25\% trimmed mean) with the ADPF, disturbance sorting, and the $\rho_u=0.9$; (4) the MPM (25\% trimmed mean) with the ADPF, disturbance sorting, and $\rho_u=0$, (5) the MPM (25\% trimmed mean) with the ADPF, state sorting, and $\rho_u=0.9$; (6) the delayed acceptance MPM (DA-MPM) (25\% trimmed mean) with ADPF, disturbance sorting, and the $\rho_u=0.9$; (7) 
the MPM (50\% trimmed mean) with the ADPF, disturbance sorting, and the $\rho_u=0.9$;
(8) the MPM (0\% trimmed mean) with the bootstrap particle filter, disturbance sorting, and the $\rho_u=0.9$; (9) the 
correlated PMMH with $\rho_u=0.9$, the bootstrap particle filter, and disturbance sorting. Each sampler ran for $25000$ iterations, with the initial $5000$ iterations discarded as burn-in. We use the adaptive random walk proposal of \citet{Roberts:2009} for $q\left(\theta^{'}|\theta\right)$  and the adaptive scaling approach of \citet{Garthwaite:2015} to tune the Metropolis-Hastings acceptance probability to $20\%$. Section~\ref{Description of SmallScale Model} of the online supplement gives details on model specifications and model parameters. We include measurement error in the observation equation. The measurement error variances are estimated together with other parameters. All the samplers ran on a high performance computer cluster with 20 CPU cores. The real dataset is obtained from \citet{Herbst2019}, using data from 1983Q1 to 2013Q4, which includes the Great Recession with a total of $124$ observations for each series.

%We also compare the performance of the auxiliary disturbance particle filter (ADPF) to the tempered particle
%filter (TPF) proposed by \citet{Herbst2019}; however, we note that \citet{Herbst2019} only apply the TPF to linear (first order) DSGE models.

Table~\ref{SmallScaleTable} reports  the relative time normalised inefficiency factor ($\widehat{\textrm{RTNIF}}$) of a PMMH sampler relative to the DA-MPM (25\% trimmed mean) with the ADPF, disturbance sorting, and $\rho_u=0.9$ for the non-linear
small scale models with $T=124$ observations. 
The computing time reported in the table is the time to run a single particle filter for the CPM and $S$ particle filters for the MPM approach. 
The table shows that: (1)~The
MPM sampler (0\%  trimmed mean) with the ADPF, disturbance sorting, $\rho_u=0.9$, and $N=250$ particles is more
efficient than the MPM sampler (0\% trimmed mean) with the bootstrap particle filter, disturbance sorting, $\rho_u=0.9$, and $N=250$ particles with the MCMC chain obtained by the latter getting stuck. (2) The running time taken 
the MPM method  with $S=100$ particles filters, each with $N=100$ particles and disturbance sorting
 is $9.73$ times faster than the CPM with $N=20000$ particles.  
(3)~In terms of $\widehat{\textrm{RTNIF}}_{\textrm{MAX}}$, the performance of MPM samplers with 0\%, 10\%, 25\%, and 50\% trimmed means, ADPF, disturbance sorting, and $\rho_u=0.9$ are $3.15$, $90.77$, $103.04$, and $91.50$ times more efficient respectively, than the correlated PMMH with $N=20000$ particles.
(4)~In terms of $\widehat{\textrm{RTNIF}}_{\textrm{MAX}}$, the MPM (25\% trimmed mean) with 
disturbance sorting, and $\rho_u=0.9$ is $1.38$ times more efficient than the MPM (25\% trimmed mean) method with state sorting method and performs similarly to the MPM (25\% trimmed mean) approach with ADPF, disturbance sorting, and $\rho_u=0$. (5)~The delayed acceptance MPM (25\% trimmed mean) is slightly more efficient than the MPM (25\% trimmed mean) approach. The delayed acceptance algorithm is $5$ times faster on average because the target Metropolis-Hastings acceptance probability is set to $20\%$ using the \citet{Garthwaite:2015} approach, but it has higher maximum $\widehat{\textrm{IF}}$ value. Table~\ref{SmallScaleTableSupplement} in 
section~\ref{additionaltablesfiguressmallscale} of the online supplement gives the details. (6)~It is possible to use the tempered particle filter of \citet{Herbst2019} within the MPM algorithm. However, the TPF is computationally expensive because of the tempering iterations and the random walk Metropolis-Hastings mutation steps.  In summary, the best sampler with the smallest $\widehat{\textrm{RTNIF}}_{\textrm{MAX}}$ and $\widehat{\textrm{RTNIF}}_{\textrm{MEAN}}$ is  the delayed acceptance MPM (DA-MPM) (25\% trimmed mean) with ADPF, disturbance sorting, and $\rho_u=0.9$.

\begin{table}[H]
\caption{Comparing the performance of different PMMH samplers with different number
of particle filters $S$ and different number of particles $N$ in each particle filter for estimating small scale DSGE model using a real dataset with $T = 124$ observations. Sampler
I: MPM (0\% trimmed mean, $\rho_{u}=0.9$, disturbance sorting,
ADPF). Sampler II: MPM (10\% trimmed mean, $\rho_{u}=0.9$,
disturbance sorting, ADPF). Sampler III: MPM (25\% trimmed mean, $\rho_{u}=0.9$, disturbance sorting, ADPF). Sampler
IV: MPM (25\% trimmed mean, $\rho_{u}=0$, disturbance
sorting, ADPF). Sampler V: MPM (25\% trimmed
mean, $\rho_{u}=0.9$, state sorting, ADPF). Sampler VI: DA-MPM (25\% trimmed mean, $\rho_{u}=0.9$, disturbance sorting, ADPF).
Sampler VII: MPM (50\% trimmed mean, $\rho_{u}=0.9$, disturbance sorting, ADPF).
Sampler VIII: MPM (0\% trimmed mean, $\rho_{u}=0.9$, disturbance
sorting, bootstrap). Sampler IX: Correlated PMMH of \citet{Deligiannidis2018}. Time denotes the time taken in seconds for
one iteration of the method. 
The $\widehat{\textrm{IF}}_{\psi,\textrm{MAX}}$, $\widehat{\textrm{RTNIF}}_{\textrm{MAX}}$, $\widehat{\textrm{IF}}_{\psi,\textrm{MEAN}}$, and $\widehat{\textrm{RTNIF}}_{\textrm{MEAN}}$  entries in columns headed I to IX are relative to the entries in column VI. The entries in column "VI actual" are the actual $\widehat{\textrm{IF}}_{\psi,\textrm{MAX}}$, $\widehat{\textrm{RTNIF}}_{\textrm{MAX}}$, $\widehat{\textrm{IF}}_{\psi,\textrm{MEAN}}$, and $\widehat{\textrm{RTNIF}}_{\textrm{MEAN}}$ for estimator VI. The numbers in bold are for the sampler with similar values of the $\widehat{\textrm{RTNIF}}_{\textrm{MAX}}$ and $\widehat{\textrm{RTNIF}}_{\textrm{MEAN}}$. 
\label{SmallScaleTable}}

\centering{}%
\begin{tabular}{|c|c|c|c|c|c|c|c|c|c|c|}
\hline 
{\footnotesize{}} & {\footnotesize{}I} & {\footnotesize{}II} & {\footnotesize{}III} & {\footnotesize{}IV} & {\footnotesize{}V } & {\footnotesize{}VI} & {\footnotesize{}VII} & {\footnotesize{}VIII} & {\footnotesize{}IX} & {\footnotesize{}VI actual}\tabularnewline
\hline 
{\footnotesize{}S} & {\footnotesize{}100} & {\footnotesize{}100} & {\footnotesize{}100} & {\footnotesize{}100} & {\footnotesize{}100} & {\footnotesize{}100} & {\footnotesize{}100} & {\footnotesize{}100} & {\footnotesize{}1} & {\footnotesize{}100}\tabularnewline
{\footnotesize{}N} & {\footnotesize{}250} & {\footnotesize{}100} & {\footnotesize{}100} & {\footnotesize{}100} & {\footnotesize{}100} & {\footnotesize{}100} & {\footnotesize{}100} & {\footnotesize{}250} & {\footnotesize{}20000} & {\footnotesize{}100}\tabularnewline
\hline 
{\footnotesize{}$\widehat{\textrm{IF}}_{\psi,\textrm{MAX}}$} & {\footnotesize{}$3.73$} & {\footnotesize{}$0.25$} & {\footnotesize{}$0.22$} & {\footnotesize{}$0.25$} & {\footnotesize{}$0.28$} & {\footnotesize{}$1.00$} & {\footnotesize{}$0.39$} & {\footnotesize{}NA} & {\footnotesize{}$2.42$} & {\footnotesize{}$604.72$}\tabularnewline
{\footnotesize{}$\widehat{\textrm{RTNIF}}_{\textrm{MAX}}$} & {\footnotesize{}$36.27$} & {\footnotesize{}$\textbf{1.26}$} & {\footnotesize{}$\textbf{1.11}$} & {\footnotesize{}$\textbf{1.25}$} & {\footnotesize{}$1.54$} & {\footnotesize{}$\textbf{1.00}$} & {\footnotesize{}$1.96$} & {\footnotesize{}NA} & {\footnotesize{}$114.37$} & {\footnotesize{}${66.52}$}\tabularnewline
\hline 
{\footnotesize{}$\widehat{\textrm{IF}}_{\psi,\textrm{MEAN}}$} & {\footnotesize{}$5.88$} & {\footnotesize{}$0.49$} & {\footnotesize{}$0.50$} & {\footnotesize{}$0.49$} & {\footnotesize{}$0.61$} & {\footnotesize{}$1.00$} & {\footnotesize{}$0.59$} & {\footnotesize{}NA} & {\footnotesize{}$3.32$} & {\footnotesize{}$190.74$}\tabularnewline
{\footnotesize{}$\widehat{\textrm{TNIF}}_{\textrm{MEAN}}$} & {\footnotesize{}$57.23$} & {\footnotesize{}$\textbf{2.47}$} & {\footnotesize{}$\textbf{2.49}$} & {\footnotesize{}$\textbf{2.44}$} & {\footnotesize{}$3.41$} & {\footnotesize{}$\textbf{1.00}$} & {\footnotesize{}$2.93$} & {\footnotesize{}NA} & {\footnotesize{}$157.02$} & {\footnotesize{}${20.98}$}\tabularnewline
\hline 
{\footnotesize{}Time} & {\footnotesize{}9.73} & {\footnotesize{}5.00} & {\footnotesize{}5.00} & {\footnotesize{}5.00} & {\footnotesize{}5.55} & {\footnotesize{}{1.00}} & {\footnotesize{}$5.00$} & {\footnotesize{}6.82} & {\footnotesize{}47.27} & {\footnotesize{}{0.11}}\tabularnewline
\hline 
\end{tabular}
\end{table}

\subsubsection{Nonlinear Medium Scale DSGE Model\label{mediumscaleDSGEmodelexample}}
This section applies the MPM samplers with 50\% trimmed mean for estimating the parameters of the second-order medium-scale DSGE model of \citet{gust2017empirical} discussed in Section \ref{Description of MediumScale Model} of the online supplement. The medium scale DSGE model is similar to the models in \citet{CEE} and \citet{SW} with nominal rigidities in both wages and prices, habit formation in consumption, investment adjustment costs, along with {five} structural shocks. 
We do not consider the zero lower bound (ZLB) on the nominal interest rate because 
estimating such a model requires solving the fully nonlinear model using global methods instead of local perturbation techniques. The time taken to solve the fully nonlinear model depends on the choice of solution technique and can vary across methods such as value function iteration and time-iteration methods. Since the paper focuses on demonstrating the strength of the estimation algorithm, we restrict our sample to the period before the zero lower bound constraint on the nominal interest rate binds and the model can be solved (at a second order approximation) using perturbation methods. The estimation algorithm, however, can be applied to fully nonlinear models as well since the estimation step requires decision rules produced by the model solution (global or local methods).

%In Section \ref{mediumscaleDSGEmodelexample}, we further test the effectiveness of the estimation algorithm by considering a medium scale DSGE model in the spirit of \citet{CEE} and \citet{SW} with nominal rigidities in both wages and prices, habit formation in consumption, investment adjustment costs along with four structural shocks. There are altogether 12 state variables, 18 control variables (30  variables combining the state and control variables) and 5 exogenous shocks in this model.  We estimate this model using 5 observables; that means that the dimensions of $y_t$, the state vector and the disturbance vector are $5$, $30$, and $5$, respectively.

%The model is solved using dynare.

The baseline model in \citet{gust2017empirical} consists of 12 state variables, 18 control variables (30  variables combining the state and control variables) and 5 exogenous shocks. We estimate this model using 5 observables; that means that the dimensions of $y_t$, the state vector and the disturbance vector are $5$, $30$, and $5$, respectively.
We include measurement error in the observation equation and consider two cases. The measurement error variances are: (1) estimated and (2) fixed at the 25\% of the sample variance of the observables over the estimation period.  We use quarterly US data from 1983Q1 to 2007Q4, with a total of $100$ observations for each of the five series.  

The PMMH samplers considered are: (1) the MPM (50\% trimmed mean) with the auxiliary disturbance particle filter (ADPF), disturbance sorting, and the correlation between random numbers used in estimating the log of estimated likelihoods set to $\rho_u=0.9$; (2) the 
correlated PMMH of \citet{Deligiannidis2018}. Each sampler ran for $60000$ iterations, with the initial $30000$ iterations discarded as burn-in. We use the adaptive random walk proposal of \citet{Roberts:2009} for $q\left(\theta^{'}|\theta\right)$.  All the samplers ran on a high performance computer cluster with $20$ CPU cores.

Table \ref{medscaleTNIF} shows the $\widehat{\textrm{RTNIF}}$ of the correlated PMMH relative to the MPM sampler with 50\% trimmed mean (ADPF, disturbance sorting, and $\rho_u=0.9$) for the non-linear medium scale DSGE model for the cases where the measurement error variances are estimated and fixed at 25\% sample variance of the observables.  
The computing time reported in the table is the time to run a single particle filter for the CPM and $S=250$ particle filters for the MPM approach. The table shows that in terms of $\widehat{\textrm{RTNIF}}_{\textrm{MAX}}$, the MPM sampler is $26$ and $14$ times more efficient than the correlated PMMH for the cases where the measurement error variances are estimated and fixed at 25\% sample variance of the observables, respectively.   
The full details on the inefficiency factors for each parameter are given in table \ref{fullmediumscaleRTNIF} in section \ref{mediumscaletable} of the online supplement.

Table \ref{tab:Mean,-,-and some parameters medium scale-1} gives the posterior means, $2.5\%$, and $97.5\%$ quantile estimates of some of the parameters with standard errors in brackets for the medium scale DSGE model estimated
using the MPM sampler with $50\%$ trimmed mean ($\rho_{u}=0.9$,
disturbance sorting, ADPF). The measurement error variances are fixed to $25\%$ of
the sample variance of the observables. The table shows that the standard errors are relatively small for all parameters indicating that the parameters are estimated accurately. Table \ref{tab:Mean,-,-and all parameters medium scale} in Section \ref{mediumscaletable} of the online supplement gives the posterior means, $2.5\%$, and $97.5\%$ quantile estimates of all parameters in the medium scale DSGE model.

\begin{figure}[H]
\caption{Kernel density estimates of the posterior density of some of the parameters of the medium scale DSGE model  for the US quarterly dataset from 1983Q1
to 2007Q4 estimated using: (1) MPM with 50\% trimmed mean ($\rho_{u}=0.9$,
disturbance sorting, ADPF) with $S=250$ and $N=100$; (2) correlated PMMH with $N=50000$. The measurement error variances are fixed to $25\%$ of
the variance of the observables.
\label{medscalekerneldensity}}

\centering{}\includegraphics[width=15cm,height=8cm]{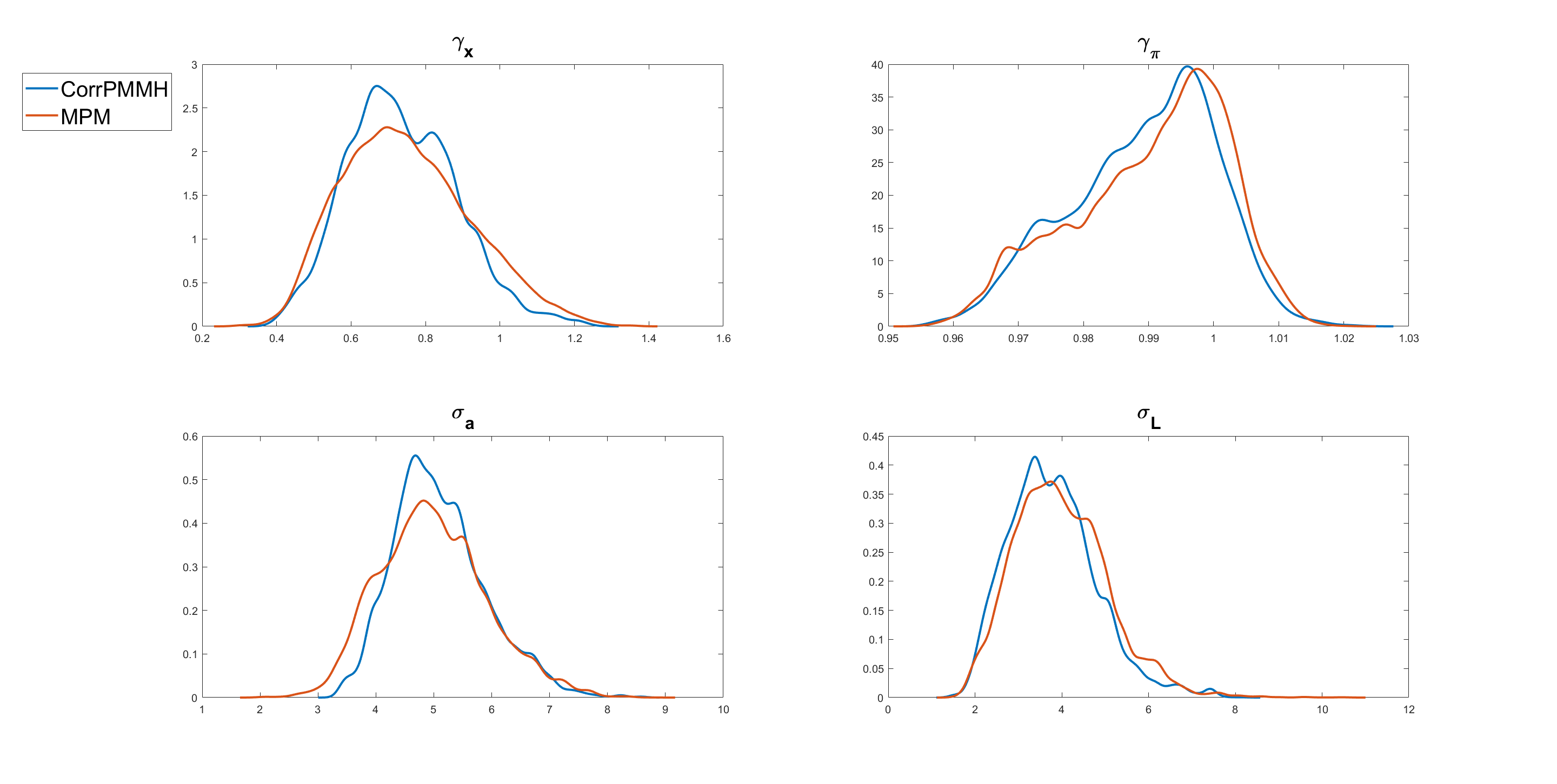}
\end{figure}

Figure \ref{medscalekerneldensity} shows the kernel density estimates of some of the parameters of the medium scale DSGE model estimated using the MPM with 50\% trimmed mean (ADPF, disturbance sorting, and $\rho_u=0.9$) and the correlated PMMH. The figure shows that the approximate posterior obtained using the MPM with 50\% trimmed
mean is very close to the exact posterior obtained using the correlated PMMH.

We also consider a variation of the medium scale DSGE model of \citet{gust2017empirical} described in section \ref{Description of MediumScale Model} of the online supplement. 
To study how well our approach performs with increasing state dimension, we estimate the model with the canonical definition of the output gap defined as the difference between output in the model with nominal rigidities and output in the model with flexible prices and wages. This variation is identical to the baseline model in  \citet{gust2017empirical} except for the Taylor rule specification, which now uses the canonical definition of the output gap. This extension of the medium scale DSGE model consists of 21 state variables, 30 control variables (51 variables combining the state and control variables) and 5 exogenous shocks.  We estimate this model using 5 observables; that means that the dimensions of $y_t$, the state vector and the disturbance vector are $5$, $51$, and $5$, respectively. The measurement error variances are fixed at 25\% of the variance of the observables over the estimation period.

Table \ref{tab:Mean,-,-and all parameters medium scale-extended model} of the online supplement shows the posterior means, 2.5\% and 97.5\% quantiles of the parameters of the extended medium scale DSGE model estimated using the MPM algorithm with a 50\% trimmed mean (ADPF, disturbance sorting, and $\rho_u=0.9$) with $S=250$ and $N=100$. The table shows that the $\widehat{\textrm{IF}}$s of the parameters are similar to the simpler medium scale DSGE model, indicating the performance of the MPM algorithm does not deteriorate with the extended model with the higher state dimensions. 

%In \citet{gust2017empirical}, the authors proxy the logarithm of the output gap in the Taylor rule using a weighted average of the deviations of utilization and labor from their non-stochastic steady state values. The model environment in \citet{gust2017empirical} with the modified definition of the output gap consists of 12 state variables, 18 control variables (30 variables combining the state and control variables) and 5 exogenous shocks.  We estimate this model using 5 observables; that means that the dimensions of $y_t$, the state vector and the disturbance vector are $5$, $25$, and $5$, respectively.

\begin{table}[H]
\caption{Comparing the performance of different PMMH samplers with different
numbers of particle filters $S$ and different numbers of particles
$N$ in each particle filter for estimating the medium scale DSGE
model using the US quarterly dataset from 1983Q1 to 2007Q4. Sampler
I: MPM (50\% trimmed mean, $\rho_{u}=0.9$, disturbance sorting,
ADPF). Sampler II: Correlated PMMH. The first three columns are for
the case when the measurement error variances are estimated. 
The $\widehat{\textrm{IF}}_{\textrm{MAX}}$, $\widehat{\textrm{RTNIF}}_{\textrm{MAX}}$, $\widehat{\textrm{IF}}_{\textrm{MEAN}}$, and $\widehat{\textrm{RTNIF}}_{\textrm{MEAN}}$  entries in columns headed I to II are relative to the entries in column I. The entries in column "I actual" are the actual $\widehat{\textrm{IF}}_{\psi,\textrm{MAX}}$, $\widehat{\textrm{RTNIF}}_{\textrm{MAX}}$, $\widehat{\textrm{IF}}_{\psi,\textrm{MEAN}}$, and $\widehat{\textrm{RTNIF}}_{\textrm{MEAN}}$ for estimator I.
The last
three columns are for the case when the measurement error variances
are fixed at 25\%
of the sample variance of the observables. The $\widehat{\textrm{IF}}_{\textrm{MAX}}$, $\widehat{\textrm{RTNIF}}_{\textrm{MAX}}$, $\widehat{\textrm{IF}}_{\textrm{MEAN}}$, and $\widehat{\textrm{RTNIF}}_{\textrm{MEAN}}$  entries in columns headed I to II are relative to the entries in column I. The entries in column "I actual" are the actual $\widehat{\textrm{IF}}_{\psi,\textrm{MAX}}$, $\widehat{\textrm{RTNIF}}_{\textrm{MAX}}$, $\widehat{\textrm{IF}}_{\psi,\textrm{MEAN}}$, and $\widehat{\textrm{RTNIF}}_{\textrm{MEAN}}$ for estimator I.
\label{medscaleTNIF}}

\centering{}%
\begin{tabular}{ccccccc}
\hline 
 & I & II & I actual & I & II & I actual\tabularnewline
\hline 
{\footnotesize{}$\widehat{\textrm{IF}}_{\textrm{MAX}}$} & $1.00$ & $2.74$ & $796.82$  & $1.00$ & $1.48$ & $383.31$\tabularnewline
{\footnotesize{}$\widehat{\textrm{RTNIF}}_{\textrm{MAX}}$} & $1.00$ & $26.51$ & $1338.66$ & $1.00$ & $14.37$ & $ 643.96$\tabularnewline
\hline 
{\footnotesize{}$\widehat{\textrm{IF}}_{\textrm{MEAN}}$} & $1.00$ & $2.78$ & $428.71 $& $1.00$ & $0.96$ & $ 275.34$\tabularnewline
{\footnotesize{}$\widehat{\textrm{RTNIF}}_{\textrm{MEAN}}$} & $1.00$ & $26.92$ & $720.22$ & $1.00$ & $9.31$ & $462.57$\tabularnewline
\hline 
{\footnotesize{}Time} & $1.00$ & $9.68$ & $1.68$ & $1.00$ & $9.68$ & $1.68$\tabularnewline
\hline
\end{tabular}
\end{table}

\begin{table}[H]
\caption{Posterior means, $2.5\%$, and $97.5\%$ quantiles 
of some of the parameters in the medium scale DSGE model estimated
using the MPM sampler with $50\%$ trimmed mean ($\rho_{u}=0.9$,
disturbance sorting, ADPF) for the US quarterly dataset from 1983Q1
to 2007Q4. The measurement error variances are fixed to $25\%$ of
the variance of the observables. The standard errors in brackets are calculated from $10$ independent runs of the MPM sampler. \label{tab:Mean,-,-and some parameters medium scale-1}}

\centering{}%
\begin{tabular}{cccc}
\hline 
Param. & Estimates & $2.5\%$ & $97.5\%$\tabularnewline
\hline 
$\rho_{R}$ & $\underset{\left(0.0178\right)}{0.4329}$ & $\underset{\left(0.0365\right)}{0.2097}$ & $\underset{\left(0.0220\right)}{0.6490}$\tabularnewline
$\rho_{g}$ & $\underset{\left(0.0008\right)}{0.9883}$ & $\underset{\left(0.0020\right)}{0.9770}$ & $\underset{\left(0.0005\right)}{0.9965}$\tabularnewline
$\rho_{\mu}$ & $\underset{\left(0.0191\right)}{0.9883}$ & $\underset{\left(0.0309\right)}{0.4136}$ & $\underset{\left(0.0410\right)}{0.7598}$\tabularnewline
$100\sigma_{g}$ & $\underset{\left(0.0104\right)}{0.2188}$ & $\underset{\left(0.0129\right)}{0.0973}$ & $\underset{\left(0.0182\right)}{0.3992}$\tabularnewline
$100\sigma_{\mu}$ & $\underset{\left(0.0978\right)}{4.1832}$ & $\underset{\left(0.2905\right)}{2.1935}$ & $\underset{\left(0.2697\right)}{7.0727}$\tabularnewline
$100\sigma_{\eta}$ & $\underset{\left(0.0096\right)}{0.3614}$ & $\underset{\left(0.0082\right)}{0.2706}$ & $\underset{\left(0.0151\right)}{0.4837}$\tabularnewline
$100\sigma_{Z}$ & $\underset{\left(0.0118\right)}{0.6240}$ & $\underset{\left(0.0168\right)}{0.4707}$ & $\underset{\left(0.0121\right)}{0.7954}$\tabularnewline
$100\sigma_{R}$ & $\underset{\left(0.0032\right)}{0.0462}$ & $\underset{\left(0.0002\right)}{0.0059}$ & $\underset{\left(0.0123\right)}{0.1570}$\tabularnewline
\hline 
\end{tabular}
\end{table}

\section{Summary and conclusions \label{sec:Conclusions}}
The article proposes a general  particle marginal approach (MPM) for estimating the posterior density of the parameters of complex and high-dimensional state-space models. It is especially useful when the single bootstrap filter is inefficient, while the more efficient auxiliary particle filter cannot be used because the state transition density is computationally intractable.
The MPM method is a general extension of the PMMH method and makes four innovations:  (a) it is based on the mean or trimmed mean of unbiased likelihood estimators; (b)~it uses a novel block version of PMMH that works with multiple particle filters;  (c) an auxiliary disturbance particle filter sampler is proposed to estimate the likelihood which is especially useful when the dimension of the states is much larger than the dimension of the disturbances; (d) a fast Euclidean sorting algorithm is proposed to preserve the correlation between the logs of the estimated likelihoods at the current and proposed parameter values. The MPM samplers are  applied to complex multivariate stochastic volatility and DSGE models and shown to work well.
Future research will consider developing
better proposals for the parameter $\theta$.

%(d) a delayed acceptance proposal based on the central difference Kalman filter is used to speed up the computation for the DSGE examples.

%The empirical results suggest that: (1)~the  auxiliary disturbance particle filter (ADPF) is much more efficient than the standard bootstrap particle filter (BPF); (2)~the MPM maintains the correlation between logs of the estimated likelihoods in successive iterates much better than the CPM and BPM; (3)~the delayed acceptance version of the MPM sampler is much more efficient than the standard MPM sampler; (iv)~the MPM with disturbance sorting is more efficient than the MPM with state sorting; (4)~the performance of the MPM just using block sampling is as efficient as that using both block sampling and correlated sampling.

%Finally, we believe that the methods in the paper will be very useful for many other models
%where particle alternative sophisticated methods, such as the particle Gibbs, are either
%inefficient or impossible to use; e.g., partially observed diffusions and large panel data models. 

\pagebreak
\renewcommand{\thealgorithm}{S\arabic{algorithm}}
\renewcommand{\theequation}{S\arabic{equation}}
\renewcommand{\thesection}{S\arabic{section}}
\renewcommand{\thepage}{S\arabic{page}}
\renewcommand{\thetable}{S\arabic{table}}
\renewcommand{\thefigure}{S\arabic{figure}}
\setcounter{page}{1}
\setcounter{section}{0}
\setcounter{equation}{0}
\setcounter{algorithm}{0}
\setcounter{table}{0}
\setcounter{figure}{0}

\section*{Online supplement for \textquotedblleft The Block-Correlated Pseudo Marginal Sampler for State Space Models\textquotedblright}
%\author{David Gunawan\textsuperscript{$\star$}, Pratiti Chatterjee\textsuperscript{$\ddagger$} , and Robert Kohn\textsuperscript{$\star\star$}}

\if1\blind%%
{
   %\title {\bf Flexible Density Tempering Approaches for State Space Models with an Application to Factor Stochastic Volatility Models}
     \title{Online supplement for \textquotedblleft The Block-Correlated Pseudo Marginal Sampler for State Space Models\textquotedblright}		
\author{}
\maketitle
} \fi

%\section*{Online supplement for \textquotedblleft Efficient Pseudo Marginal Method for State Space Models\textquotedblright}
\maketitle
\blfootnote{
	\textsuperscript{$\ddagger$}
	Level 4, West Lobby, School of Economics, University of New South Wales Business School -- Building E-12, Kensington Campus, UNSW Sydney -- 2052, \textit{Email:} {pratiti.chatterjee@unsw.edu.au}, \textit{Phone Number:} {(+61) 293852150}. Website: {http://www.pratitichatterjee.com}\\
	\textsuperscript{$\star$} 39C. 164, School of Mathematics and Applied Statistics (SMAS), University of Wollongong, Wollongong, 2522; Australian Center of Excellence for Mathematical and Statistical Frontiers (ACEMS); National Institute for Applied Statistics Research Australia (NIASRA); \textit{Email}: dgunawan@uow.edu.au. \textit{Phone Number:} {(+61) 424379015}. \\
	\textsuperscript{$\star\star$} 	Level 4, West Lobby, School of Economics, University of New South Wales Business School -- Building E-12, Kensington Campus, UNSW Sydney -- 2052, and ACEMS \textit{Email:} {r.kohn@unsw.edu.au}, \textit{Phone Number:} {(+61) 424802159}.}

\section{The Disturbance Particle Filter Algorithm \label{Disturbance particle filter}}
This section discusses the disturbance particle filter algorithm.
Let $u$ be the random vector used to obtain the unbiased estimate
of the likelihood. It  has the two components $u_{\epsilon,1:T}^{1:N}$
and $u_{A,1:T-1}^{1:N}$; $u_{\epsilon,t}^{i}$ is the vector
random variable used to generate the particles $\epsilon_{t}^{i}$
given $\theta$. We can write
\begin{equation}
\epsilon_{1}^{i}\sim m\left(\epsilon_{1}^{i}|u_{\epsilon,1}^{i},\theta\right),z_{1}^{i}=F
\left(z_{0},\epsilon_{1}^{i};\theta\right)\;\textrm{and}\;\epsilon_{t}^{i}\sim m\left(\epsilon_{t}^{i}|u_{\epsilon,t}^{i},\theta\right),z_{t}^{i}=
F\left(z_{t-1}^{a_{t-1}^{i}},\epsilon_{t}^{i};\theta\right),t\geq2,
\label{eq:transformationrandomnumbers}
\end{equation}
where $m\left(\epsilon_{t}^{i}|u_{\epsilon,t}^{i},\theta\right)$ is the proposal density to generate $\epsilon_{t}^{i}$, and $z_0$ is the initial state vector with density $p(z_0|\theta)$. Denote
the distribution of $u_{\epsilon,t}^{i}$ as $\psi_{\epsilon t}\left(\cdot\right)$.
For $t\geq2$, let $u_{A,t-1}$ be the vector of random variables
used to generate the ancestor indices $a_{t-1}^{1:N}$ using the resampling
scheme $M\left(a_{t-1}^{1:N}|\overline{w}_{t-1}^{1:N},z_{t-1}^{1:N}\right)$
and define $\psi_{At-1}\left(\cdot\right)$ as the distribution of
$u_{A,t-1}$. Common choices for $\psi_{\epsilon t}\left(\cdot\right)$
and $\psi_{At-1}\left(\cdot\right)$ are iid $N\left(0,1\right)$
and i.i.d. $U\left(0,1\right)$ random variables, respectively.

Algorithm \ref{alg:The-Disturbance-Particle filter} takes the number
of particles $N$, the parameters $\theta$, the random variables
used to generate the disturbance particles $u_{\epsilon,1:T}^{1:N}$,
and the random variables used in the resampling steps $u_{A,t-1}^{1:N}$
as the inputs; it outputs the set of state particles $z_{1:T}^{1:N}$,
disturbance particles $\epsilon_{1:T}^{1:N}$, ancestor indices $A_{1:T-1}^{1:N}$,
and the weights $\overline{w}_{1:T}^{1:N}$. At $t=1$,
the disturbance particles $\epsilon_{1}^{1:N}$ are obtained as a function of the
random numbers $u_{\epsilon,1}^{1:N}$ using \Eqref{eq:transformationrandomnumbers}
in step (1a) and the state particles are obtained from $z_{1}^{i}=F\left(z_{0},\epsilon_{1}^{i};\theta\right)$, for $i=1,...,N$;
the weights for all particles are then computed in steps (1b) and (1c).

Step (2a) sorts the state or disturbance particles from smallest to largest using
the Euclidean sorting procedure described in section \ref{Multidimensional Sorting} to
obtain the sorted disturbance particles, sorted state particles and
weights. Algorithm \ref{alg:Multinomial-Resampling-Algorithm} resamples
the particles using the correlated multinomial resampling to obtain the sorted ancestor
indices $\widetilde{a}_{t-1}^{1:N}$. Step (2d) generates the disturbance particles $\epsilon_{t}^{1:N}$
as a function of the random numbers $u_{\epsilon,t}^{1:N}$ using
\Eqref{eq:transformationrandomnumbers} and the state particles are obtained from  $z_{t}^{i}=F\left(z_{t-1}^{a_{t-1}^{i}},\epsilon_{t}^{i};\theta\right)$, for
$i=1,...,N$; we then compute the weights for all particles in step
(2e) and (2f).

The disturbance particle filter provides the unbiased estimate of the likelihood
 $$\widehat{p}_{N}\left(y|\theta,u\right) :=\prod_{t=1}^{T}\left(N^{-1}\sum_{i=1}^{N}w_{t}^{i}\right),$$
%\]
where
\begin{equation}
w_{t}^{i}=\frac{p\left(y_{t}|z^{i}_{t},\theta\right)p\left(\epsilon^{i}_{t}\right)}
{m\left(\epsilon_{t}|u_{\epsilon,t},\theta\right)}\;\textrm{for }t=1,...,T,
\textrm{  and  } \overline{w}_{t}^{i}=\frac{w_{t}^{i}}{\sum_{j=1}^{N}w_{t}^{j}}.
\label{importanceweights}
\end{equation}
%It is possible to sort the disturbance particles instead
%of the state particles in step (2a).

\begin{algorithm}[H]
\caption{The Correlated Disturbance Particle Filter \label{alg:The-Disturbance-Particle filter}}

Input: $u_{\epsilon,1:T}^{1:N}$, $u_{A,t-1}^{1:N}$, $\theta$ and
$N$

Output: $\epsilon_{1:T}^{1:N}$, $z_{1:T}^{1:N}$, $A_{1:T-1}^{1:N}$,
and $\overline{w}_{1:T}^{1:N}$

For $t=1$
\begin{itemize}
\item (1a) Generate $\epsilon_{1}^{i}$ from $m\left(\epsilon_{1}^{i}|u_{\epsilon,1}^{i},\theta\right)$
and set $z_{1}^{i}=F\left(z_{0},\epsilon_{1}^{i};\theta\right)$
for $i=1,...,N$
\item (1b) Compute the unnormalised weights $w_{1}^{i}$, for $i=1,...,N$
\item (1c) Compute the normalised weights $\overline{w}_{1}^{i}$ for $i=1,...,N$.
\end{itemize}
For $t\geq2$
\begin{itemize}
\item (2a) Sort the state particles $z_{t-1}^{i}$ or disturbance particles $\epsilon_{t-1}^{i}$ using the Euclidean sorting
method of \citet{Choppala2016} and obtain the sorted index $\zeta_{i}$
for $i=1,...,N$, and the sorted state particles, disturbance particles,
and weights $\widetilde{z}_{t-1}^{i}=z_{t-1}^{\zeta_{i}}$, $\widetilde{\epsilon}_{t-1}^{i}=\epsilon_{t-1}^{\zeta_{i}}$
and $\widetilde{\overline{w}}_{t}^{i}=\overline{w}_{t-1}^{i}$, for
$i=1,...,N$.
\item (2b) Obtain the ancestor indices based on the sorted state particles
$\widetilde{a}_{t-1}^{1:N}$ using the correlated multinomial resampling
in Algorithm \ref{alg:Multinomial-Resampling-Algorithm}.
\item (2c) Obtain the ancestor indices based on original order of the particles
$a_{t-1}^{i}$ for $i=1,...,N$.
\item (2d) Generate $\epsilon_{t}^{i}$ from $m\left(\epsilon_{t}^{i}|u_{\epsilon,t}^{i},\theta\right)$
and set $z_{t}^{i}=F\left(z_{t-1}^{a_{t-1}^{i}},\epsilon_{t}^{i};\theta\right)$, for
$i=1,...,N$
\item (2e) Compute the unnormalised weights $w_{t}^{i}$, for $i=1,...,N$
\item (2f) Compute the normalised weights $\overline{w}_{t}^{i}$ for $i=1,...,N$.
\end{itemize}
\end{algorithm}

\begin{algorithm}[H]
\caption{Multinomial Resampling Algorithm \label{alg:Multinomial-Resampling-Algorithm}}

Input: $u_{A,t-1}$, sorted states $\widetilde{z}_{t-1}^{1:N}$, sorted
disturbances $\widetilde{\epsilon}_{t-1}^{1:N}$, and sorted weights $\widetilde{\overline{w}}_{t-1}^{1:N}$

Output: $\widetilde{a}_{t-1}^{1:N}$
\begin{enumerate}
\item Compute the cumulative weights
\[
\widehat{F}_{t-1}^{N}\left(j\right)=\sum_{i=1}^{j}\widetilde{\overline{w}}_{t-1}^{i}
\]
 based on the sorted state particles $\left\{ \widetilde{z}_{t-1}^{1:N},\widetilde{\overline{w}}_{t-1}^{1:N}\right\} $ or the
sorted disturbance particles $\left\{ \widetilde{\epsilon}_{t-1}^{1:N},\widetilde{\overline{w}}_{t-1}^{1:N}\right\}$
\item set $\widetilde{a}_{t-1}^{i}=\underset{j}{\min} \widehat{F}_{t-1}^{N}\left(j\right)\geq u_{A,t-1}^{i}$
for $i=1,...N$. Note that $\widetilde{a}_{t-1}^{i}$ for $i=1,...,N$
is the ancestor index based on the sorted states or the sorted disturbances.
\end{enumerate}
\end{algorithm}

\section{Delayed Acceptance Multiple PMMH (DA-MPM)
\label{subsec:Delayed-Acceptance-Multiple}}

Delayed acceptance MCMC can used to speed up the computation
for models with expensive likelihoods \citep{Christen2005}. The motivation  in
delayed acceptance is to avoid  computing of the expensive likelihood
if it is likely that the proposed draw will ultimately
be rejected. A first accept-reject stage is applied with the cheap
(or deterministic) approximation substituted for the expensive likelihood
in the Metropolis-Hastings acceptance ratio. Then, only for a proposal that is
accepted in the first stage, the computationally expensive likelihood
is calculated with a second accept-reject stage to ensure detailed
balance is satisfied with respect to the true posterior. This section
discusses the delayed-acceptance mixed PMMH (DA-MPM)
algorithm.

%COMMENT
%
%please give a reference to DA above
%
%END COMMENT

Given the current parameter value $\theta$, the delayed acceptance
Metropolis-Hastings (MH) algorithm proposes a new value $\theta^{'}$
from the proposal density $q\left(\theta^{'}|\theta\right)$ and uses
a cheap approximation  $\widehat{p}^{c}\left(y|\theta^{'}\right)$ to the likelihood
%in the first stage. At stage one, $\widehat{p}^{c}\left(y|\theta^{'}\right)$
%is substituted for the expensive likelihood $\overline{\widehat{p}}_{N}\left(y|\theta^{'},\widetilde{u}^{'}\right)$
in the  MH acceptance probability
\begin{equation}
\alpha_{1}\left(\theta,\widetilde{u};\theta^{'},\widetilde{u}^{'}\right)=\min\left(1,\frac{\widehat{p}^{c}\left(y|\theta^{'}\right)p\left(\theta^{'}\right)q\left(\theta|\theta^{'}\right)}{\widehat{p}^{c}\left(y|\theta\right)p\left(\theta\right)q\left(\theta^{'}|\theta\right)}\right). \label{eq:firststageMH}
\end{equation}
As an alternative to particle filtering, \citet{Norgaard2000}
develop the central difference Kalman filter (CDKF) for estimating
 the state in general non-linear and non-Gaussian state-space models.
\citet{Andreasen2013} shows that the CDKF frequently outperforms the extended Kalman Filter (EKF) for
 general non-linear and non-Gaussian state-space models.
We use the likelihood approximation, $\widehat{p}^{c}(y|\theta)$, obtained from  the  CDKF in the first stage accept-reject
in  \Eqref{eq:firststageMH}.

A second accept-reject stage is applied to a proposal that is accepted in the
first stage, with the second  acceptance probability
\begin{equation}
\alpha_{2}\left(\theta,\widetilde{u};\theta^{'},\widetilde{u}^{'}\right)=\min\left(1,\frac{\overline{\widehat{p}}_{N}\left(y|\theta^{'},\widetilde{u}^{'}\right)\widehat{p}^{c}\left(y|\theta\right)}{\overline{\widehat{p}}_{N}\left(y|\theta,\widetilde{u}\right)\widehat{p}^{c}\left(y|\theta^{'}\right)}\right).\label{eq:secondstageMH}
\end{equation}
The overall acceptance probability $\alpha_{1}\left(\theta,\widetilde{u};\theta^{'},\widetilde{u}^{'}\right)\alpha_{2}\left(\theta,\widetilde{u};\theta^{'},\widetilde{u}^{'}\right)$
ensures that detailed balance is satisfied with respect to the
true posterior. If a rejection occurs at the first stage, then
the expensive evaluation of the likelihood at the second stage is
avoided \citep{Sherlock2017a}. Algorithm \ref{alg:The-Delayed-Acceptance MPM} describes
the Delayed Acceptance MPM algorithm.

\begin{algorithm}[H]

\caption{The Delayed Acceptance Multiple PMMH (DA-MPM) algorithm \label{alg:The-Delayed-Acceptance MPM}}

\begin{itemize}
\item Set the initial values of $\theta^{\left(0\right)}$ arbitrarily
\item Sample $u_s\sim N\left(0,I\right)$ for $s=1,...,S$,  run $S$ (disturbance) particle filters to
 estimate the likelihood $\overline{\widehat{p}}_{N}\left(y|\theta,\widetilde{u}\right)$ as the trimmed mean of $\widehat{p}_{N}\left(y|\theta,u_{s}\right), s=1, \dots, S$; a 0\% is the mean and a 50\% trimmed mean is the median, and run the algorithm \ref{alg:Constructing-proposal-for ADPF} in Section \ref{subsec:Disturbance-Particle-Filter} to construct the (initial) proposal for the auxiliary disturbance particle filter.
%\item Sample $u_s\sim N\left(0,I\right)$ for $s=1,...,S$, and run $S$ particle filters to compute
%an estimate of likelihood $\overline{\widehat{p}}_{N}\left(y|\theta,\widetilde{u}\right)=\frac{1}{S}\sum_{s=1}^{S}\widehat{p}_{N}\left(y|\theta,u_{s}\right)$, and run ancestral tracing algorithm in Section~\ref{sec:AncestralTracing} after each particle filter algorithm to obtain the initial $S$ trajectories of $\epsilon_{s,1:T}$. The mean $\widehat{\mu}_{t}$ and the covariance matrix $\widehat{\Sigma}_{t}$ of the proposal defined in Section \ref{subsec:Disturbance-Particle-Filter} are set as the sample mean and the sample covariance matrix of these $S$ trajectories of $\epsilon_{s,1:T}$ at each time $t$.
\item For each MCMC iteration $p$, $p=1,...,P$,
\begin{itemize}
\item (1) Sample $\theta^{'}$ from the proposal density $q\left(\theta^{'}|\theta\right)$.
\item (2) Compute the likelihood approximation $\widehat{p}^{c}\left(y|\theta^{'}\right)$
using the central difference Kalman filter (CDKF).
\item (3) Accept the first stage proposal with the acceptance probability
in Equation~\eqref{eq:firststageMH}. If the proposal is accepted, then go to step (4),
otherwise go to step (8)
\item (4) Choose index $s$ with probability $1/S$, sample $\eta_{u}\sim N\left(0,I\right)$,
and set $u_{s}^{'}=\rho_{u}u_{s}+\sqrt{1-\rho_{u}^{2}}\eta_{u}$.
\item (5) Run $S$ particle filters to compute an  estimate of likelihood
$\overline{\widehat{p}}_{N}\left(y|\theta^{'},\widetilde{u}^{'}\right)$.
\item (6) Run the algorithm \ref{alg:Constructing-proposal-for ADPF} in Section \ref{subsec:Disturbance-Particle-Filter} to construct the proposal for the auxiliary disturbance particle filter for iteration $p+1$.
\item (7) Accept the second stage Metropolis-Hastings with acceptance probability
in Equation \eqref{eq:secondstageMH}. If the proposal is accepted, then set $\overline{\widehat{p}}_{N}\left(y|\theta,\widetilde{u}\right)^{\left(p\right)}=\overline{\widehat{p}}_{N}\left(y|\theta^{'},\widetilde{u}^{'}\right)$,
$\widehat{p}^{c}\left(y|\theta\right)^{\left(p\right)}=\widehat{p}^{c}\left(y|\theta^{'}\right)$,
$\widetilde{u}^{\left(i\right)}=\widetilde{u}^{'}$, and $\theta^{\left(p\right)}=\theta^{'}$.
Otherwise, go to step (8).
\item (8) Otherwise, $\overline{\widehat{p}}_{N}\left(y|\theta,\widetilde{u}\right)^{\left(p\right)}=\overline{\widehat{p}}_{N}\left(y|\theta,\widetilde{u}\right)^{\left(p-1\right)}$,
$\widehat{p}^{c}\left(y|\theta\right)^{\left(p\right)}=\widehat{p}^{c}\left(y|\theta\right)^{\left(p-1\right)}$,
$\widetilde{u}^{\left(p\right)}=\widetilde{u}^{\left(p-1\right)}$, and $\theta^{\left(p\right)}=\theta^{\left(p-1\right)}$
%\item (9) The mean $\widehat{\mu}_{t}$ and the covariance matrix $\widehat{\Sigma}_{t}$ are set as the sample mean and the sample covariance matrix of these $S$ trajectories of $\epsilon_{s,1:T}^{(i)}$ at each time $t$.

\end{itemize}
\end{itemize}
\end{algorithm}

\section{Overview of DSGE models\label{Supp: overview of DSGE models}}
This section briefly overviews  DSGE models. The equilibrium conditions for a wide variety of DSGE models
can be summarized by
\begin{equation}\label{equation:dsge1}
E_tG(z_{t+1},z_{t},\epsilon_{t+1} ; \theta)=0;
\end{equation}
 $E_t$ is the expectation conditional on date $t$ information and $G$$:\mathbb{R}^{2n+m}\mapsto\mathbb{R}^n$; $z_{t}$ is an $n \times 1$ vector containing all
variables known at time $t$; $\epsilon_{t+1}$ is an $m \times 1$ vector of serially independent
innovations.
The solution to  \Eqref{equation:dsge1} for $z_{t+1}$ can be written as
\begin{equation}\label{equation:dsge2}
z_{t+1}=F(z_{t},\epsilon_{t+1} ,\zeta; \theta)
\end{equation}
such that \[E_tG(F(z_{t},\epsilon_{t+1} ,\zeta; \theta),z_{t},\epsilon_{t+1} ; \theta)=0
\,\, \text{for any}\,\, t. \]
% Eq~\eqref{equation:dsge1} is satisfied $\forall \text{\space} z_{t}$.
For our applications, we assume that $\epsilon_{t} \sim
N(0,\zeta^2\Sigma_\epsilon)$,  where $\zeta$ is a scalar perturbation parameter.
%To describe the solution we follow the notation in \citet{Kollmann2015}.
Under suitable differentiablity assumptions, and using the notation in \citet{Kollmann2015}, we now describe the first and second order accurate solutions of DSGE models.

For most applications, the function $F(\cdot)$  in \Eqref{equation:dsge2} is analytically  intractable and %\Eqref{equation:dsge1paper}
is approximated using local solution techniques. We use first and second order  Taylor series  approximations around the \textit{deterministic} steady state with $\zeta=0$ to  approximate $F$. The \textit{deterministic} steady state $z^s$  satisfies $z^s=F(z^s,0,0; \theta)$.

%\begin{equation*}
%%z_{t+1}=F_1 (\theta) z_{t}	+F_2(\theta) \epsilon_{t+1}, \quad z_0 = 0;
%z_{t+1} - z = F_1 (\theta) ( z_{t}- z ),
%\end{equation*}
%where $z_0 -z= 0$. In terms of deviation from deterministic steady state with $z_t^d=z_t -z$:

A first order-accurate approximation, around the \textit{deterministic} steady state is
\begin{equation}\label{equation:dsge3paper}
%z_{t+1}=F_1 (\theta) z_{t}	+F_2(\theta) \epsilon_{t+1}, \quad z_0 = 0;
z_{t+1}^d=F_1 (\theta) z_{t}^d	+F_2(\theta) \epsilon_{t+1}, \quad z_0^d = 0.
\end{equation}
For \Eqref{equation:dsge3paper} to be stable, it is necessary for all the eigenvalues of $F_1(\theta)$ to be less than 1 in absolute value.  The initial value $z_0^d=0$ because we work with approximations around the \textit{deterministic} steady state. For a given set of parameters $\theta$, the matrices $F_1(\theta)$ and  $F_2(\theta)$ can be solved using existing software;  our applications use Dynare.

The second order accurate approximation (around the \textit{deterministic} steady state) is
\begin{equation}\label{equation:dsge2paper}
%z_{t+1}=F_{0}(\theta) \zeta^{2}+F_{1}(\theta) z_{t}+F_2(\theta) \epsilon_{t+1}+F_{11}(\theta) \vech2(z_t)+F_{12}(\theta)(z_t\otimes\epsilon_{t+1})+F_{22}(\theta) \vech2(\epsilon_{t+1});
z_{t+1}^d=F_{0}(\theta) \zeta^{2}+F_{1}(\theta) z_{t}^d+F_2(\theta) \epsilon_{t+1}+F_{11}(\theta) \vechsq(z_t^d)+F_{12}(\theta)(z_t^d\otimes\epsilon_{t+1})+F_{22}(\theta) \vechsq(\epsilon_{t+1});
\end{equation}
where $z_{t}^d=( z_{t}- z^s)$ and for any vector $x:=(x_1, \dots, x_m)^\top$, we define $\vechsq(x) := \vech(xx^\top),$
where $\vech(xx^\top)$ is the strict upper triangle and diagonal of $xx^\top$,
such that $\vech(xx^\top):=(x_1^2, x_1x_2 , x_2^2, x_1 x_3, ... , x_m^2)^\top$.  The term $F_{0}(\theta) \zeta^{2}$ captures the level correction due to uncertainty from taking a second-order approximation. For our analysis we normalize the perturbation parameter $\zeta$ to 1. As before, for a given set of parameters $\theta$, the matrices $F_1(\theta), F_2(\theta),F_{11}(\theta), F_{12}(\theta),F_{22}(\theta)$ can be solved using Dynare.

The measurement (observation) equation for the DSGE model and its approximations is
\begin{align*}
y_{t} & = Hz_{t}+\eta_{t},\;\eta_{t}\sim N\left(0,\Sigma_{\eta}\right), %\label{eq:measurementeqn}
\end{align*}
where $H$ is a known matrix;
$\{\eta_{t}\}$ is an independent $N\left(0,\Sigma_{\eta}\right)$ and sequence and it is also independent  of the  $\{\epsilon_t \}$  sequence. The matrix $\Sigma_{\eta}$ is usually unknown and is estimated from the data.

%\section*{COMMENT}
%please replace the above paragraph by the corresponding paragraph in the main article.
%\section*{END COMMENT}

%For a vector $x=(x_1, \dots, x_q)^\top$
%\begin{equation*}
%P(x):= \vech(xx^\top)
% =(x_1^2, x_1 x_2,..,x_1 x_q, x_2^2, x_2 x_3,..,x_2 x_q,..,x_{q-1}^2, x_{q-1} x_q,..,x_q^2 ).
%\end{equation*}

\subsection{Pruned State Space Representation of DSGE Models \label{DescriptionOfPrunedDSGEModel}}
To obtain a stable solution we use the pruning approach recommended in \citet{Kim2008};  for details of how pruning is implemented in DSGE models, see  \citet{Kollmann2015} and  \citet{AFVVRR}. The pruned second order solution preserves only second order accurate terms by using the first order accurate solution to calculate $P(z^{(2)}_t)$ and $(z^{(2)}_t\otimes\epsilon_{t+1})$; where $P(\cdot)$ is defined below.
%\footnote{To see why pruning preserves only the second order effects consider the following. For any variable $a_t$ we can show that $a_t=a_t^{(n)}+R^{(n+1)}$ where $a_t^{(n)}$ denotes the $n^{th}$ order accurate solution and $R^{(n+1)}$ summarizes higher order effects. Consider the $i^{th}$ and $j^{th}$ variables in $z_t$. Consider  $z_t^i z_t^j=({z_t^i}^{(2)}+R^{(3)})({z_t^j}^{(2)}+R^{(3)})$. In the absence of pruning, $z_t^i z_t^j={z_t^j}^{(2)}{z_t^i}^{(2)} +{z_t^j}^{(2)}R^{(3)} +{z_t^i}^{(2)}R^{(3)} +R^{(6)}.
%$ Pruning uses the first order accurate solution to compute the cross products of variables captured in $P(z_t)$ i.e.  $z_t^i z_t^j={z_t^j}^{(1)}{z_t^i}^{(1)} +R^{(3)}$.}
That is, we obtain the solution preserving \textit{only} the second order effects by using $P(z_t^{(1)})$ instead of $P(z^{(2)}_t)$ and $(z_t^{(1)}\otimes\epsilon_{t+1})$ instead of $(z^{(2)}_t\otimes\epsilon_{t+1})$ in \Eqref{equation:dsge2paper}. We note
that the variables are still expressed as deviations from steady state; however, for notational simplicity, we drop the superscript $d$. 

The evolution of  $P(z_t^{(1)})$ is
\begin{equation*}
P(z_{t+1}^{(1)})=K_{11} P(z_t^{(1)})+K_{12}(z_t^{(1)}\otimes\epsilon_{t+1})+K_{22}P(\epsilon_{t+1}),
\end{equation*}with $K_{11},K_{12}, K_{22}$ being functions of $F_1$ and $F_2$, respectively. A consequence of using pruning in preserving only second order effects is that it increases the dimension of the state space. The pruning-augmented solution of the DSGE model is given by
{\small{\begin{multline*}
\begin{bmatrix}
z^{(2)}_{t+1}
\\ P(z_{t+1}^{(1)})
\\ z_{t+1}^{(1)}
\end{bmatrix} = \begin{bmatrix}F_0 \zeta^2
\\0
\\ 0
\end{bmatrix}+\begin{bmatrix}F_1
& F_{11} & 0 \\
0 & K_{11} & 0\\
0&  0& F_1
\end{bmatrix} \begin{bmatrix}z^{(2)}_{t}
\\  P(z_{t}^{(1)})
\\ z_{t}^{(1)}
\end{bmatrix}+\begin{bmatrix}F_2
\\ 0
\\ F_2
\end{bmatrix}\epsilon_{t+1}+\begin{bmatrix}F_{12}
\\ K_{12}
\\ 0
\end{bmatrix}(z_{t}^{(1)}\otimes\epsilon_{t+1})+\\\begin{bmatrix}F_{22}
\\ K_{22}
\\ 0	
\end{bmatrix}P(\epsilon_{t+1}).
\end{multline*}}}

The augmented state representation of the pruned second order accurate system becomes
\begin{equation*}
\widetilde{z}_{t+1}=g_0+G_1 \widetilde{z}_t+G_2\epsilon_{t+1}+G_{12}(z_t^{(1)}\otimes\epsilon_{t+1})+G_{22}P(\epsilon_{t+1}),
\end{equation*}
where $\widetilde{z}_t=[z^{(2)}_{t},P(z_{t}^{(1)}),z_{t}^{(1)}]^\top$.
We now summarize the state transition and the measurement equations for the pruned second order accurate system as
\begin{eqnarray}
\begin{aligned}\label{equation:dsge4}
&y_{t+1} = H \widetilde{z}_{t+1}+\eta_{t+1},\;\eta_{t+1}\sim N\left(0,\Sigma_{\eta}\right)\label{eq:measurementeqn}\\
&\widetilde{z}_{t+1}=g_0+G_1 \widetilde{z}_t+G_2\epsilon_{t+1}+G_{12}(z_t^{(1)}\otimes\epsilon_{t+1})+G_{22}P(\epsilon_{t+1}), \; \epsilon_{t+1}\sim N\left(0,\zeta^2\Sigma_{\epsilon}\right)
\end{aligned}
\end{eqnarray}

\def\dim{{\rm dim}}
The matrix $H$ selects the observables from the pruning augmented state space representation of the system. If $\dim(z_t^{(2)})=n\times1$ and $\dim(\epsilon_t)=m\times1$ then $\dim(P(z_t^{(1)})=n(n+1)/2\times1$ and
$\dim(\widetilde{z}_{t})=[n+n(n+1)/2+n]\times1$. If $n_y \leq n$ denotes the number of observables then
the selection matrix $[H=[H_{ny};\mathbf{0}]]$, where $\dim(H_{ny})=n_y\times [n+n(n+1)/2+n])$ and $\dim(\mathbf{0})=(n-n_y)\times [n+n(n+1)/2+n])$. The selection matrix $H_{ny}$ consists only of  zeros and ones.

\section{Additional tables and figures for the linear Gaussian state space model example\label{LGSSexample}}

This section gives additional tables and figures for the linear Gaussian state space model in Section \ref{subsec:Linear-Gaussian-State Space Model}.

Tables \ref{table1dimT200} and \ref{table1dimT300} show the variance of the log of the estimated likelihood obtained by using  five different approaches for the $1$ dimensional linear Gaussian state space model with $T=200$ and $T=300$ time periods, respectively. The table shows that for all approaches there is a substantial reduction in the variance of the log of the estimated likelihood when the number of particle filters $S$ increases and the number of particles within each particle filter $N$ increases. 

Tables~\ref{table5dimT200} -- \ref{table10dimT200}  show the variance of the log of the estimated likelihood obtained by using the five estimators of the likelihood for the $d=5$ dimensional linear Gaussian state space model with $T=200$ and $T=300$ time periods and $d=10$ with $T=200$ time period. The table shows that: (1) there is no substantial reduction in variance of the log of the estimated likelihood for the 0\% trimmed mean. (2)~The 5\%, 10\%, 25\%, and 50\% trimmed means decrease the variance substantially as  $S$ and/or $N$ increase. The  25\% and 50\% trimmed means have the smallest variance for all cases.

\begin{table}[H]
\caption{Comparing the variance of the log of the estimated likelihood for five different estimators of  the likelihood: I: Averaging the likelihood (0\% trimmed mean), II: Averaging
the likelihood (5\% trimmed mean), III: Averaging the likelihood (10\%
trimmed mean), IV: Averaging the likelihood (25\% trimmed mean), and
V: Averaging the likelihood (50\% trimmed mean) for $d=1$ dimension
linear Gaussian state space model with $T=200$. The variance of the
log of the estimated likelihood of a single particle filter is reported
under the column ``Single''. The results are based on $1000$ independent
runs. \label{table1dimT200}}

\centering{}%
\begin{tabular}{cccccccc}
\hline 
$N$ & $S$ & $I$ & $II$ & $III$ & $IV$ & $V$ & Single\tabularnewline
\hline 
100 & 1 &  &  &  &  &  & $2.881$\tabularnewline
 & 20 & $0.271$ & $0.191$ & $0.175$ & $0.166$ & $0.191$ & \tabularnewline
 & 50 & $0.131$ & $0.077$ & $0.068$ & $0.064$ & $0.080$ & \tabularnewline
 & $100$ & $0.081$ & $0.040$ & $0.036$ & $0.035$ & $0.043$ & \tabularnewline
 & 500 & $0.018$ & $0.007$ & $0.007$ & $0.007$ & $0.008$ & \tabularnewline
 & 1000 & $0.011$ & $0.004$ & $0.003$ & $0.003$ & $0.004$ & \tabularnewline
\hline 
250 & 1 &  &  &  &  &  & $1.066$\tabularnewline
 & 20 & $0.073$ & $0.061$ & $0.058$ & $0.061$ & $0.077$ & \tabularnewline
 & 50 & $0.030$ & $0.024$ & $0.022$ & $0.023$ & $0.030$ & \tabularnewline
 & 100 & $0.016$ & $0.012$ & $0.011$ & $0.012$ & $0.015$ & \tabularnewline
 & 500 & $0.003$ & $0.002$ & $0.002$ & $0.002$ & $0.003$ & \tabularnewline
 & 1000 & $0.002$ & $0.001$ & $0.001$ & $0.001$ & $0.002$ & \tabularnewline
\hline 
500 & 1 &  &  &  &  &  & $0.493$\tabularnewline
 & 20 & $0.032$ & $0.029$ & $0.028$ & $0.030$ & $0.037$ & \tabularnewline
 & 50 & $0.013$ & $0.012$ & $0.012$ & $0.013$ & $0.017$ & \tabularnewline
 & 100 & $0.007$ & $0.006$ & $0.006$ & $0.006$ & $0.008$ & \tabularnewline
 & 500 & $0.001$ & $0.001$ & $0.001$ & $0.001$ & $0.002$ & \tabularnewline
 & 1000 & $0.001$ & $0.001$ & $0.001$ & $0.001$ & $0.001$ & \tabularnewline
\hline 
1000 & 1 &  &  &  &  &  & $0.262$\tabularnewline
 & 20 & $0.014$ & $0.014$ & $0.014$ & $0.016$ & $0.020$ & \tabularnewline
 & 50 & $0.006$ & $0.006$ & $0.006$ & $0.006$ & $0.008$ & \tabularnewline
 & 100 & $0.003$ & $0.003$ & $0.003$ & $0.003$ & $0.004$ & \tabularnewline
 & 500 & $0.001$ & $0.001$ & $0.001$ & $0.001$ & $0.001$ & \tabularnewline
 & 1000 & $0.000$ & $0.000$ & $0.000$ & $0.000$ & $0.000$ & \tabularnewline
\hline 
\end{tabular}
\end{table}

\begin{table}[H]
\caption{Comparing the variance of the log of the estimated likelihood for five different estimators of  the likelihood: I: Averaging the likelihood (0\% trimmed mean), II: Averaging
the likelihood (5\% trimmed mean), III: Averaging the likelihood (10\%
trimmed mean), IV: Averaging the likelihood (25\% trimmed mean), and
V: Averaging the likelihood (50\% trimmed mean), for $d=1$ dimension
linear Gaussian state space model with $T=300$. The variance of the
log of estimated likelihood of a single particle filter is reported
under the column ``Single''. The results are based on $1000$ independent
runs. \label{table1dimT300}}

\centering{}%
\begin{tabular}{cccccccc}
\hline 
$N$ & $S$ & $I$ & $II$ & $III$ & $IV$ & $V$ & Single\tabularnewline
\hline 
100 & 1 &  &  &  &  &  & $3.277$\tabularnewline
 & 20 & $0.455$ & $0.298$ & $0.255$ & $0.229$ & $0.275$ & \tabularnewline
 & 50 & $0.251$ & $0.123$ & $0.098$ & $0.088$ & $0.109$ & \tabularnewline
 & $100$ & $0.149$ & $0.056$ & $0.048$ & $0.043$ & $0.054$ & \tabularnewline
 & 500 & $0.041$ & $0.013$ & $0.011$ & $0.010$ & $0.012$ & \tabularnewline
 & 1000 & $0.023$ & $0.005$ & $0.005$ & $0.004$ & $0.005$ & \tabularnewline
\hline 
250 & 1 &  &  &  &  &  & $1.509$\tabularnewline
 & 20 & $0.121$ & $0.096$ & $0.091$ & $0.092$ & $0.110$ & \tabularnewline
 & 50 & $0.054$ & $0.038$ & $0.035$ & $0.036$ & $0.047$ & \tabularnewline
 & 100 & $0.027$ & $0.019$ & $0.018$ & $0.018$ & $0.023$ & \tabularnewline
 & 500 & $0.006$ & $0.004$ & $0.003$ & $0.004$ & $0.005$ & \tabularnewline
 & 1000 & $0.003$ & $0.002$ & $0.002$ & $0.002$ & $0.002$ & \tabularnewline
\hline 
500 & 1 &  &  &  &  &  & $0.686$\tabularnewline
 & 20 & $0.048$ & $0.041$ & $0.040$ & $0.041$ & $0.050$ & \tabularnewline
 & 50 & $0.021$ & $0.018$ & $0.018$ & $0.018$ & $0.023$ & \tabularnewline
 & 100 & $0.011$ & $0.009$ & $0.009$ & $0.009$ & $0.012$ & \tabularnewline
 & 500 & $0.002$ & $0.002$ & $0.002$ & $0.002$ & $0.002$ & \tabularnewline
 & 1000 & $0.001$ & $0.001$ & $0.001$ & $0.001$ & $0.001$ & \tabularnewline
\hline 
1000 & 1 &  &  &  &  &  & $0.344$\tabularnewline
 & 20 & $0.020$ & $0.019$ & $0.019$ & $0.021$ & $0.027$ & \tabularnewline
 & 50 & $0.008$ & $0.008$ & $0.008$ & $0.008$ & $0.010$ & \tabularnewline
 & 100 & $0.004$ & $0.004$ & $0.004$ & $0.004$ & $0.006$ & \tabularnewline
 & 500 & $0.001$ & $0.001$ & $0.001$ & $0.001$ & $0.001$ & \tabularnewline
 & 1000 & $0.000$ & $0.000$ & $0.000$ & $0.000$ & $0.001$ & \tabularnewline
\hline 
\end{tabular}
\end{table}

\begin{table}[H]
\caption{Comparing the variance of the log of the estimated likelihood for five different estimators of  the likelihood: I: Averaging the likelihood (0\% trimmed mean), II: Averaging
the likelihood (5\% trimmed mean), III: Averaging the likelihood (10\%
trimmed mean), IV: Averaging the likelihood (25\% trimmed mean), and
V: Averaging the likelihood (50\% trimmed mean), for $d=5$ dimension
linear Gaussian state space model with $T=200$. The variance of the
log of estimated likelihood of a single particle filter is reported
under the column ``Single''. The results are based on $1000$ independent
runs. \label{table5dimT200}}

\centering{}%
\begin{tabular}{cccccccc}
\hline 
$N$ & $S$ & $I$ & $II$ & $III$ & $IV$ & $V$ & Single\tabularnewline
\hline 
100 & 1 &  &  &  &  &  & $44.224$\tabularnewline
 & 20 & $11.611$ & $5.939$ & $4.797$ & $3.612$ & $3.399$ & \tabularnewline
 & 50 & $8.280$ & $2.935$ & $1.88$ & $1.523$ & $1.515$ & \tabularnewline
 & $100$ & $6.681$ & $1.408$ & $0.997$ & $0.712$ & $0.725$ & \tabularnewline
 & 500 & $4.871$ & $0.238$ & $0.177$ & $0.136$ & $0.148$ & \tabularnewline
 & 1000 & $4.358$ & $0.134$ & $0.099$ & $0.068$ & $0.068$ & \tabularnewline
\hline 
250 & 1 &  &  &  &  &  & $21.175$\tabularnewline
 & 20 & $4.465$ & $2.627$ & $2.099$ & $1.556$ & $1.517$ & \tabularnewline
 & 50 & $3.470$ & $1.212$ & $0.781$ & $0.551$ & $0.588$ & \tabularnewline
 & 100 & $2.677$ & $0.527$ & $0.411$ & $0.295$ & $0.290$ & \tabularnewline
 & 500 & $1.652$ & $0.111$ & $0.081$ & $0.059$ & $0.063$ & \tabularnewline
 & 1000 & $1.511$ & $0.054$ & $0.039$ & $0.029$ & $0.032$ & \tabularnewline
\hline 
500 & 1 &  &  &  &  &  & $9.981$\tabularnewline
 & 20 & $2.415$ & $1.342$ & $1.023$ & $0.804$ & $0.819$ & \tabularnewline
 & 50 & $1.607$ & $0.550$ & $0.371$ & $0.294$ & $0.326$ & \tabularnewline
 & 100 & $1.216$ & $0.244$ & $0.191$ & $0.154$ & $0.175$ & \tabularnewline
 & 500 & $0.691$ & $0.053$ & $0.039$ & $0.030$ & $0.034$ & \tabularnewline
 & 1000 & $0.516$ & $0.026$ & $0.020$ & $0.015$ & $0.018$ & \tabularnewline
\hline 
1000 & 1 &  &  &  &  &  & $6.090$\tabularnewline
 & 20 & $1.127$ & $0.631$ & $0.497$ & $0.410$ & $0.463$ & \tabularnewline
 & 50 & $0.661$ & $0.255$ & $0.181$ & $0.151$ & $0.167$ & \tabularnewline
 & 100 & $0.502$ & $0.119$ & $0.094$ & $0.076$ & $0.085$ & \tabularnewline
 & 500 & $0.201$ & $0.023$ & $0.020$ & $0.016$ & $0.019$ & \tabularnewline
 & 1000 & $0.141$ & $0.011$ & $0.009$ & $0.007$ & $0.009$ & \tabularnewline
\hline 
\end{tabular}
\end{table}

\begin{table}[H]
\caption{Comparing the variance of the log of the estimated likelihood for five different estimators of  the likelihood: I: Averaging the likelihood (0\% trimmed mean), II: Averaging
the likelihood (5\% trimmed mean), III: Averaging the likelihood (10\%
trimmed mean), IV: Averaging the likelihood (25\% trimmed mean), and
V: Averaging the likelihood (50\% trimmed mean), for $d=5$ dimension
linear Gaussian state space model with $T=300$. The variance of the
log of estimated likelihood of a single particle filter is reported
under the column ``Single''. The results are based on $1000$ independent
runs.\label{table5dimT300}}

\centering{}%
\begin{tabular}{cccccccc}
\hline 
$N$ & $S$ & $I$ & $II$ & $III$ & $IV$ & $V$ & Single\tabularnewline
\hline 
100 & 1 &  &  &  &  &  & $56.287$\tabularnewline
 & 20 & $16.252$ & $8.660$ & $7.130$ & $4.796$ & $4.477$ & \tabularnewline
 & 50 & $12.211$ & $4.524$ & $2.866$ & $2.153$ & $1.953$ & \tabularnewline
 & $100$ & $10.481$ & $2.174$ & $1.456$ & $0.956$ & $0.970$ & \tabularnewline
 & 500 & $7.432$ & $0.402$ & $0.285$ & $0.203$ & $0.192$ & \tabularnewline
 & 1000 & $5.960$ & $0.196$ & $0.137$ & $0.096$ & $0.108$ & \tabularnewline
\hline 
250 & 1 &  &  &  &  &  & $28.276$\tabularnewline
 & 20 & $6.416$ & $3.659$ & $2.894$ & $1.986$ & $1.994$ & \tabularnewline
 & 50 & $4.719$ & $1.718$ & $1.182$ & $0.831$ & $0.840$ & \tabularnewline
 & 100 & $3.897$ & $0.744$ & $0.564$ & $0.401$ & $0.419$ & \tabularnewline
 & 500 & $2.653$ & $0.161$ & $0.114$ & $0.078$ & $0.081$ & \tabularnewline
 & 1000 & $2.092$ & $0.077$ & $0.057$ & $0.040$ & $0.044$ & \tabularnewline
\hline 
500 & 1 &  &  &  &  &  & $14.327$\tabularnewline
 & 20 & $3.261$ & $1.881$ & $1.326$ & $0.990$ & $0.996$ & \tabularnewline
 & 50 & $2.222$ & $0.768$ & $0.512$ & $0.393$ & $0.422$ & \tabularnewline
 & 100 & $1.685$ & $0.338$ & $0.259$ & $0.188$ & $0.219$ & \tabularnewline
 & 500 & $1.057$ & $0.070$ & $0.054$ & $0.039$ & $0.045$ & \tabularnewline
 & 1000 & $0.784$ & $0.036$ & $0.027$ & $0.020$ & $0.022$ & \tabularnewline
\hline 
1000 & 1 &  &  &  &  &  & $7.269$\tabularnewline
 & 20 & $1.432$ & $0.782$ & $0.662$ & $0.544$ & $0.586$ & \tabularnewline
 & 50 & $0.887$ & $0.332$ & $0.257$ & $0.219$ & $0.239$ & \tabularnewline
 & 100 & $0.648$ & $0.154$ & $0.124$ & $0.103$ & $0.117$ & \tabularnewline
 & 500 & $0.291$ & $0.030$ & $0.024$ & $0.019$ & $0.022$ & \tabularnewline
 & 1000 & $0.255$ & $0.016$ & $0.012$ & $0.010$ & $0.012$ & \tabularnewline
\hline 
\end{tabular}
\end{table}

\begin{table}[H]
\caption{Comparing the variance of the log of the estimated likelihood for five different estimators of  the likelihood: I: Averaging the likelihood (0\% trimmed mean), II: Averaging
the likelihood (5\% trimmed mean), III: Averaging the likelihood (10\%
trimmed mean), IV: Averaging the likelihood (25\% trimmed mean), and
V: Averaging the likelihood (50\% trimmed mean), for $d=10$ dimension
linear Gaussian state space model with $T=200$. The variance of the
log of estimated likelihood of a single particle filter is reported
under the column ``Single''. The results are based on $1000$ independent
runs. \label{table10dimT200}}

\centering{}%
\begin{tabular}{cccccccc}
\hline 
$N$ & $S$ & $I$ & $II$ & $III$ & $IV$ & $V$ & Single\tabularnewline
\hline 
100 & 1 &  &  &  &  &  & $253.715$\tabularnewline
 & 20 & $63.021$ & $37.828$ & $28.792$ & $21.095$ & $19.567$ & \tabularnewline
 & 50 & $48.295$ & $19.969$ & $12.351$ & $7.951$ & $7.221$ & \tabularnewline
 & $100$ & $43.519$ & $9.194$ & $6.272$ & $3.771$ & $3.664$ & \tabularnewline
 & 500 & $31.729$ & $1.882$ & $1.310$ & $0.830$ & $0.725$ & \tabularnewline
 & 1000 & $27.667$ & $0.940$ & $0.586$ & $0.382$ & $0.362$ & \tabularnewline
\hline 
250 & 1 &  &  &  &  &  & $128.193$\tabularnewline
 & 20 & $35.400$ & $20.403$ & $15.420$ & $10.949$ & $9.358$ & \tabularnewline
 & 50 & $28.430$ & $10.005$ & $6.600$ & $4.623$ & $4.031$ & \tabularnewline
 & 100 & $24.810$ & $4.921$ & $3.435$ & $2.154$ & $1.899$ & \tabularnewline
 & 500 & $16.666$ & $0.889$ & $0.599$ & $0.423$ & $0.387$ & \tabularnewline
 & 1000 & $15.061$ & $0.475$ & $0.312$ & $0.219$ & $0.204$ & \tabularnewline
\hline 
500 & 1 &  &  &  &  &  & $77.094$\tabularnewline
 & 20 & $22.465$ & $12.918$ & $8.891$ & $6.612$ & $6.269$ & \tabularnewline
 & 50 & $17.785$ & $6.131$ & $3.533$ & $2.465$ & $2.450$ & \tabularnewline
 & 100 & $14.073$ & $2.767$ & $1.861$ & $1.276$ & $1.271$ & \tabularnewline
 & 500 & $10.376$ & $0.576$ & $0.415$ & $0.270$ & $0.243$ & \tabularnewline
 & 1000 & $8.640$ & $0.290$ & $0.209$ & $0.131$ & $0.121$ & \tabularnewline
\hline 
1000 & 1 &  &  &  &  &  & $50.319$\tabularnewline
 & 20 & $12.620$ & $6.955$ & $5.598$ & $3.803$ & $3.663$ & \tabularnewline
 & 50 & $9.230$ & $3.436$ & $2.406$ & $1.663$ & $1.515$ & \tabularnewline
 & 100 & $8.070$ & $1.645$ & $1.144$ & $0.759$ & $0.738$ & \tabularnewline
 & 500 & $5.825$ & $0.308$ & $0.220$ & $0.147$ & $0.150$ & \tabularnewline
 & 1000 & $5.255$ & $0.153$ & $0.116$ & $0.076$ & $0.076$ & \tabularnewline
\hline 
\end{tabular}
\end{table}

\begin{table}[H]
\caption{Comparing the performance of different PMMH samplers with different numbers of particle filters $S$ and different numbers of particles $N$ in each particle filter for estimating linear Gaussian state space model using a simulate dataset with $T=200$
and $d=10$ dimensions.  Sampler I: Correlated
PMMH of \citet{Deligiannidis2018}. Sampler II: MPM with averaging
likelihood (5\% trimmed mean). Sampler III: MPM with averaging likelihood
(10\% trimmed mean). Sampler IV: MPM with averaging likelihood (25\%
trimmed mean). Time denotes the time taken in seconds for
one iteration of the method.\label{tabledim10T200IF}}

\begin{centering}
\begin{tabular}{ccccccccccc}
\hline 
   & {\footnotesize{}I} & \multicolumn{3}{c}{{\footnotesize{}II}} & \multicolumn{3}{c}{{\footnotesize{}III}} & \multicolumn{3}{c}{{\footnotesize{}IV}}\tabularnewline
\hline 
{\footnotesize{}N}  & {\footnotesize{}20000} & {\footnotesize{}100} & {\footnotesize{}250} & {\footnotesize{}500} & {\footnotesize{}100} & {\footnotesize{}250} & {\footnotesize{}500} & {\footnotesize{}100} & {\footnotesize{}250} & {\footnotesize{}500}\tabularnewline
{\footnotesize{}S} & {\footnotesize{}1} & {\footnotesize{}100} & {\footnotesize{}100} & {\footnotesize{}100} & {\footnotesize{}100} & {\footnotesize{}100} & {\footnotesize{}100} & {\footnotesize{}100} & {\footnotesize{}100} & {\footnotesize{}100}\tabularnewline
{\footnotesize{}$\widehat{\textrm{IF}}$} & {\footnotesize{}42.076} & {\footnotesize{}NA} & {\footnotesize{}137.725} & {\footnotesize{}22.507} & {\footnotesize{}253.251} & {\footnotesize{}17.065} & {\footnotesize{}10.658} & {\footnotesize{}18.901} & {\footnotesize{}9.748} & {\footnotesize{}7.899}\tabularnewline
{\footnotesize{}CT} & {\footnotesize{}2.44} & {\footnotesize{}0.32} & {\footnotesize{}0.71} & {\footnotesize{}1.21} & {\footnotesize{}0.32} & {\footnotesize{}0.71} & {\footnotesize{}1.21} & {\footnotesize{}0.32} & {\footnotesize{}0.71} & {\footnotesize{}1.21}\tabularnewline
{\footnotesize{}$\widehat{\textrm{TNIF}}$} & {\footnotesize{}102.665} & {\footnotesize{}NA} & {\footnotesize{}97.785} & {\footnotesize{}27.234} & {\footnotesize{}81.040} & {\footnotesize{}12.116} & {\footnotesize{}12.896} & {\footnotesize{}6.048} & {\footnotesize{}6.921} & {\footnotesize{}9.558}\tabularnewline
{\footnotesize{}$\widehat{\textrm{RTNIF}}$}  & {\footnotesize{}1} & {\footnotesize{}NA} & {\footnotesize{}0.952} & {\footnotesize{}0.265} & {\footnotesize{}0.789} & {\footnotesize{}0.119} & {\footnotesize{}0.126} & {\footnotesize{}0.059} & {\footnotesize{}0.067} & {\footnotesize{}0.093}\tabularnewline
\hline 
\end{tabular}
\par\end{centering}
\end{table}

\begin{figure}[H]
\caption{Left: Kernel density estimates of the parameter $\theta$ estimated
using (1) Metropolis-Hastings algorithm with the (exact) Kalman filter method;
(2) Correlated PMMH with $N=20000$ particles; (3) MPM with averaging the
likelihood (5\% trimmed mean, $S=100$, $N=250$); (4) MPM with averaging
the likelihood (5\% trimmed mean, $S=100$, $N=500$). Middle: Kernel
density estimates of the parameter $\theta$ estimated using: (1)
Metropolis-Hastings algorithm with exact Kalman filter method; (2)
Correlated PMMH with 20000 particles; (3) MPM with averaging the likelihood
(10\% trimmed mean, $S=100$, $N=250$); (4) MPM with averaging the
likelihood (10\% trimmed mean, $S=100$, $N=500$). Right: Kernel
density estimates of the parameter $\theta$ estimated using: (1)
Metropolis-Hastings algorithm with exact Kalman filter method; (2)
Correlated PMMH with 20000 particles; (3) MPM with averaging the likelihood
(25\% trimmed mean, $S=100$, $N=250$); (4) MPM with averaging the
likelihood (25\% trimmed mean, $S=100$, $N=500$) for estimating
$d=10$ dimensions linear Gaussian state space model with $T=200$. \label{kerneldensitydim10T200}}

\centering{}\includegraphics[width=15cm,height=6cm]{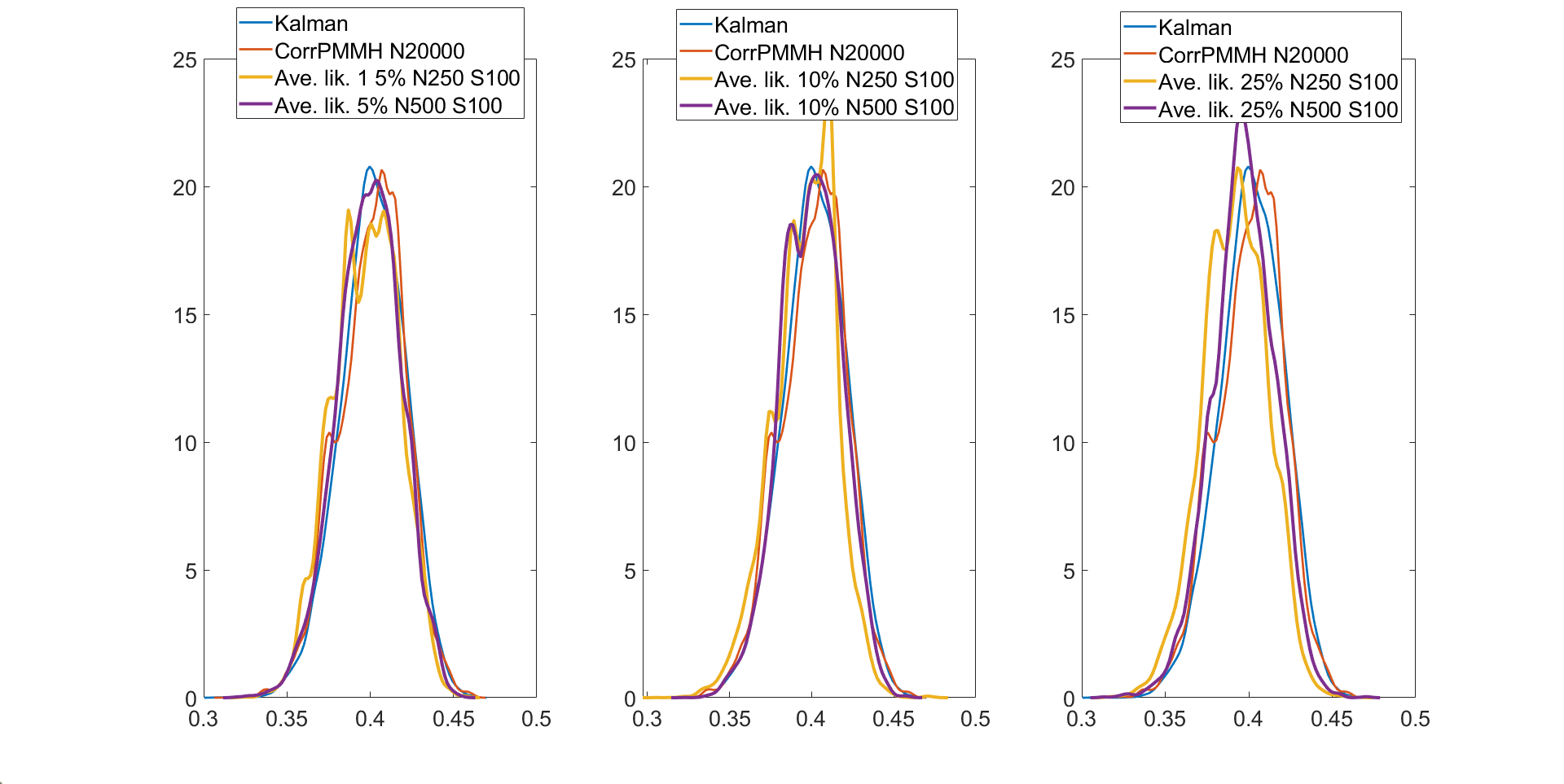}
\end{figure}

Table \ref{tabledim10T200IF} reports the $\widehat{\textrm{IF}}$, $\widehat{\textrm{TNIF}}$, and $\widehat{\textrm{RTNIF}}$ values for the parameter $\theta$ in the linear Gaussian state space model with $d=10$ dimensions and $T=200$ time periods estimated using the four different MCMC samplers: (1) the correlated PMMH of \cite{Deligiannidis2018}, (2) the MPM  with 5\% trimmed mean, (3) the MPM with 10\% trimmed mean, and (4) the MPM with 25\% trimmed mean.
The computing time reported in the table is the time to run a single particle filter for the CPM and $S$ particle filters for the MPM approach.
The table shows that: (1) The correlated PMMH requires more than $20000$ particles to improve the mixing of the MCMC chain for the parameter $\theta$; (2) the CPU time for running a single particle filter with $N=20000$ particles is $3.43$ times higher than running multiple particle filters with $S=100$ and $N=250$. The MPM method can be much faster than the CPM method if it
is run using high-performance computing with a large number of cores; (3) the
MPM allows us to use a much smaller number of particles for each independent
PF and these multiple PFs can be run independently; (4) in terms of $\widehat{\textrm{RTNIF}}$, the 5\%, 10\%, and 25\% trimmed means with $S=100$ and $N=250$ are $1.05$, $8.40$, and $14.93$ times smaller than the correlated PMMH with $N=20000$ particles; (6) the best sampler for this example is the 25\% trimmed mean approach with $S=100$ and $N=100$ particles. Figure \ref{kerneldensitydim10T200} shows the kernel density estimates of the parameter $\theta$ estimated using Metropolis-Hastings algorithm with the (exact) Kalman filter method and the MPM algorithm  with 5\%, 10\%, and 25\% trimmed means. The figure shows that the approximate posteriors obtained by various approaches using the trimmed means are very close to the true posterior. Figure \ref{kerneldensitydim10T200} also shows that the approximate posterior obtained using the MPM with the trimmed mean of the likelihood is very close to the exact posterior obtained using the correlated PMMH.

\begin{figure}[H]
\caption{The estimated correlation of log of the estimated likelihoods evaluated at $\theta=0.4=\theta^{'}$ (top), $\theta=0.4$ and $\theta^{'}=0.399$ (middle), and $\theta=0.4$ and $\theta^{'}=0.385$ (bottom) for the 10 dimensional linear Gaussian state space model with $T=300$ using the five estimators: (1) MPM (Blocking) with 0\% trimmed mean of the likelihood, (2) MPM (Blocking) with 5\% trimmed mean of the likelihood, (3) MPM (Blocking) with 10\% trimmed mean of the likelihood, (4) MPM (Blocking) with 25\% trimmed mean of the likelihood, (5) MPM (Blocking) with 50\% trimmed mean of the likelihood. 
\label{corr_LGSS_sim}}

\centering{}\includegraphics[width=15cm,height=8cm]{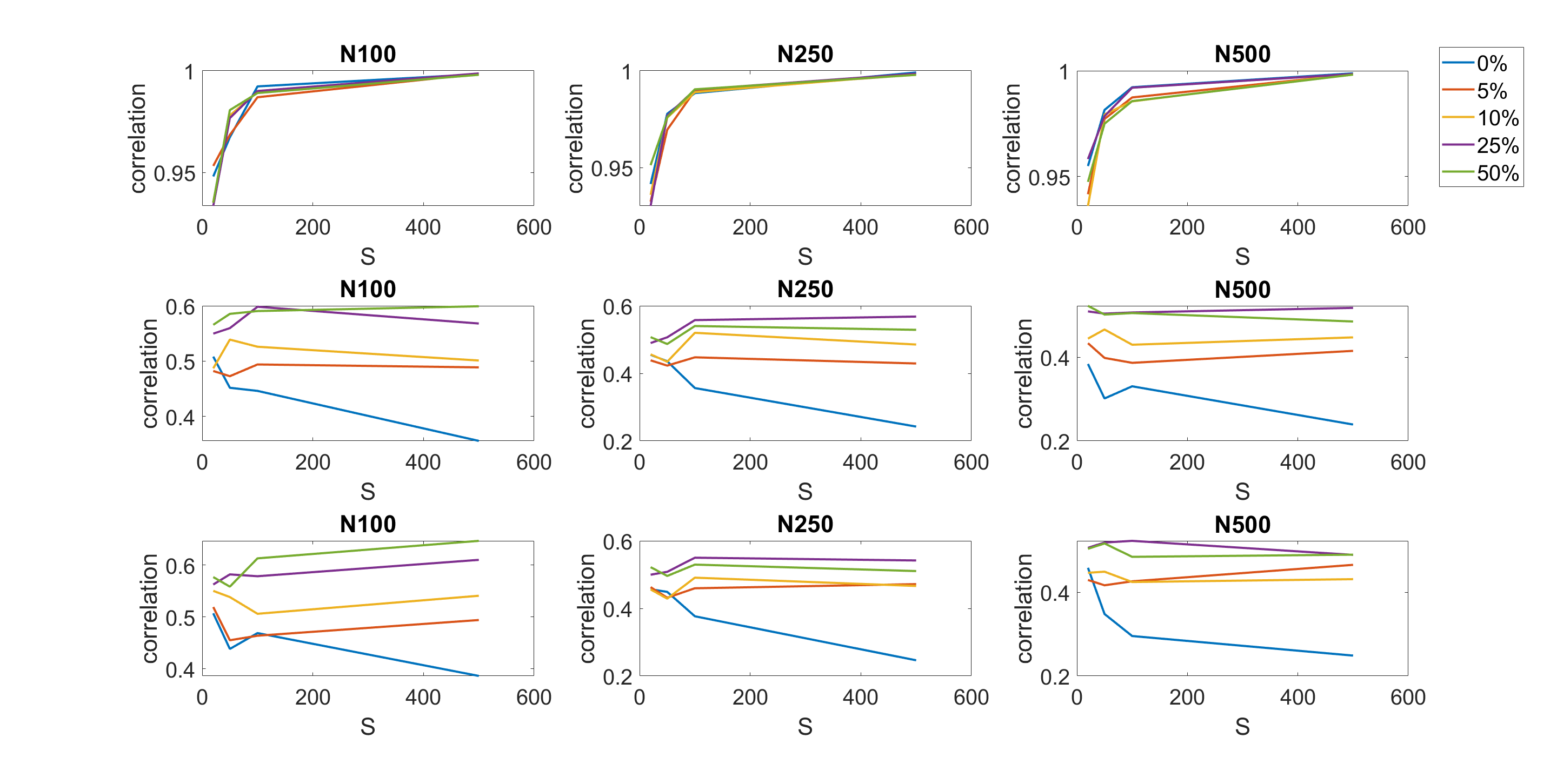}
\end{figure}

Figure~\ref{corr_LGSS_sim}
reports the correlation estimates at the current and proposed values of $\theta$ of the log of the estimated likelihood obtained using different MPM approaches. The figure
show that: (1) When the current and proposed values of the parameters are equal to $0.4$, all MPM methods		maintain a high correlation between the logs of the estimated likelihoods. The estimated
correlations between logs of the estimated likelihoods when the number
of particle filters $S$ is 100 are about 0.99.  (2) When the current parameter $\theta$ and the proposed parameter $\theta^{'}$ are (slightly) different, the MPM methods can still maintain some of the correlation between the log of the estimated likelihoods. The MPM methods with 25\% and 50\% trimmed means of the likelihood perform the best to maintain the correlation between the logs of the estimated likelihoods. 

%This
%supports  the result in \citeauthor{Tran2016} that shows that the
%correlation of the logs of the estimated likelihood at the current
%and proposed values is close to $1-1/S=1-1/100=0.99$;

\section{Additional tables and figures for the multivariate stochastic volatility in mean model example \label{SVexample}}

This section gives additional tables and figures for the multivariate stochastic volatility in mean example in Section \ref{subsec:Linear-Gaussian-State Space Model}.

\begin{figure}[H]
\caption{Kernel density estimates of the parameters of the multivariate stochastic
volatility in mean model estimated using the MPM with averaging likelihood
(25\% trimmed mean) for a simulated dataset (data 2) with $d=20$
dimensions and $T=100$. \label{diffusiondata2}}

\centering{}\includegraphics[width=15cm,height=8cm]{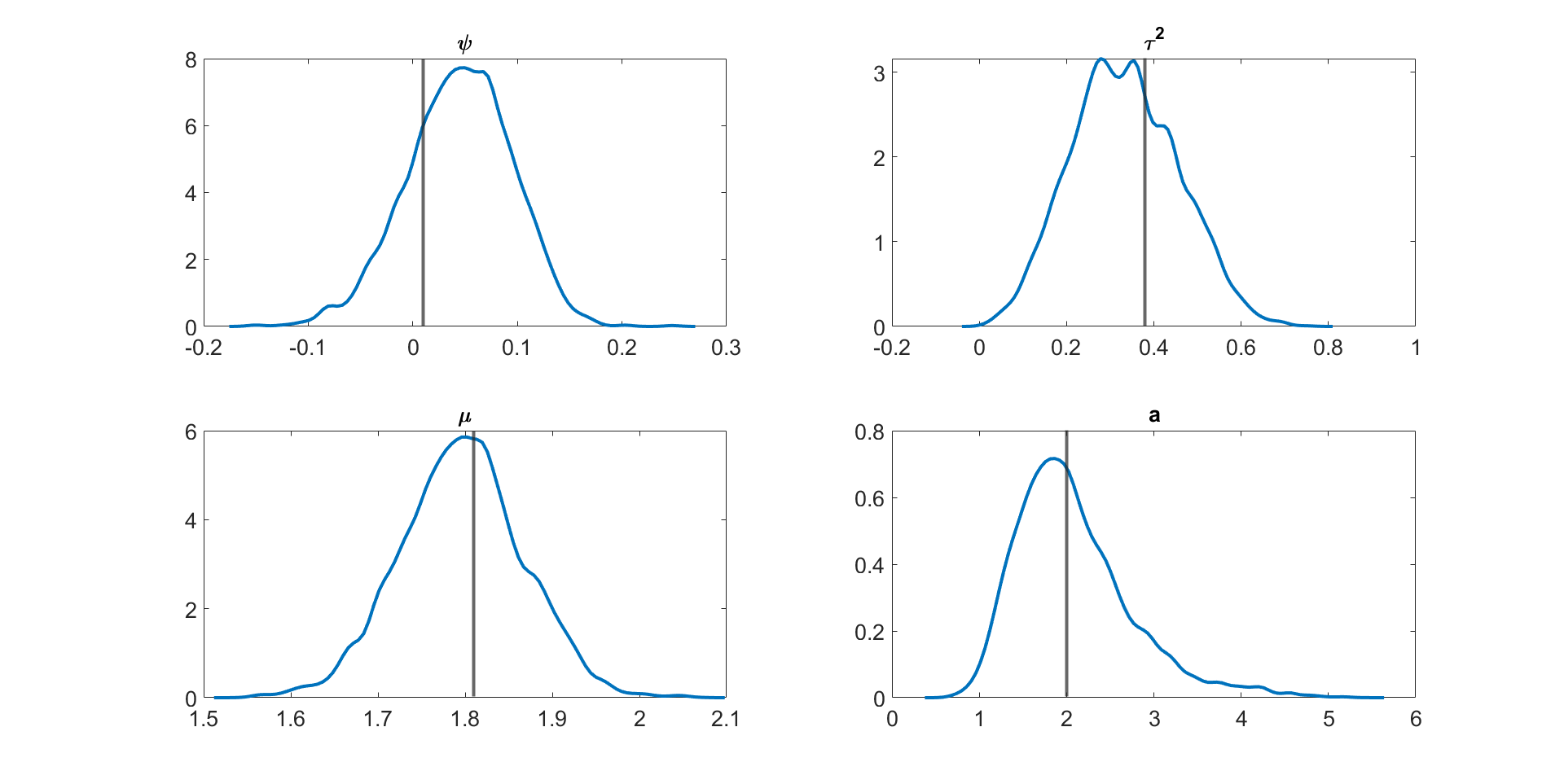}
\end{figure}

\begin{figure}[H]
\caption{Kernel density estimates of the parameters of the multivariate stochastic
volatility in mean model estimated using the MPM with averaging likelihood
(25\% trimmed mean) for a simulated dataset (data 10) with $d=20$
dimensions and $T=100$.\label{diffusiondata10}}

\centering{}\includegraphics[width=15cm,height=8cm]{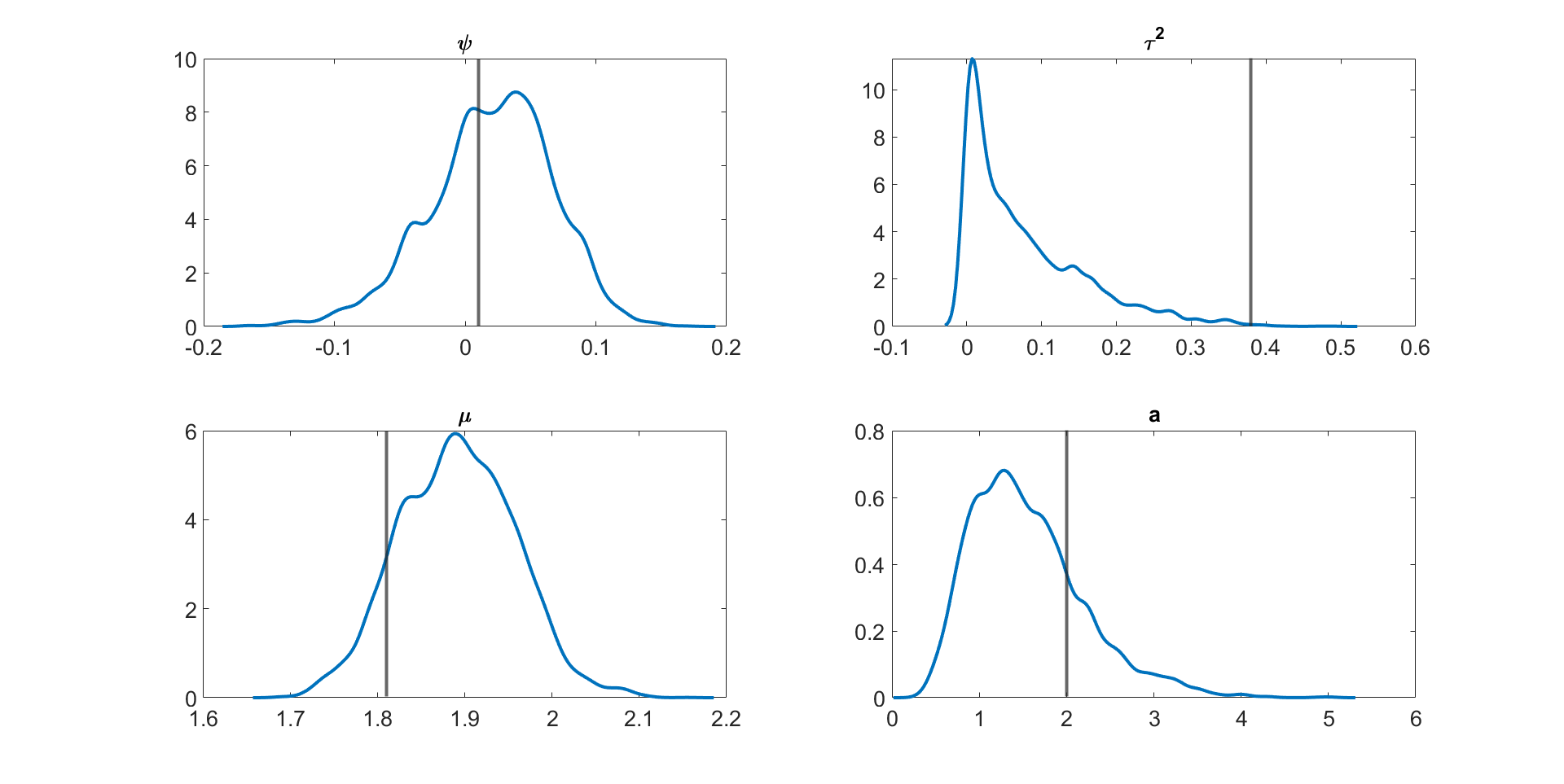}
\end{figure}

Figures~\ref{diffusiondata2} to \ref{diffusiondata10} show the kernel density estimates of the parameters of the multivariate stochastic volatility in mean model described above estimated using the MPM sampler ($S=100$ and $N=250$) using a 25\% trimmed mean with the vertical lines showing the true parameter values. The figures show that the model parameters are estimated well.

\begin{figure}[H]
\caption{Kernel density estimates of the parameters of the multivariate stochastic
volatility in mean model estimated using the MPM with averaging likelihood
(25\% trimmed mean) for a simulated dataset (data 3) with $d=20$
dimensions and $T=100$.}

\centering{}\includegraphics[width=15cm,height=8cm]{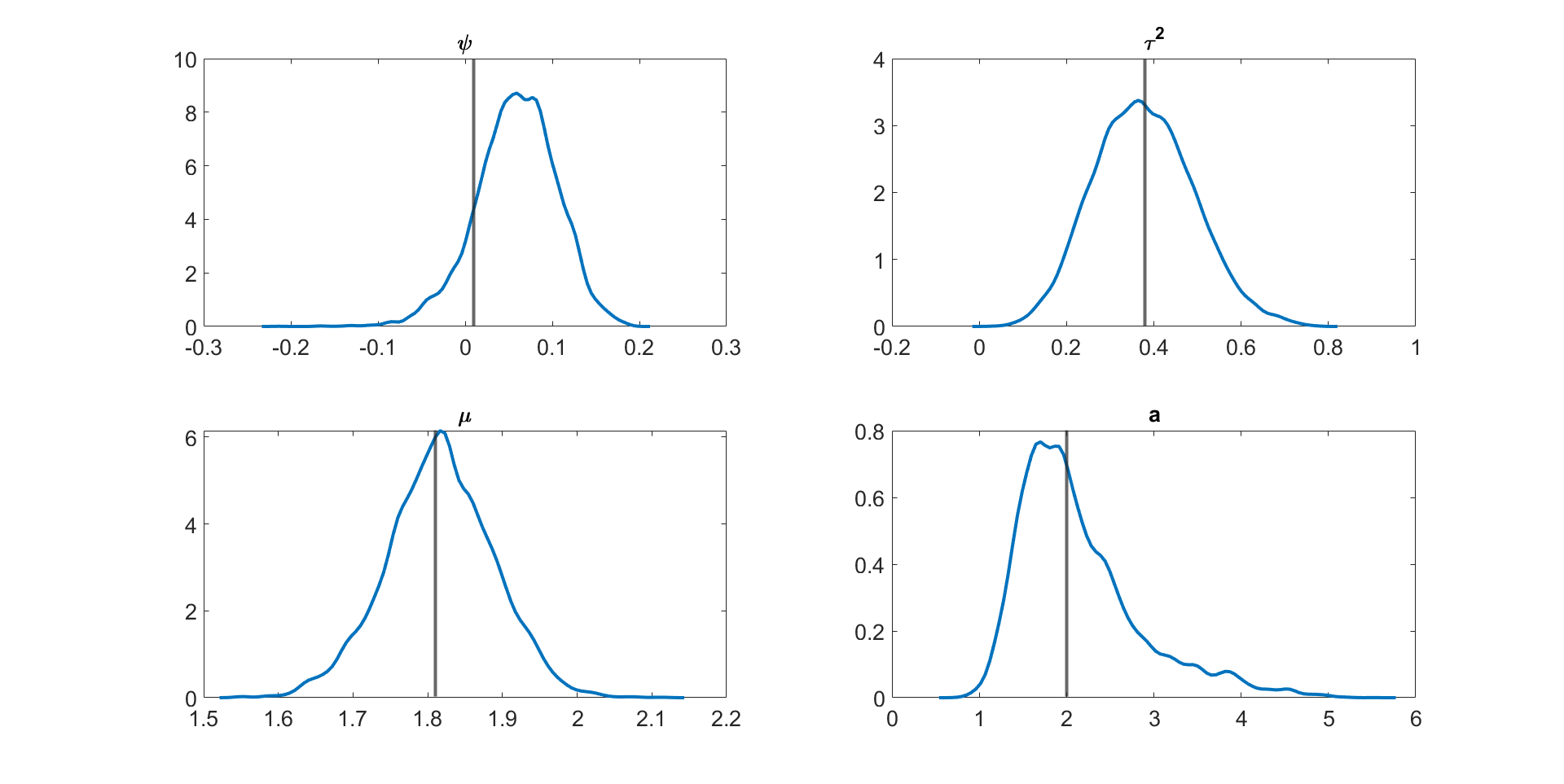}
\end{figure}

\begin{figure}[H]
\caption{Kernel density estimates of the parameters of the multivariate stochastic
volatility in mean model estimated using the MPM with averaging likelihood
(25\% trimmed mean) for a simulated dataset (data 4) with $d=20$
dimensions and $T=100$.}

\centering{}\includegraphics[width=15cm,height=8cm]{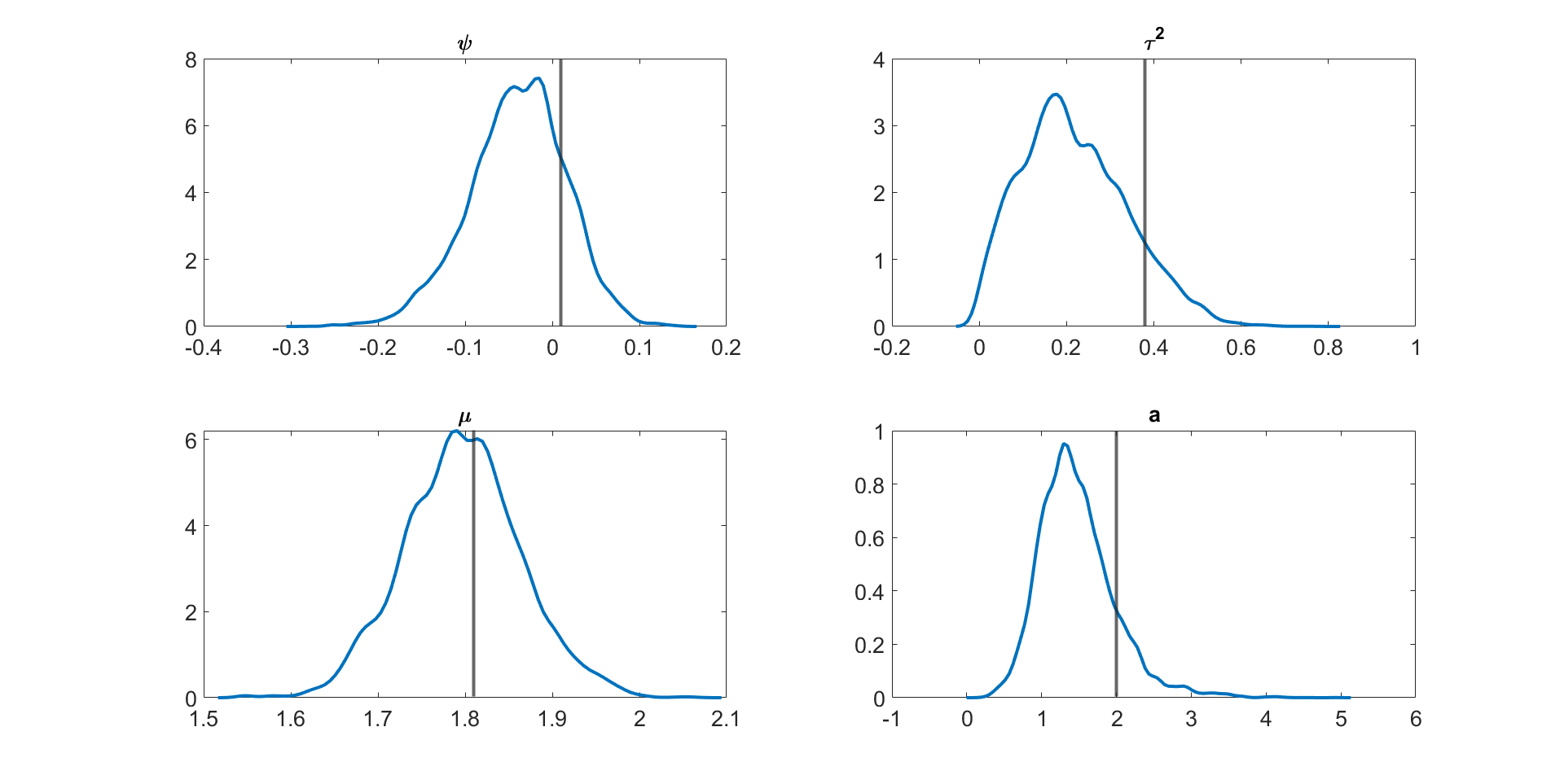}
\end{figure}

\begin{figure}[H]
\caption{Kernel density estimates of the parameters of the multivariate stochastic
volatility in mean model estimated using the MPM with averaging likelihood
(25\% trimmed mean) for a simulated dataset (data 5) with $d=20$
dimensions and $T=100$.}

\centering{}\includegraphics[width=15cm,height=8cm]{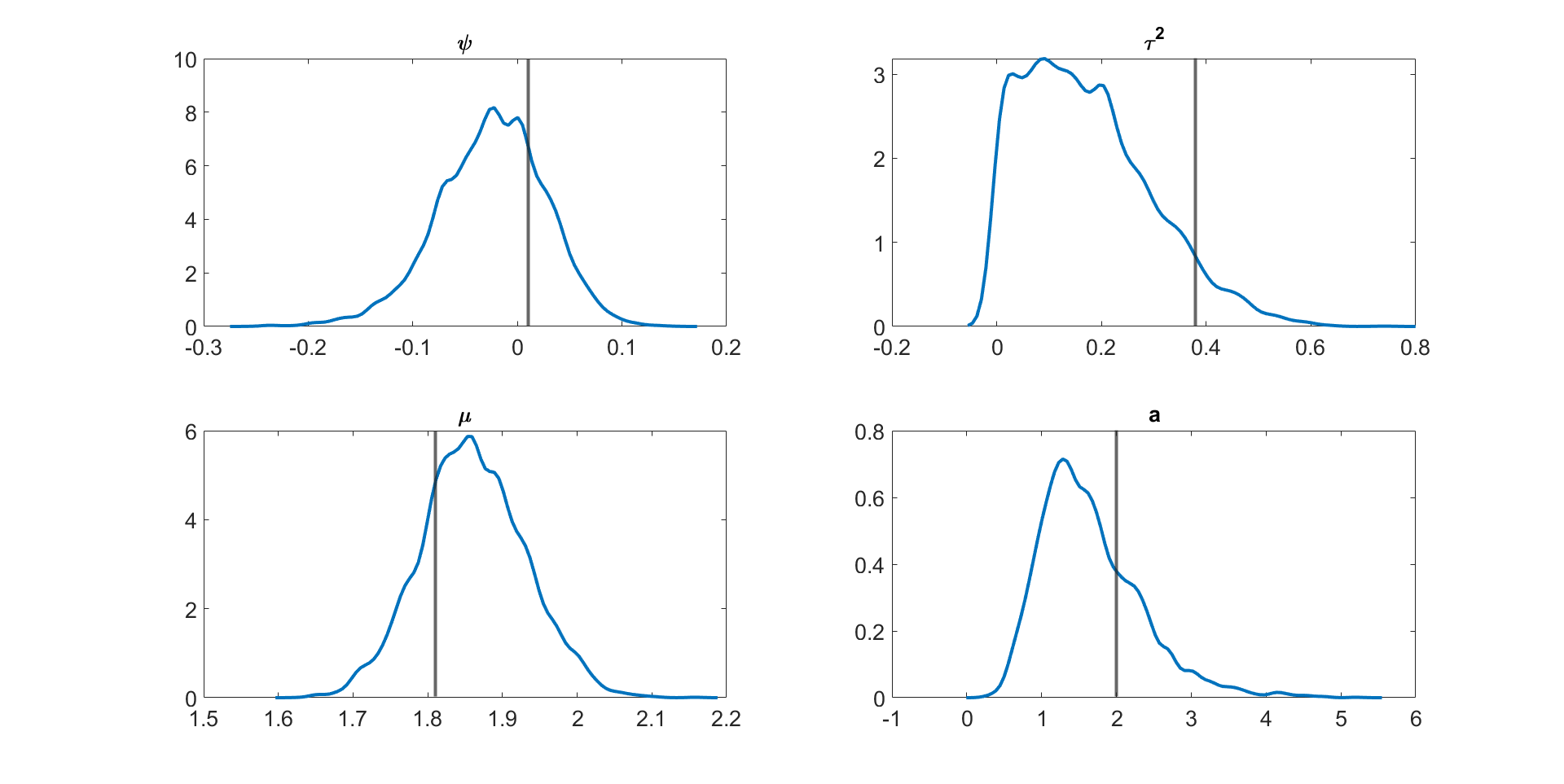}
\end{figure}

\begin{figure}[H]
\caption{Kernel density estimates of the parameters of the multivariate stochastic
volatility in mean model estimated using the MPM with averaging likelihood
(25\% trimmed mean) for a simulated dataset (data 6) with $d=20$
dimensions and $T=100$.}

\centering{}\includegraphics[width=15cm,height=8cm]{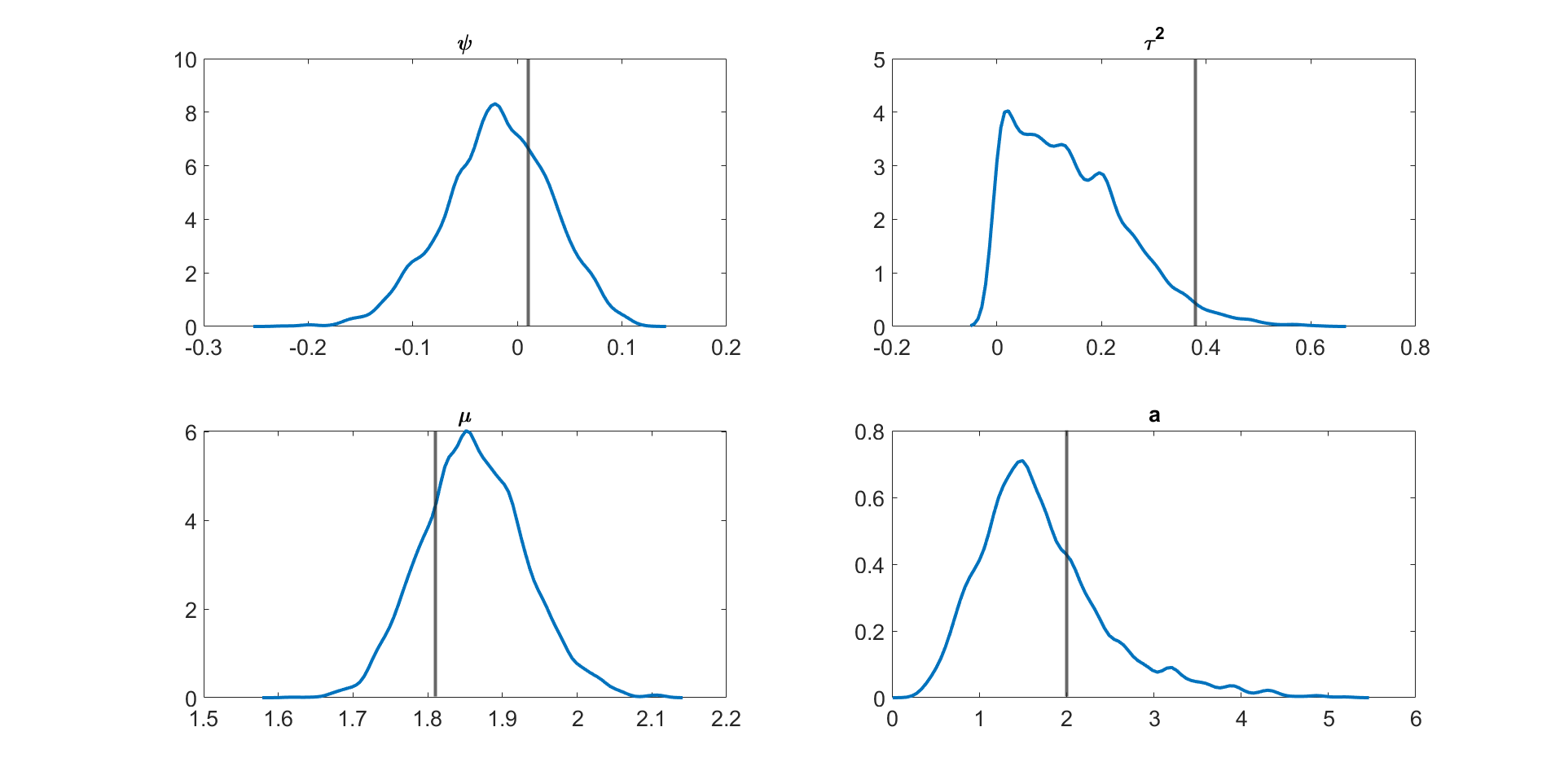}
\end{figure}

\begin{figure}[H]
\caption{Kernel density estimates of the parameters of the multivariate stochastic
volatility in mean model estimated using the MPM with averaging likelihood
(25\% trimmed mean) for a simulated dataset (data 7) with $d=20$
dimensions and $T=100$.}

\centering{}\includegraphics[width=15cm,height=8cm]{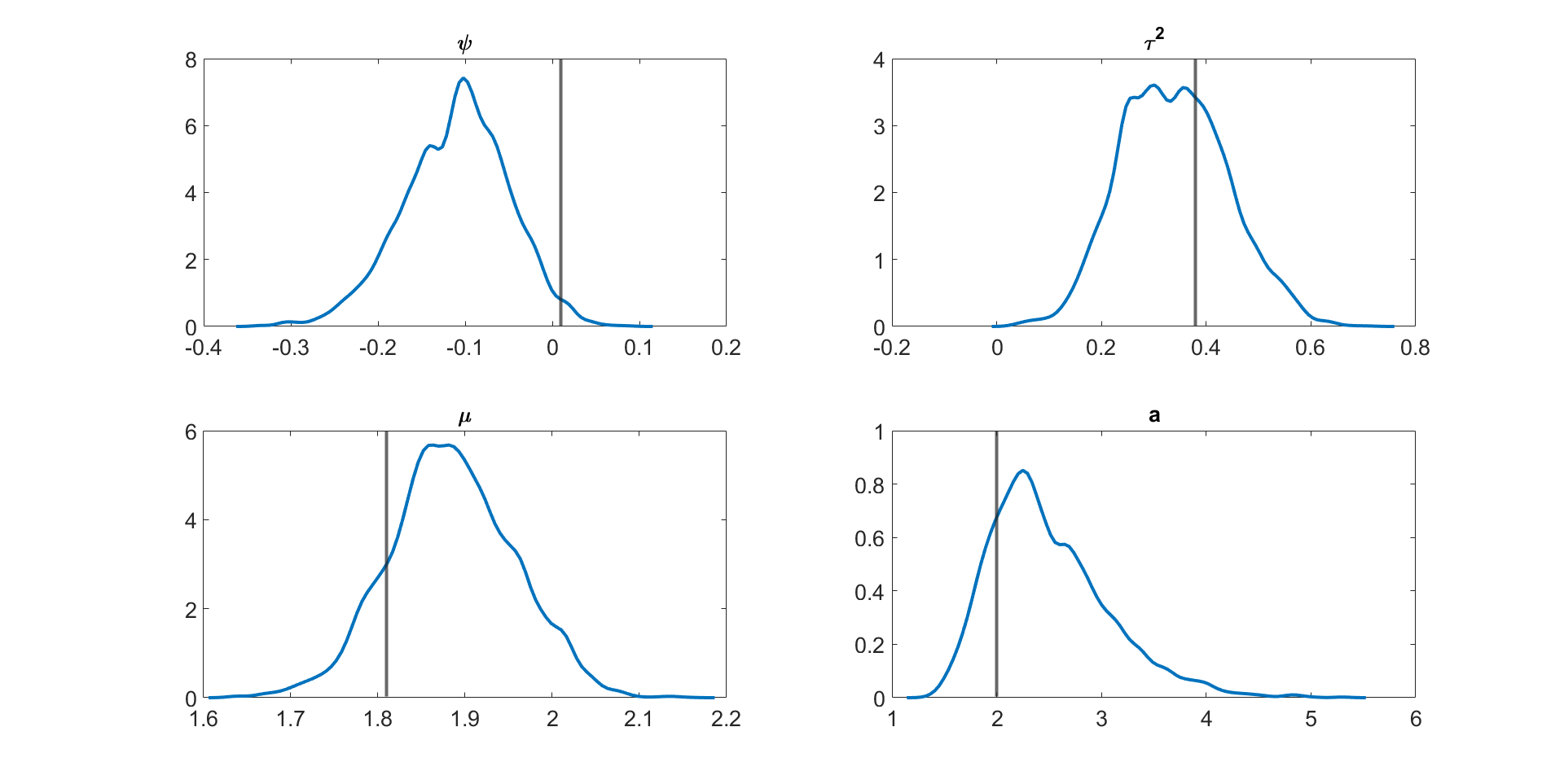}
\end{figure}

\begin{figure}[H]
\caption{Kernel density estimates of the parameters of the multivariate stochastic
volatility in mean model estimated using the MPM with averaging likelihood
(25\% trimmed mean) for a simulated dataset (data 8) with $d=20$
dimensions and $T=100$.}

\centering{}\includegraphics[width=15cm,height=8cm]{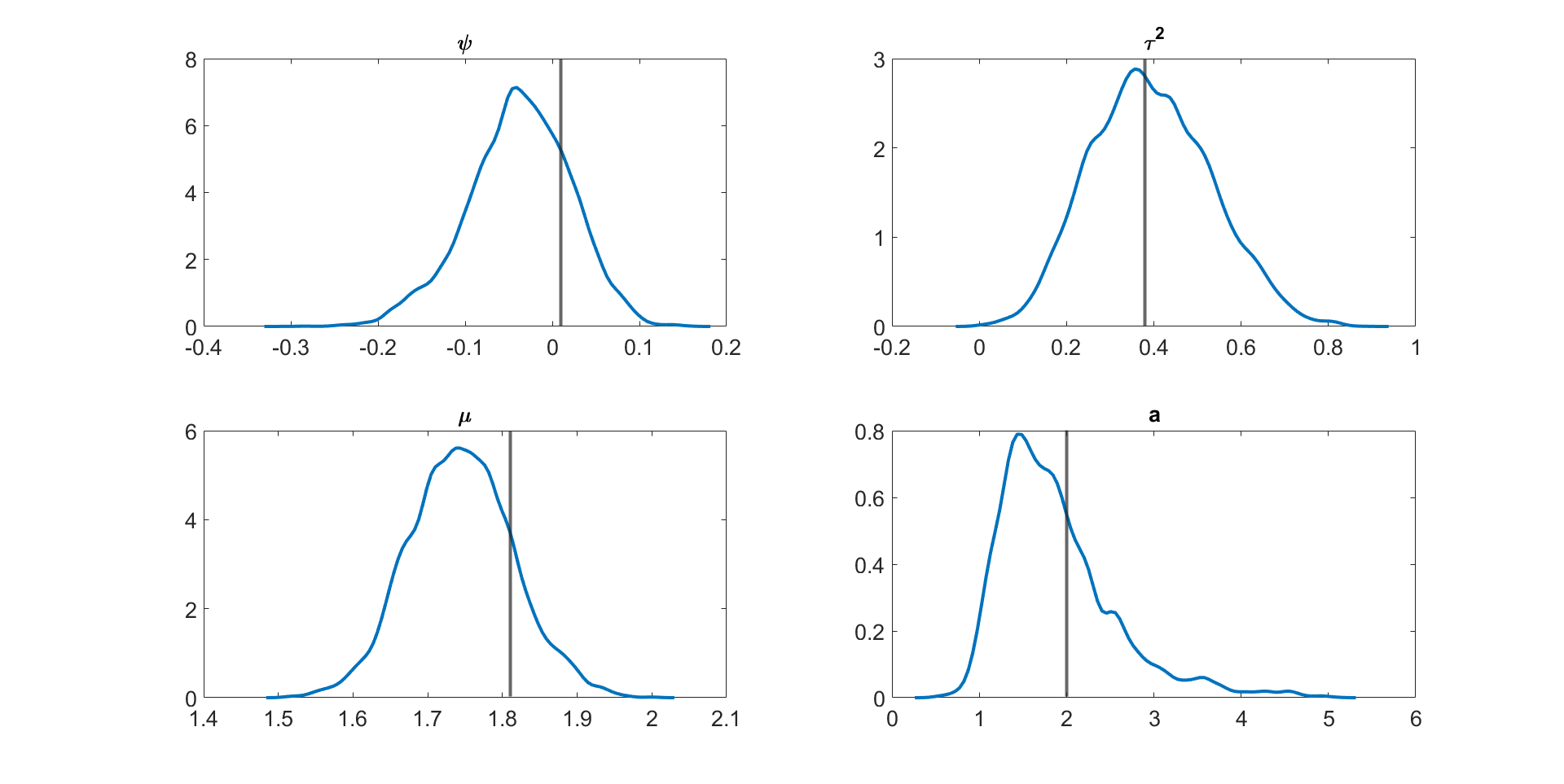}
\end{figure}

\begin{figure}[H]
\caption{Kernel density estimates of the parameters of the multivariate stochastic
volatility in mean model estimated using the MPM with averaging likelihood
(25\% trimmed mean) for a simulated dataset (data 9) with $d=20$
dimensions and $T=100$.}

\centering{}\includegraphics[width=15cm,height=8cm]{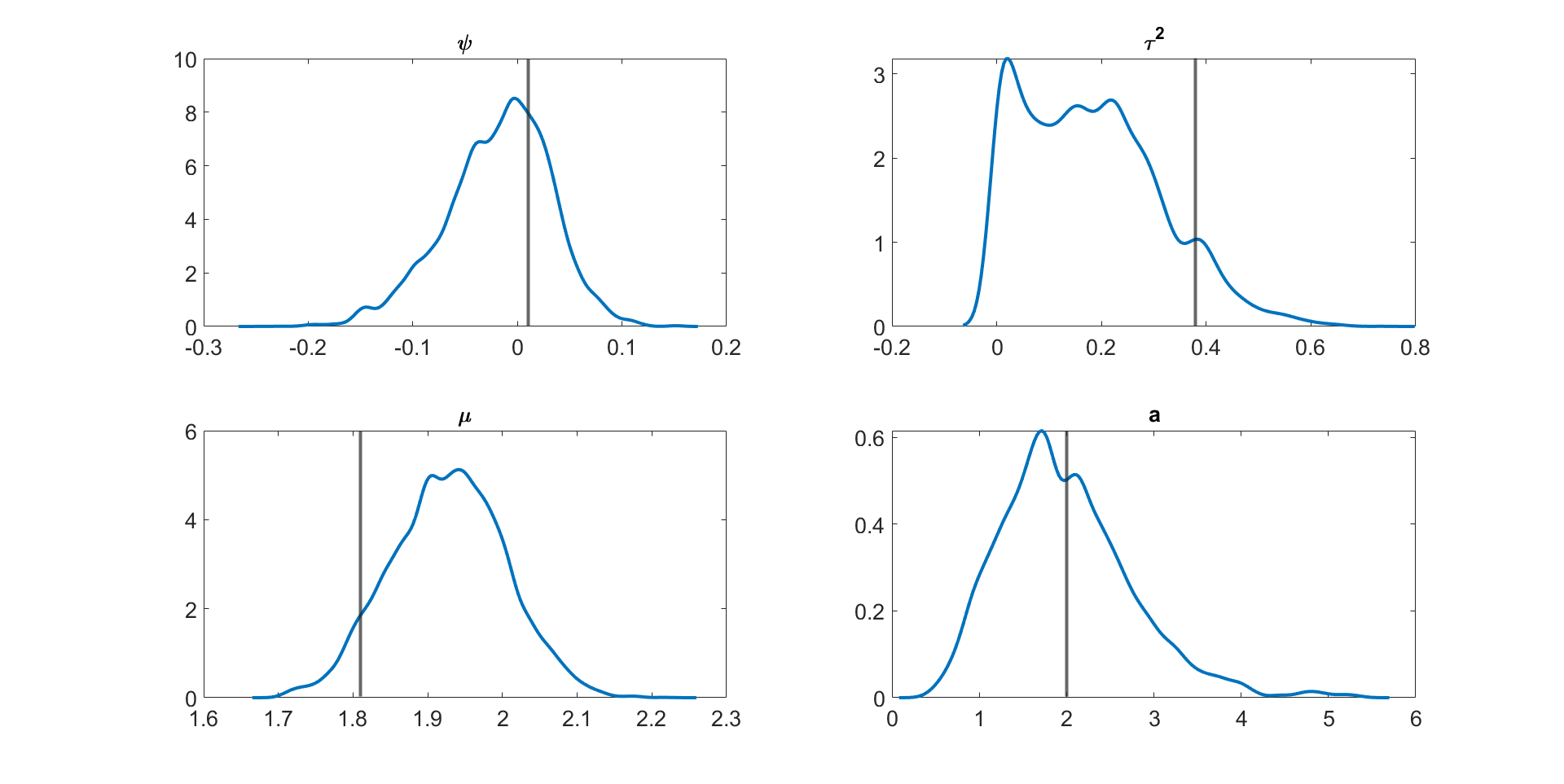}
\end{figure}

\section{Additional tables and figures for the small scale DSGE model example \label{additionaltablesfiguressmallscale}}

This section gives additional tables and figures for the small scale DSGE model example in Section \ref{subsec:SecondOrderSmallScale}.

Table \ref{priordistsmallscale} gives the prior  distributions for each  model parameter for the non-linear small
scale DSGE model in section \ref{subsec:SecondOrderSmallScale}. Table \ref{SmallScaleTableSupplement} reports the 
inefficiency factors for each parameter and
the relative time normalised inefficiency factor (RTNIF) of a PMMH sampler relative to the MPM (0\% trimmed mean) with the ADPF, disturbance sorting, and $\rho_u=0.99$ for the non-linear small scale models with $T=124$ time periods. 
Section \ref{subsec:SecondOrderSmallScale} discusses the results.

\begin{table}[H]
\caption{Marginal prior  distributions for each  model parameter for the non-linear small
scale DSGE model in Section \ref{subsec:SecondOrderSmallScale}.  The
Param (1) and Param (2) columns list the mean
and the standard deviation for the truncated normal (TN) and Normal distribution;
$v$ and $s$ for the inverse gamma (IG) distribution. The domain $R^{+}:=(0,\infty)$. $\textrm{TN}_{\left(0,\infty\right)}$ means the normal distribution truncated  to the interval $(0,\infty)$.
$\textrm{TN}_{\left(0,1\right)}$ is the normal distribution truncated  to the interval $(0,1)$. $\sigma^{m}_{r}$, $\sigma^{m}_{g}$, and $\sigma^{m}_{m}$ are the measurement error variances.\label{priordistsmallscale}}

\centering{}%
\begin{tabular}{ccccc}
\hline
Param. & Domain & Density & Param (1) & Param (2)\tabularnewline
\hline
$r^{\left(A\right)}$ & $R^{+}$ & $\textrm{TN}_{\left(0,\infty\right)}$ & 0.7 & 0.5\tabularnewline
$\pi^{\left(A\right)}$ & $R^{+}$ & $\textrm{TN}_{\left(0,\infty\right)}$ & 3.12 & 0.5\tabularnewline
$\gamma^{\left(Q\right)}$ & $R$ & N & 0.60 & 0.5\tabularnewline
$\tau$ & $R^{+}$ & $\textrm{TN}_{\left(0,\infty\right)}$ & 1.50 & 0.5\tabularnewline
$\psi_{1}$ & $R^{+}$ & $\textrm{TN}_{\left(0,\infty\right)}$ & 2.41 & 0.5\tabularnewline
$\psi_{2}$ & $R^{+}$ & $\textrm{TN}_{\left(0,\infty\right)}$ & 0.39 & 0.5\tabularnewline
$\rho_{r}$ & $\left[0,1\right]$ & $\textrm{TN}_{\left(0,1\right)}$ & 0.5 & 0.2\tabularnewline
$\rho_{g}$ & $\left[0,1\right]$ & $\textrm{TN}_{\left(0,1\right)}$ & 0.5 & 0.2\tabularnewline
$\rho_{m}$ & $\left[0,1\right]$ & $\textrm{TN}_{\left(0,1\right)}$ & 0.5 & 0.2\tabularnewline
$\sigma_{r}$ & $R^{+}$ & IG & 5 & 0.5\tabularnewline
$\sigma_{g}$ & $R^{+}$ & IG & 5 & 0.5\tabularnewline
$\sigma_{m}$ & $R^{+}$ & IG & 5 & 0.5\tabularnewline
$\sigma_{r}^{m}$ & $R^{+}$ & $\textrm{G}$ & 1 & 1\tabularnewline
$\sigma_{g}^{m}$ & $R^{+}$ & $\textrm{G}$ & 1 & 1\tabularnewline
$\sigma_{m}^{m}$ & $R^{+}$ & $\textrm{G}$ & 1 & 1\tabularnewline
\hline
\end{tabular}
\end{table}

\begin{table}[H]

\caption{Comparing the performance of different PMMH samplers with different
number of particle filters $S$ and different number of particles
$N$ in each particle filter for estimating the small scale DSGE model for the real dataset obtained from \citet{Herbst2019} with $T = 124$. Sampler
I: MPM (averaging likelihood, $\rho_{u}=0.99$, disturbance sorting,
ADPF). Sampler II: MPM (averaging likelihood 10\% trimmed mean, $\rho_{u}=0.99$,
disturbance sorting, ADPF). Sampler III: MPM (averaging likelihood
25\% trimmed mean, $\rho_{u}=0.99$, disturbance sorting, ADPF). Sampler
IV: MPM (averaging likelihood 25\% trimmed mean, $\rho_{u}=0$, disturbance
sorting, ADPF). Sampler V: MPM (averaging likelihood 25\% trimmed
mean, $\rho_{u}=0.99$, state sorting, ADPF). Sampler VI: DA-MPM (averaging
likelihood 25\% trimmed mean, $\rho_{u}=0.99$, state sorting, ADPF).
Sampler VII: MPM (averaging likelihood
50\% trimmed mean, $\rho_{u}=0.99$, disturbance sorting, ADPF).
Sampler VIII: MPM (averaging likelihood, $\rho_{u}=0.99$, disturbance
sorting, bootstrap). Sampler IX: Correlated PMMH. Time is the time in seconds for
one iteration of the method. \label{SmallScaleTableSupplement}}

\centering{}%
\begin{tabular}{|c|c|c|c|c|c|c|c|c|c|}
\hline 
{\footnotesize{}Param} & {\footnotesize{}I} & {\footnotesize{}II} & {\footnotesize{}III} & {\footnotesize{}IV} & {\footnotesize{}V } & {\footnotesize{}VI} & {\footnotesize{}VII} & {\footnotesize{}VIII} & {\footnotesize{}IX}\tabularnewline
\hline 
{\footnotesize{}S} & {\footnotesize{}100} & {\footnotesize{}100} & {\footnotesize{}100} & {\footnotesize{}100} & {\footnotesize{}100} & {\footnotesize{}100} & {\footnotesize{}100} & {\footnotesize{}100} & {\footnotesize{}1}\tabularnewline
{\footnotesize{}N} & {\footnotesize{}250} & {\footnotesize{}100} & {\footnotesize{}100} & {\footnotesize{}100} & {\footnotesize{}100} & {\footnotesize{}100} & {\footnotesize{}100} & {\footnotesize{}250} & {\footnotesize{}20000}\tabularnewline
{\footnotesize{}$\pi^{\left(A\right)}$} & {\footnotesize{}$1477.673$} & {\footnotesize{}$87.796$} & {\footnotesize{}$87.531$} & {\footnotesize{}$86.956$} & {\footnotesize{}$97.641$} & {\footnotesize{}$127.836$} & {\footnotesize{}$95.566$} & {\footnotesize{}NA} & {\footnotesize{}$184.817$}\tabularnewline
{\footnotesize{}$\tau$} & {\footnotesize{}$689.320$} & {\footnotesize{}$57.260$} & {\footnotesize{}$59.924$} & {\footnotesize{}$60.839$} & {\footnotesize{}$92.847$} & {\footnotesize{}$108.051$} & {\footnotesize{}$77.129$} & {\footnotesize{}NA} & {\footnotesize{}$486.328$}\tabularnewline
{\footnotesize{}$\psi_{1}$} & {\footnotesize{}$1074.308$} & {\footnotesize{}$84.835$} & {\footnotesize{}$83.481$} & {\footnotesize{}$82.317$} & {\footnotesize{}$112.780$} & {\footnotesize{}$157.470$} & {\footnotesize{}$95.780$} & {\footnotesize{}NA} & {\footnotesize{}$313.848$}\tabularnewline
{\footnotesize{}$\psi_{2}$} & {\footnotesize{}$645.963$} & {\footnotesize{}$90.750$} & {\footnotesize{}$96.507$} & {\footnotesize{}$105.948$} & {\footnotesize{}$148.919$} & {\footnotesize{}$234.819$} & {\footnotesize{}$116.554$} & {\footnotesize{}NA} & {\footnotesize{}$931.106$}\tabularnewline
{\footnotesize{}$\gamma^{\left(Q\right)}$} & {\footnotesize{}$662.909$} & {\footnotesize{}$74.881$} & {\footnotesize{}$134.589$} & {\footnotesize{}$105.032$} & {\footnotesize{}$131.219$} & {\footnotesize{}$128.182$} & {\footnotesize{}$88.105$} & {\footnotesize{}NA} & {\footnotesize{}$138.769$}\tabularnewline
{\footnotesize{}$r^{\left(A\right)}$} & {\footnotesize{}$617.848$} & {\footnotesize{}$110.671$} & {\footnotesize{}$108.337$} & {\footnotesize{}$79.477$} & {\footnotesize{}$118.001$} & {\footnotesize{}$156.412$} & {\footnotesize{}$84.812$} & {\footnotesize{}NA} & {\footnotesize{}$1463.057$}\tabularnewline
{\footnotesize{}$\rho_{r}$} & {\footnotesize{}$651.307$} & {\footnotesize{}$116.273$} & {\footnotesize{}$113.495$} & {\footnotesize{}$93.525$} & {\footnotesize{}$135.521$} & {\footnotesize{}$194.771$} & {\footnotesize{}$118.165$} & {\footnotesize{}NA} & {\footnotesize{}$386.218$}\tabularnewline
{\footnotesize{}$\rho_{g}$} & {\footnotesize{}$1145.901$} & {\footnotesize{}$152.711$} & {\footnotesize{}$123.128$} & {\footnotesize{}$99.316$} & {\footnotesize{}$167.833$} & {\footnotesize{}$604.722$} & {\footnotesize{}$132.060$} & {\footnotesize{}NA} & {\footnotesize{}$1064.070$}\tabularnewline
{\footnotesize{}$\rho_{m}$} & {\footnotesize{}$864.234$} & {\footnotesize{}$62.183$} & {\footnotesize{}$110.929$} & {\footnotesize{}$151.060$} & {\footnotesize{}$96.787$} & {\footnotesize{}$200.407$} & {\footnotesize{}$236.684$} & {\footnotesize{}NA} & {\footnotesize{}$1407.105$}\tabularnewline
{\footnotesize{}$\sigma_{r}$} & {\footnotesize{}$2254.695$} & {\footnotesize{}$88.552$} & {\footnotesize{}$104.442$} & {\footnotesize{}$89.854$} & {\footnotesize{}$142.248$} & {\footnotesize{}$148.776$} & {\footnotesize{}$94.689$} & {\footnotesize{}NA} & {\footnotesize{}$556.250$}\tabularnewline
{\footnotesize{}$\sigma_{g}$} & {\footnotesize{}$1728.309$} & {\footnotesize{}$149.187$} & {\footnotesize{}$91.283$} & {\footnotesize{}$114.929$} & {\footnotesize{}$140.916$} & {\footnotesize{}$212.246$} & {\footnotesize{}$148.838$} & {\footnotesize{}NA} & {\footnotesize{}$676.259$}\tabularnewline
{\footnotesize{}$\sigma_{m}$} & {\footnotesize{}$1213.429$} & {\footnotesize{}$87.724$} & {\footnotesize{}$80.627$} & {\footnotesize{}$72.743$} & {\footnotesize{}$120.432$} & {\footnotesize{}$195.319$} & {\footnotesize{}$97.566$} & {\footnotesize{}NA} & {\footnotesize{}$213.740$}\tabularnewline
{\footnotesize{}$\sigma_{r}^{m}$} & {\footnotesize{}$1242.351$} & {\footnotesize{}$82.957$} & {\footnotesize{}$94.455$} & {\footnotesize{}$81.547$} & {\footnotesize{}$83.689$} & {\footnotesize{}$122.662$} & {\footnotesize{}$101.957$} & {\footnotesize{}NA} & {\footnotesize{}$199.629$}\tabularnewline
{\footnotesize{}$\sigma_{g}^{m}$} & {\footnotesize{}$1078.089$} & {\footnotesize{}$82.678$} & {\footnotesize{}$66.751$} & {\footnotesize{}$76.764$} & {\footnotesize{}$81.131$} & {\footnotesize{}$98.441$} & {\footnotesize{}$79.810$} & {\footnotesize{}NA} & {\footnotesize{}$104.288$}\tabularnewline
{\footnotesize{}$\sigma_{m}^{m}$} & {\footnotesize{}$1487.408$} & {\footnotesize{}$85.196$} & {\footnotesize{}$69.188$} & {\footnotesize{}$97.438$} & {\footnotesize{}$87.323$} & {\footnotesize{}$170.904$} & {\footnotesize{}$110.041$} & {\footnotesize{}NA} & {\footnotesize{}$1377.649$}\tabularnewline
\hline 
{\footnotesize{}$\widehat{\textrm{IF}}_{\psi,\textrm{MAX}}$} & {\footnotesize{}$2254.695$} & {\footnotesize{}$152.711$} & {\footnotesize{}$134.589$} & {\footnotesize{}$151.060$} & {\footnotesize{}$167.833$} & {\footnotesize{}$604.723$} & {\footnotesize{}$236.684$} & {\footnotesize{}NA} & {\footnotesize{}$1463.057$}\tabularnewline
{\footnotesize{}$\widehat{\textrm{TNIF}}_{\textrm{MAX}}$} & {\footnotesize{}$2412.524$} & {\footnotesize{}$83.991$} & {\footnotesize{}$74.024$} & {\footnotesize{}$83.083$} & {\footnotesize{}$102.378$} & {\footnotesize{}$66.520$} & {\footnotesize{}$130.176$} & {\footnotesize{}NA} & {\footnotesize{}$7607.896$}\tabularnewline
{\footnotesize{}$\widehat{\textrm{RTNIF}}_{\textrm{MAX}}$} & {\footnotesize{}$1$} & {\footnotesize{}$0.035$} & {\footnotesize{}$0.031$} & {\footnotesize{}$0.034$} & {\footnotesize{}$0.042$} & {\footnotesize{}$0.028$} & {\footnotesize{}$0.054$} & {\footnotesize{}NA} & {\footnotesize{}$3.154$}\tabularnewline
\hline 
{\footnotesize{}$\widehat{\textrm{IF}}_{\psi,\textrm{MEAN}}$} & {\footnotesize{}$1122.250$} & {\footnotesize{}$94.244$} & {\footnotesize{}$94.978$} & {\footnotesize{}$93.183$} & {\footnotesize{}$117.153$} & {\footnotesize{}$190.735$} & {\footnotesize{}$111.852$} & {\footnotesize{}NA} & {\footnotesize{}$633.542$}\tabularnewline
{\footnotesize{}$\widehat{\textrm{TNIF}}_{\textrm{MEAN}}$} & {\footnotesize{}$1200.808$} & {\footnotesize{}$51.834$} & {\footnotesize{}$52.238$} & {\footnotesize{}$51.251$} & {\footnotesize{}$71.463$} & {\footnotesize{}$20.981$} & {\footnotesize{}$61.519$} & {\footnotesize{}NA} & {\footnotesize{}$3294.418$}\tabularnewline
{\footnotesize{}$\widehat{\textrm{RTNIF}}_{\textrm{MEAN}}$} & {\footnotesize{}$1$} & {\footnotesize{}$0.043$} & {\footnotesize{}$0.044$} & {\footnotesize{}$0.043$} & {\footnotesize{}$0.060$} & {\footnotesize{}$0.017$} & {\footnotesize{}$0.051$} & {\footnotesize{}NA} & {\footnotesize{}$2.744$}\tabularnewline
\hline 
{\footnotesize{}Time} & {\footnotesize{}1.07} & {\footnotesize{}0.55} & {\footnotesize{}0.55} & {\footnotesize{}0.55} & {\footnotesize{}0.61} & {\footnotesize{}0.11} & {\footnotesize{}$0.55$} & {\footnotesize{}0.75} & {\footnotesize{}5.20}\tabularnewline
\hline 
\end{tabular}
\end{table}

\section{Additional tables and figures for the medium scale DSGE model example \label{mediumscaletable}}

This section gives additional tables and figures for the medium scale DSGE model example in Section \ref{mediumscaleDSGEmodelexample}. Table \ref{priordistmediumscale} gives the prior  distributions for each  model parameter for the non-linear medium
scale DSGE model in Section \ref{mediumscaleDSGEmodelexample}.
Table \ref{tab:Mean,-,-and all parameters medium scale} gives the mean, $2.5\%$, and $97.5\%$ quantiles estimates of each parameter in the medium scale DSGE model estimated
using the MPM sampler with a $50\%$ trimmed mean ($\rho_{u}=0.99$,
disturbance sorting, ADPF)) for the US quarterly dataset from 1983Q1
to 2007Q4. The measurement error variances are fixed to $25\%$ of
the variance of the observables. The standard errors are relatively small for all parameters indicating that the parameters are estimated accurately. 

Table \ref{fullmediumscaleRTNIF} gives the inefficiency factor for each parameter in the medium scale DSGE model estimated using the MPM with 50\% trimmed mean (ADPF, disturbance sorting, $\rho_u=0.9$) and the correlated PMMH samplers. Table \ref{tab:Mean,-,-and all parameters medium scale-extended model} gives the mean, $2.5\%$, and $97.5\%$ quantiles estimates of each parameter in the extended version of the medium scale DSGE model estimated
using the MPM sampler with a $50\%$ trimmed mean ($\rho_{u}=0.99$,
disturbance sorting, ADPF)) for the US quarterly dataset from 1983Q1
to 2007Q4. The measurement error variances are fixed to $25\%$ of
the variance of the observables.

\begin{table}[H]
\caption{Prior distribution for each model parameter for the non-linear medium
scale DSGE model. The Param (1) and Param (2) columns list the mean
and the standard deviation for the truncated normal (TN) and Normal
distribution; $v$ and $s$ for the shape and scale parameters of
the inverse gamma (IG) and gamma (G) distribution. The domain $R^{+}:=(0,\infty)$.
$\textrm{TN}_{\left(0,\infty\right)}$ means the normal distribution
truncated to the interval $(0,\infty)$. $\textrm{TN}_{\left(0,1\right)}$
is the normal distribution truncated to the interval $(0,1)$. $\sigma_{r}^{m}$,
$\sigma_{g}^{m}$, and $\sigma_{m}^{m}$ are the measurement error
variances.\label{priordistmediumscale}}

\centering{}%
\begin{tabular}{ccccc}
\hline 
Param. & Domain & Density & Param. (1) & Param. (2)\tabularnewline
\hline 
$\rho_{R}$ & $\left[0,1\right]$ & $\textrm{TN}_{\left[0,1\right]}$ & $0.6$ & $0.2$\tabularnewline
$\rho_{g}$ & $\left[0,1\right]$ & $\textrm{TN}_{\left[0,1\right]}$ & $0.6$ & $0.2$\tabularnewline
$\rho_{\mu}$ & $\left[0,1\right]$ & $\textrm{TN}_{\left[0,1\right]}$ & $0.6$ & $0.2$\tabularnewline
$100\sigma_{g}$ & $R^{+}$ & $\textrm{G}$ & $0.5$ & $0.5$\tabularnewline
$100\sigma_{\mu}$ & $R^{+}$ & $\textrm{G}$ & $0.5$ & $0.5$\tabularnewline
$100\sigma_{\eta}$ & $R^{+}$ & $\textrm{G}$ & $0.5$ & $0.5$\tabularnewline
$100\sigma_{Z}$ & $R^{+}$ & $\textrm{G}$ & $0.5$ & $0.5$\tabularnewline
$100\sigma_{R}$ & $R^{+}$ & $\textrm{G}$ & $0.5$ & $0.5$\tabularnewline
$100\sigma_{gdp}$ & $R^{+}$ & $\textrm{G}$ & $0.5$ & $0.5$\tabularnewline
$100\sigma_{con}$ & $R^{+}$ & $\textrm{G}$ & $0.5$ & $0.5$\tabularnewline
$100\sigma_{inv}$ & $R^{+}$ & $\textrm{G}$ & $0.5$ & $0.5$\tabularnewline
$100\sigma_{inf}$ & $R^{+}$ & $\textrm{G}$ & $0.5$ & $0.5$\tabularnewline
$100\sigma_{ffr}$ & $R^{+}$ & $\textrm{G}$ & $0.5$ & $0.5$\tabularnewline
$\gamma_{g}$ & $R$ & $\textrm{N}$ & $0.4$ & $0.2$\tabularnewline
$\gamma_{\Pi}$ & $R$ & $\textrm{N}$ & $1.7$ & $0.3$\tabularnewline
$\gamma_{x}$ & $R$ & $\textrm{N}$ & $0.4$ & $0.3$\tabularnewline
$\gamma$ & $\left[0,1\right]$ & $\textrm{TN}_{\left[0,1\right]}$ & $0.6$ & $0.1$\tabularnewline
$\sigma_{a}$ & $R^{+}$ & $\textrm{G}$ & $25$ & $0.2$\tabularnewline
$\varphi_{p}/100$ & $R$ & $\textrm{N}$ & $1$ & $0.25$\tabularnewline
$\varphi_{w}/1000$ & $R$ & $\textrm{N}$ & $3$ & $5$\tabularnewline
$\varphi_{I}$ & $R^{+}$ & $\textrm{G}$ & $16$ & $0.25$\tabularnewline
$\sigma_{L}$ & $R^{+}$ & $\textrm{G}$ & $5.34$ & $0.37$\tabularnewline
$1-a$ & $\left[0,1\right]$ & $\textrm{TN}_{\left[0,1\right]}$ & $0.5$ & $0.15$\tabularnewline
$1-a_{w}$ & $\left[0,1\right]$ & $\textrm{TN}_{\left[0,1\right]}$ & $0.5$ & $0.15$\tabularnewline
$100\left(\beta^{-1}-1\right)$ & $R^{+}$ & $\textrm{G}$ & $0.625$ & $4$\tabularnewline
$100\log\left(G_{z}\right)$ & $R$ & $\textrm{N}$ & $0.5$ & $0.03$\tabularnewline
$100\left(\overline{\Pi}-1\right)$ & $R$ & $\textrm{N}$ & $0.62$ & $0.1$\tabularnewline
$\alpha$ & $R$ & $\textrm{N}$ & $0.3$ & $0.05$\tabularnewline
\hline 
\end{tabular}
\end{table}

\begin{table}[H]
\caption{Posterior means, $2.5\%$, and $97.5\%$ quantiles of the posterior distributions
of each of the parameters in the medium scale DSGE model estimated
using the MPM sampler with $50\%$ trimmed mean ($\rho_{u}=0.99$,
disturbance sorting, ADPF)) for the US quarterly dataset from 1983Q1
to 2007Q4. The measurement error variances are fixed to $25\%$ of
the variance of the observables. The
standard errors in brackets are calculated from $10$ independent runs of the MPM sampler \label{tab:Mean,-,-and all parameters medium scale}}

\centering{}%
\begin{tabular}{cccc}
\hline 
Param. & Estimates & $2.5\%$ & $97.5\%$\tabularnewline
\hline 
$\rho_{R}$ & $\underset{\left(0.0178\right)}{0.4329}$ & $\underset{\left(0.0365\right)}{0.2097}$ & $\underset{\left(0.0220\right)}{0.6490}$\tabularnewline
$\rho_{g}$ & $\underset{\left(0.0008\right)}{0.9883}$ & $\underset{\left(0.0020\right)}{0.9770}$ & $\underset{\left(0.0005\right)}{0.9965}$\tabularnewline
$\rho_{\mu}$ & $\underset{\left(0.0191\right)}{0.5975}$ & $\underset{\left(0.0309\right)}{0.4136}$ & $\underset{\left(0.0410\right)}{0.7598}$\tabularnewline
$100\sigma_{g}$ & $\underset{\left(0.0104\right)}{0.2188}$ & $\underset{\left(0.0129\right)}{0.0973}$ & $\underset{\left(0.0182\right)}{0.3992}$\tabularnewline
$100\sigma_{\mu}$ & $\underset{\left(0.0978\right)}{4.1832}$ & $\underset{\left(0.2905\right)}{2.1935}$ & $\underset{\left(0.2697\right)}{7.0727}$\tabularnewline
$100\sigma_{\eta}$ & $\underset{\left(0.0096\right)}{0.3614}$ & $\underset{\left(0.0082\right)}{0.2706}$ & $\underset{\left(0.0151\right)}{0.4837}$\tabularnewline
$100\sigma_{Z}$ & $\underset{\left(0.0118\right)}{0.6240}$ & $\underset{\left(0.0168\right)}{0.4707}$ & $\underset{\left(0.0121\right)}{0.7954}$\tabularnewline
$100\sigma_{R}$ & $\underset{\left(0.0032\right)}{0.0462}$ & $\underset{\left(0.0002\right)}{0.0059}$ & $\underset{\left(0.0123\right)}{0.1570}$\tabularnewline
$\gamma_{g}$ & $\underset{\left(0.0162\right)}{0.3051}$ & $\underset{\left(0.0130\right)}{0.0435}$ & $\underset{\left(0.0541\right)}{0.7360}$\tabularnewline
$\gamma_{\Pi}$ & $\underset{\left(0.0039\right)}{0.9897}$ & $\underset{\left(0.0146\right)}{0.9671}$ & $\underset{\left(0.0016\right)}{1.0072}$\tabularnewline
$\gamma_{x}$ & $\underset{\left(0.0115\right)}{0.7649}$ & $\underset{\left(0.0106\right)}{0.4726}$ & $\underset{\left(0.0428\right)}{1.0829}$\tabularnewline
$\gamma$ & $\underset{\left(0.0086\right)}{0.4606}$ & $\underset{\left(0.0131\right)}{0.3506}$ & $\underset{\left(0.0128\right)}{0.5727}$\tabularnewline
$\sigma_{a}$ & $\underset{\left(0.1244\right)}{4.9188}$ & $\underset{\left(0.1777\right)}{3.4442}$ & $\underset{\left(0.2509\right)}{6.7495}$\tabularnewline
$\varphi_{p}$ & $\underset{\left(3.6050\right)}{86.7032}$ & $\underset{\left(15.1263\right)}{37.4987}$ & $\underset{\left(3.4207\right)}{138.1528}$\tabularnewline
$\varphi_{w}$ & $\underset{\left(289.8829\right)}{2879.123}$ & $\underset{\left(110.8468\right)}{1230.612}$ & $\underset{\left(856.0024\right)}{5076.731}$\tabularnewline
$\varphi_{I}$ & $\underset{\left(0.0798\right)}{2.4537}$ & $\underset{\left(0.0887\right)}{1.4166}$ & $\underset{\left(0.2252\right)}{3.9636}$\tabularnewline
$\sigma_{L}$ & $\underset{\left(0.1389\right)}{3.8518}$ & $\underset{\left(0.1179\right)}{2.0824}$ & $\underset{\left(0.2504\right)}{6.1554}$\tabularnewline
$1-a$ & $\underset{\left(0.0140\right)}{0.4386}$ & $\underset{\left(0.0351\right)}{0.2153}$ & $\underset{\left(0.0355\right)}{0.7024}$\tabularnewline
$1-a_{w}$ & $\underset{\left(0.0119\right)}{0.6493}$ & $\underset{\left(0.0236\right)}{0.3875}$ & $\underset{\left(0.0074\right)}{0.8285}$\tabularnewline
$100\left(\beta^{-1}-1\right)$ & $\underset{\left(0.0187\right)}{0.0651}$ & $\underset{\left(0.0034\right)}{0.0052}$ & $\underset{\left(0.0866\right)}{0.2034}$\tabularnewline
$100\log\left(G_{z}\right)$ & $\underset{\left(0.0035\right)}{0.5144}$ & $\underset{\left(0.0039\right)}{0.4650}$ & $\underset{\left(0.0042\right)}{0.5708}$\tabularnewline
$100\left(\overline{\Pi}-1\right)$ & $\underset{\left(0.0069\right)}{0.6150}$ & $\underset{\left(0.0132\right)}{0.4213}$ & $\underset{\left(0.0164\right)}{0.7977}$\tabularnewline
$\alpha$ & $\underset{\left(0.0032\right)}{0.1738}$ & $\underset{\left(0.0064\right)}{0.1284}$ & $\underset{\left(0.0034\right)}{0.2151}$\tabularnewline
\hline
\end{tabular}
\end{table}

\begin{table}[H]
\caption{Comparing the performance of different PMMH samplers with different
number of particle filters $S$ and different number of particles
$N$ in each particle filter for estimating the medium scale DSGE
model using the US quarterly dataset from 1983Q1 to 2007Q4. Sampler
I: MPM (50\% trimmed mean, $\rho_{u}=0.9$, disturbance sorting,
ADPF). Sampler II: Correlated PMMH. The first two columns are the inefficiency factors of each parameter for
the case when the measurement error variances are estimated. The last
two columns are the inefficiency factor of each parameter for the case when the measurement error variances
are fixed at 25\% of the variance of each observable over the estimation period. \label{fullmediumscaleRTNIF}}

\centering{}%
\begin{tabular}{ccccc}
\hline 
Param. & I & II & I & II\tabularnewline
\hline 
$\rho_{R}$ & $450.704$ & $1886.791 $ & $318.305$ & $277.221$\tabularnewline
$\rho_{g}$ & $523.660$ & $653.350 $ & $205.736$ & $292.167$\tabularnewline
$\rho_{\mu}$ & $618.928$ & $1354.582 $ & $233.298$ & $238.901$\tabularnewline
$100\sigma_{g}$ & $593.827$ & $1731.764 $ & $286.744$ & $302.512$\tabularnewline
$100\sigma_{\mu}$ & $396.348$ & $1239.142 $ & $335.987$ & $173.760$\tabularnewline
$100\sigma_{\eta}$ & $380.984$ & $634.356 $ & $234.383$ & $203.419$\tabularnewline
$100\sigma_{Z}$ & $456.971$ & $1035.444 $ & $263.295$ & $226.536$\tabularnewline
$100\sigma_{R}$ & $412.131$ & $1008.377 $ & $247.562$ & $214.989$\tabularnewline
$100\sigma_{gdp}$ & $372.244$ & $621.6010 $ & NA & NA\tabularnewline
$100\sigma_{con}$ & $364.200$ & $1272.367 $ & NA & NA\tabularnewline
$100\sigma_{inv}$ & $287.053$ & $1024.054 $ & NA & NA\tabularnewline
$100\sigma_{inf}$ & $353.456$ & $1395.761 $ & NA & NA\tabularnewline
$100\sigma_{ffr}$ & $292.713$ & $1309.101 $ & NA & NA\tabularnewline
$\gamma_{g}$ & $279.772$ & $1512.739 $ & $243.631$ & $244.901$\tabularnewline
$\gamma_{\Pi}$ & $794.587$ & $842.370  $ & $367.625$ & $441.666$\tabularnewline
$\gamma_{x}$ & $361.702$ & $1210.535 $ & $308.195$ & $280.568$\tabularnewline
$\gamma$ & $365.017$ & $1175.542 $ & $208.171$ & $269.869$\tabularnewline
$\sigma_{a}$ & $379.348$ & $1939.116 $ & $241.076$ & $247.873$\tabularnewline
$\varphi_{p}/100$ & $393.882$ & $1036.088 $ & $245.650$ & $175.856$\tabularnewline
$\varphi_{w}/1000$ & $462.727$ & $1222.371 $ & $341.178$ & $568.954$\tabularnewline
$\varphi_{I}$ & $371.357$ & $1004.829 $ & $299.040$ & $242.744$\tabularnewline
$\sigma_{L}$ & $426.799$ & $926.466 $ & $230.927$ & $242.741$\tabularnewline
$1-a$ & $434.454$ & $1566.728 $ & $251.796$ & $237.491$\tabularnewline
$1-a_{w}$ & $359.744$ & $839.328 $ & $255.983$ & $229.476$\tabularnewline
$100\left(\beta^{-1}-1\right)$ & $796.824$ & $660.065 $ & $383.310$ & $343.446$\tabularnewline
$100\log\left(G_{z}\right)$ & $336.080$ & $2181.566 $ & $252.953$ & $203.629$\tabularnewline
$100\left(\overline{\Pi}-1\right)$ & $295.468$ & $967.840 $ & $259.442$ & $202.170$\tabularnewline
$\alpha$ & $442.753$ & $1120.534 $ & $318.548$ & $224.703$\tabularnewline
\hline 
\end{tabular}
\end{table}

\begin{table}[H]
\caption{Posterior means, $2.5\%$, and $97.5\%$ quantiles of the posterior
distributions, and the inefficiency factors of each of the parameters
in the extended version of the medium scale DSGE model estimated using
the MPM sampler with $50\%$ trimmed mean ($\rho_{u}=0.99$, disturbance
sorting, ADPF)) for the US quarterly dataset from 1983Q1 to 2007Q4.
The measurement error variances are fixed to $25\%$ of the variance
of the observables. \label{tab:Mean,-,-and all parameters medium scale-extended model}}

\centering{}%
\begin{tabular}{ccccc}
\hline 
Param. & Estimates & $2.5\%$ & $97.5\%$ & $\widehat{\textrm{IF}}$\tabularnewline
\hline 
$\rho_{R}$ & $0.522$ & $0.204$ & $0.759$ & $229.752$\tabularnewline
$\rho_{g}$ & $0.608$ & $0.241$ & $0.929$ & $201.565$\tabularnewline
$\rho_{\mu}$ & $0.560$ & $0.383$ & $0.723$ & $170.293$\tabularnewline
$100\sigma_{g}$ & $0.052$ & $0.006$ & $0.157$ & $151.086$\tabularnewline
$100\sigma_{\mu}$ & $4.085$ & $2.494$ & $6.047$ & $169.246$\tabularnewline
$100\sigma_{\eta}$ & $0.350$ & $0.262$ & $0.468$ & $220.044$\tabularnewline
$100\sigma_{Z}$ & $0.628$ & $0.483$ & $0.800$ & $203.168$\tabularnewline
$100\sigma_{R}$ & $0.102$ & $0.007$ & $0.238$ & $185.973$\tabularnewline
$\gamma_{g}$ & $0.354$ & $-0.019$ & $0.856$ & $209.208$\tabularnewline
$\gamma_{\Pi}$ & $1.011$ & $1.001$ & $1.032$ & $301.381$\tabularnewline
$\gamma_{x}$ & $0.878$ & $0.506$ & $1.314$ & $266.684$\tabularnewline
$\gamma$ & $0.457$ & $0.330$ & $0.588$ & $199.630$\tabularnewline
$\sigma_{a}$ & $5.596$ & $3.798$ & $7.694$ & $167.707$\tabularnewline
$\varphi_{p}$ & $65.973$ & $17.465$ & $118.752$ & $250.439$\tabularnewline
$\varphi_{w}$ & $814.476$ & $263.520$ & $1911.136$ & $310.302$\tabularnewline
$\varphi_{I}$ & $2.292$ & $1.223$ & $3.663$ & $193.898$\tabularnewline
$\sigma_{L}$ & $5.212$ & $2.946$ & $7.784$ & $203.591$\tabularnewline
$1-a$ & $0.326$ & $0.087$ & $0.545$ & $164.680$\tabularnewline
$1-a_{w}$ & $0.613$ & $0.273$ & $0.862$ & $257.090$\tabularnewline
$100\left(\beta^{-1}-1\right)$ & $0.094$ & $0.006$ & $0.315$ & $223.610$\tabularnewline
$100\log\left(G_{z}\right)$ & $0.518$ & $0.468$ & $0.568$ & $176.914$\tabularnewline
$100\left(\overline{\Pi}-1\right)$ & $0.611$ & $0.429$ & $0.795$ & $154.344$\tabularnewline
$\alpha$ & $0.181$ & $0.137$ & $0.221$ & $165.125$\tabularnewline
\hline
\end{tabular}
\end{table}

\section{Description: Small scale DSGE model \label{Description of SmallScale Model}}
%Next, we estimate a small scale New Keynesian model separately using a first order and second order approximation.
The small scale DSGE model specification follows \citet{HS2015}, and is also
used  by \citet{Herbst2019}.  The small scale DSGE model follows the environment in \citet{Herbst2019} with nominal rigidities through price adjustment costs following \citet{Rotemberg}, and three structural shocks. There are altogether $4$ state variables, $4$ control variables ($8$ variables combining the state and control variables) and $3$ exogenous shocks in the model. We estimate this model using $3$ observables; that means that the dimensions of $y_t$, the state vector and the disturbance vector are $3$, $8$, and $3$, respectively.

\paragraph{Households}
Time is discrete, households live forever, the representative household  derives utility from consumption $C_t$ relative to a habit stock (which is approximated by the level of technology $A_t$)\footnote{This assumption ensures that the economy evolves along a balanced growth path even if the utility function is additively separable in consumption, real money balances and hours.}
and real money balances $\frac{M_t}{P_t}$ and disutility from hours worked $N_t$. The household maximizes
%chooses consumption ($C_t$), hours ($N_t$) and bond holdings ($B_t$) to maximize life-time utility
\begin{equation*}
E_0\sum_{t=0}^{\infty}\beta^{t}\Big[\frac{\big(\frac{C_t}{A_t}\big)^{1-\tau}-1}{1-\tau} +\chi_M\log\frac{M_t}{P_t} -\chi_N{N_t}\Big]
\end{equation*}
subject to the budget constraint
\begin{equation*}
P_tC_t+B_t+M_t+T_t=P_tW_tN_t+R_{t-1}B_{t-1}+M_{t-1}+P_tD_t+P_t{SC}_t ;
\end{equation*}
$T_t,D_t$ and ${SC}_t$ denote lump-sum taxes, aggregate residual profits and net cash inflow from trading a full set of
state contingent securities; $P_t$ is the aggregate price index, and  $W_t$ is the real wage; $\beta$ is the discount factor, ${\tau}$ is the coefficient of relative risk aversion, $\chi_M$ and $\chi_N$ are scale factors determining the steady state money balance holdings and hours. We set $\chi_N=1$. $A_t$ is the level of aggregate productivity.

\paragraph{Firms}
Final output is produced by a perfectly competitive representative firm  which uses a continuum of intermediate goods $Y_t(i)$ and the production function
\begin{equation*}
{Y_t}=\Bigg(\int_{0}^{1}Y_t(i)^{1-\nu}di\Bigg)^\frac{1}{1-\nu},
\end{equation*}
with $\nu < 1$. The demand for intermediate good $i$ is
\begin{equation*}
Y_t(i)=\Bigg(\frac{P_t(i)}{P_t}\Bigg)^{-\frac{1}{\nu}}Y_t ,
\end{equation*}
and the aggregate price index is
\begin{equation*}
P_t=\Bigg(\int_{0}^{1}P_t(i)^\frac{\nu-1}{\nu}di\Bigg)^\frac{\nu}{\nu-1} .
\end{equation*}
Intermediate good $i$ is produced using the linear production technology
\begin{equation*}
Y_t(i)=A_tN_t(i);
\end{equation*}
$A_t$ is an exogenous productivity process common to all firms, and $N_t(i)$ is the labor input of firm i. Intermediate firms set prices $(P_t(i))$ and labor input ($N_t(i)$) by maximizing the net present value of future profit. Nominal rigidities are introduced through price adjustment costs following \citet{Rotemberg}.
%\section*{COMMENT}
%please use a proper cite for Rotemberg (1982)
%\section*{END COMMENT}
\begin{equation*}
E_t\sum_{s=0}^{\infty}\beta^sQ_{t,t+s}\Big[\frac{P_{t+s(i)}}{P_{t+s}}Y_{t+s}(i) -W_{t+s}N_{t+s}(i) -AC_{t+s}(i)\Big]
\end{equation*}
with
\begin{equation*}
AC_{t}(i)=\frac{\phi}{2}\Bigg(\frac{P_{t}(i)}{P_{t-1}(i)} -\pi\Bigg)^2{Y_t(i)}
\end{equation*}
The parameter $\phi$ governs the extent of price rigidity in the economy and $\pi$ is the steady state inflation rate associated with the final good. In equilibrium, households and firms use the same stochastic discount factor $Q_{t,t+s}$ where
\begin{equation*}
Q_{t,t+s}=\Big(\frac{C_{t+s}}{C_{t}}\Big)^{-\tau}{\Big(\frac{A_t}{A_{t+s}}\Big)^{1-\tau}}.
\end{equation*}
In a symmetric equilibrium all firms choose the same price.

\paragraph{Monetary and Fiscal Policy}
The central bank conducts monetary policy following an interest rate feedback rule given by
\begin{equation*}
R_t={R_t^*}^{1-\rho_R}R_{t-1}^{\rho_R}\epsilon_t^R,
\end{equation*}
where $\epsilon_t^R\sim IID({0,\sigma_r})$ is an iid shock to the nominal interest rate, ${R_t^*}$ is the nominal target and $(1-\rho_R)$ captures interest rate smoothing in the conduct of policy,
\begin{equation*}
{R_t^*}=r\pi^*\Big(\frac{\pi_t}{\pi^*}\Big)^{\psi_1}\Big(\frac{Y_t}{Y_t^*}\Big)^{\psi_2};
\end{equation*}
$\pi_t:=\frac{P_t}{P_{t-1}}$ is the gross inflation rate and $\pi^*$ is the target inflation rate. $Y_t^*$ is the output in the absence of nominal rigidities. The parameters $\psi_1$ and $\psi_2$ capture the intensity with which the central bank responds to inflation and output gap in the model.
Government expenditure accounts for a fraction $\upsilon_t \in [0,1]$ of final output such that $G_t=\upsilon_tY_t$. The government budget constraint is
\begin{equation*}
P_tG_t+R_{t-1}B_{t-1}+M_{t-1}=T_t+B_t+M_t.
\end{equation*}
%\section*{COMMENT}
%\begin{itemize}
%\item
%define $\rho_R$,
%\textit{has been defined}
%\item
%what is the distribution of $\epsilon_t^R $~? I think it must be positive.
%\textit{has been defined}
%\item
%We use $\zeta$ in a different context in the article; similarly with $G$
%\textit{I Here it is $\zeta_t$.}
%\end{itemize}
%\section*{END COMMENT}

\paragraph{Aggregation}
Combining the household budget constraint with the government budget constraint  gives
\begin{equation*}
C_t+G_t+AC_{t}=Y_t
\end{equation*}
where in equilibrium, $AC_t=\frac{\phi}{2}(\pi_t -\pi)^2{Y}_t$.
\paragraph{Exogenous Processes}
Aggregate technology grows at the rate $\gamma$ and $m_t$ is the shock to aggregate demand such that
\begin{equation*}
%\log{A_t}=\log{\gamma}+ \log{A_{t-1}} +\log{z_t}
\log{A_t}=\log{\gamma}+ \log{A_{t-1}} +\log{m_t}
\end{equation*}
\begin{equation*}
%\log{z_t}=(1-\rho_z)\log{z} +\rho_z\log{z_{t-1}} +\epsilon_t^z
\log{m_t}=(1-\rho_m)\log{m} +\rho_m\log{m_{t-1}} +\epsilon_t^m
\end{equation*}
with $\epsilon_t^m\sim IID({0,\sigma_m})$.  We define $g_t:=1/(1-\upsilon_t)$ and $g_t$ evolves as
\begin{equation*}
\log{g_t}=(1-\rho_g)\log{g} +\rho_g\log{g_{t-1}} +\epsilon_t^g,
\end{equation*}
with $\epsilon_t^g\sim IID({0,\sigma_g})$.
%The small scale DSGE model therefore  consists of a consumption Euler equation, a new Keynesian Philip curve, a monetary policy rule, fiscal policy rule,  three exogenous shock processes, and 8 endogenous latent
%variables. Section \ref{section:NKMdesc} of the appendix summarizes the nonlinear equilibrium conditions.
%We estimate the model using both first and second order approximations to assess the performance of our proposed method relative to existing methods.
%As explained, the source of nonlinearity in our case stems from taking a second order approximation of the equilibrium conditions.
We summarize the nonlinear equilibrium conditions after detrending $C_t,Y_t,G_t$  by  deterministic technology, i.e define $\tilde{X_t}:=X_t/A_t$:

\begin{equation}\label{eq:startdsgemodel}
1=\beta{E_t}\Big[\bigg(\frac{\tilde{C}_{t+1}}{\tilde{C}_{t}}\bigg)^{-\tau}{z_t}\frac{R_t}{\pi_{t+1}}\Big]
\end{equation}
\begin{multline*}
1=\phi(\pi_t-\pi)\Big[\bigg(1-\frac{1}{2\nu}\bigg)\pi_t + \frac{\pi}{2\nu}\Big]-\phi{E_t}\Big[\bigg(\frac{\tilde{C}_{t+1}}
{\tilde{C}_{t}}\bigg)^{-\tau}\frac{\tilde{Y}_{t+1}}{\tilde{Y}_{t}}(\pi_{t+1} - \pi)\pi_{t+1} \Big] +\frac{1}{\nu}\Big[1 -\bigg({\tilde{C}_t}\bigg)^{\tau}\Big]
\end{multline*}
\begin{equation*}
R_t={R_t^*}^{1-\rho_R}R_{t-1}^{\rho_R}\exp{\epsilon_t^R}
\end{equation*}
\begin{equation*}
{R_t^*}=r\pi^*\Big(\frac{\pi_t}{\pi^*}\Big)^{\psi_1}\Big(\frac{\tilde{Y}_t}{\tilde{Y}_t^*}\Big)^{\psi_2}
\end{equation*}
%\begin{equation}
%\tilde{C}_t+\tilde{G}_t+AC_{t}=\tilde{Y}_t
%\end{equation}
\begin{equation*}
\tilde{C}_t+\tilde{G}_t+\frac{\phi}{2}(\pi_t -\pi)^2\tilde{Y}_t=\tilde{Y}_t
\end{equation*}

\begin{equation*}
\log{A_t}=\log{\gamma}+ \log{A_{t-1}} +\log{m_t}
\end{equation*}
\begin{equation}\label{eq:m}
\log{m_t}=(1-\rho_m)\log{m} +\rho_m\log{m_{t-1}} +\epsilon_t^m
\end{equation}
\begin{equation}\label{eq:enddsgemodel}
\log{g_t}=(1-\rho_g)\log{g} +\rho_g\log{g_{t-1}} +\epsilon_t^g.
\end{equation}

After detrending, the steady state solution of  the model is
$
\pi=\pi^*,\tilde{R}=\frac{\gamma\pi}{\beta},
\tilde{C}=(1-\nu)^\frac{1}{\tau}
\;\; \mathrm{   and   } \;\;
\tilde{Y}=g\tilde{C}=\tilde{y^*}
$.

%COMMENT
%\begin{itemize}
%	\item
%	\lq The detrended steady state of the model are given as follows:\rq{}
%	
%	should that be
%	\lq The detrended steady state of the model is given as follows:\rq{}?
%	
%	\item
%	I suggest  $$
%	\tilde{R}=\frac{\gamma\pi}{\beta},
%	\tilde{C}=(1-\nu)^\frac{1}{\tau}
%	\;\; \mathrm{   and   } \;\;
%	\tilde{Y}=g\tilde{C}=\tilde{y^*}
%	$$
%	\item
%	please explain and reference where these equations come from
%	\textit{This is standard. Simply solve for the steady state of equations 44-51. }
%\end{itemize}
%END COMMENT

Equations~\eqref{eq:startdsgemodel}-\eqref{eq:enddsgemodel} can now be rewritten as
\begin{equation*}
E_tG(z_{t+1},z_{t},\epsilon_{t+1} )=0,
\end{equation*}
where \[z_{t}=[\tilde{C}_{t},\tilde{Y}_{t},\tilde{G}_{t},R_t,R_t^*,A_t,m_t,g_t]\quad  \text{and} \quad \epsilon_{t}=[\epsilon_t^R,\epsilon_t^m,\epsilon_t^g],\] and solved using the methods in Section~ \ref{Supp: overview of DSGE models}.

We solve the model in log deviations from steady state,  i.e. $\widehat{x_t}:=\log\Big(\frac{\tilde{X_t}}{\tilde{X}}\Big)$ using Dynare.
The small scale DSGE model therefore  consists of a consumption Euler equation, a new Keynesian Philip curve, a monetary policy rule, fiscal policy rule,  three exogenous shock processes, and eight endogenous latent variables.
We estimate the model using both first and second order approximations to assess the performance of our proposed method relative to existing methods.
As explained, the source of nonlinearity in our case stems from taking a second order approximation of the equilibrium conditions.

The observables used in estimating the model consist of per capita GDP growth rate ($YGR_t$), annualized quarter on quarter inflation rate ($Infl_t$) and annualized nominal rates ($Int_t$). The observed data is measured in percentages, and, after applying a log transformation to the endogenous variables, the measurement equations for our system using the transformed endogenous variables given by (a hatted variable denotes the log transformation).
\begin{align*}
YGR_t=\gamma^{(Q)}+100(\hat{y}_t-\hat{y}_{t-1}+\hat{m}_t),\\
Infl_t=\pi^{(A)}+400\hat{\pi}_t,\\
Int_t=\pi^{(A)}+r^{(A)}+4\gamma^{(Q)}+400\hat{R}_t.\end{align*}
Here, $\hat{y}_t=\log \Big(\frac{\tilde{Y_t}}{\tilde{Y}}\Big)$, $\hat{\pi}_t=\log \Big(\frac{\pi_t}{{\pi}}\Big)$,  $\hat{R}_t=\log \Big(\frac{R_t}{{\overline{R}}}\Big)$ and $\hat{m}_t=\log \Big(\frac{m_t}{{\overline{m}}}\Big)=\log {m_t}$ since $\log(\overline{m})=0$ in steady state as evident from Equation \eqref{eq:m}.

The model has the 15 parameters $$\Big\{\upsilon,\nu,\phi,\tau,\pi^{\left(A\right)},r^{\left(A\right)},\gamma^{\left(Q\right)}, \psi_{1},
 \psi_{2},\rho_{r},\rho_{g}, \rho_{m},\sigma_{r},\sigma_{g},\sigma_{m}\Big\}.$$
  We calibrate the parameters characterizing the share of fiscal expenditure in GDP $\upsilon=0.2$, the elasticity of substitution across varieties $1/\nu=11$ and the parameter guiding the extent of nominal rigidities, $\phi=100$. The remaining parameters along with the measurement errors in the observation equation are estimated. The steady-state inflation rate, $\pi$, in the model
relates to the estimated parameter for annualized inflation $\pi^{(A)}$ such that $\pi=\pi^{(A)}/400$ and the discount factor $\beta$ to the estimated parameter for annualized interest rate $r^{(A)}$ such that $\beta=\Big ((1+r^{(A)})/400\Big)^{-1}$. The sample used for estimation is
1983Q1-2002Q4. The data set and the  variables are  identical to those  used in \citet{Herbst2019}.\footnote{Data on all three observables are
	sourced from FRED. Per capita  GDP growth rate is calculated using data on real gross domestic product (FRED mnemonic `GDPC1') and Civilian Non-institutional
	Population (FRED mnemonic `CNP16OV' / BLS series `LNS10000000'), Annualized Inflation is calculated using CPI price level (FRED mnemonic `CPIAUCSL'), the
	Federal Funds Rate (FRED mnemonic `FEDFUNDS').}

\section{Description: Medium scale DSGE model \label{Description of MediumScale Model}}
The specification of the medium-scale model is similar to \citet{SW} and follows the specification in \citet*{gust2017empirical}. The model allows for habit formation in consumption, nominal rigidities in wages and prices, indexation in wages and prices, and investment adjustment costs. The model environment consists of monopolistically competitive intermediate goods-producing firms, a perfectly competitive final good-producing firm, households, and the government conducting fiscal and monetary policy. We describe the behavior of each agent in detail below.
The baseline model in \citet{gust2017empirical} consists of 12 state variables, 18 control variables (30  variables combining the state and control variables) and 5 exogenous shocks. We estimate this model using 5 observables; that means that the dimensions of $y_t$, the state vector and the disturbance vector are $5$, $30$, and $5$, respectively.
We also consider a variation of the medium scale DSGE model of \citet{gust2017empirical}. 
To study how well our approach performs with increasing state dimension, we estimate the model with the canonical definition of the output gap defined as the difference between output in the model with nominal rigidities and output in the model with flexible prices and wages. This variation is identical to the baseline model in  \citet{gust2017empirical} except for the Taylor rule specification, which now uses the canonical definition of the output gap. This extension of the medium scale DSGE model consists of 21 state variables, 30 control variables (51 variables combining the state and control variables) and 5 exogenous shocks.  We estimate this model using 5 observables; that means that the dimensions of $y_t$, the state vector and the disturbance vector are $5$, $51$, and $5$, respectively.

\subsection{Firms}
There is a continuum of monopolistically competitive firms producing differentiated intermediate goods. 
The final good ($Y_t$) is produced by a perfectly competitive representative firm that uses a continuum of differentiated intermediate goods ($Y_t(i)$) such that
\begin{equation*}
{Y_t}=\Bigg(\int_{0}^{1}Y_t(j)^{\frac{\epsilon_p-1}{\epsilon_p}}dj\Bigg)^{\frac{\epsilon_p}{\epsilon_p-1}}.
\end{equation*}
Under the constant elasticity of substitution aggregator, the demand for good produced by the $j^{th}$ firm is 
\begin{equation*}
Y_t(j)=\Bigg(\frac{P_t(j)}{P_t}\Bigg)^{-\epsilon_p}Y_t ,
\end{equation*}
The production function for the $j^{th}$ intermediate good producer is
\begin{equation*}
{Y_t(j)}=K_t(j)^\alpha(Z_tN_t(j))^{(1-\alpha)}.
\end{equation*}
$N_t(j)$ and $K_t(j)$ denote labor and capital inputs for the  $j^{th}$ intermediate good producer and $Z_t$ denotes labor augmenting technological progress. $Z_t$	evolves as
\begin{equation*}
Z_t=Z_{t-1}G_z\exp(\epsilon_{z,t}); \text{ \space } \epsilon_{z,t} \overset{iid}{\sim} N(0,\sigma_z^2);
\end{equation*}
$G_z$ is the deterministic growth rate of technology and $\epsilon_{z,t}$ introduces deviations about this deterministic trend.  Following \citet{Rotemberg}, intermediate good producing firm $j$ face quadratic costs in adjusting nominal prices $P_t(j)$ and is given by
\begin{equation*}
\frac{\phi_p}{2}\Bigg(\frac{P_{t}(j)}{\tilde{\pi}_{t-1}P_{t-1}(j)} -1\Bigg)^2{Y_t}
\end{equation*}
with $\phi_p$ describes the size of adjustment cost. The change in price is indexed to  $\tilde{\pi}_{t-1}$  with
 \begin{equation*}
 \tilde{\pi}_{t-1}=\overline{\pi}^a{{\pi}_{t-1}^{1-a}}.
 \end{equation*} 
Here $\overline{\pi}$ is the inflation target set by the central bank and ${\pi}_{t-1}:=\frac{P_{t-1}}{P_{t-2}}$ is the lagged gross inflation in period $t-1$. The parameter $a$ quantifies the extent of indexation to the central bank inflation target or lagged gross inflation with $a\in[0,1]$. The optimal price is set by each intermediate good producing firm by maximizing the expected present discounted value of profits 
\begin{equation*}
E_t\sum_{s=0}^{\infty}\frac{\Lambda_{t+s}}{\Lambda_{t}}\Big[\Big(\frac{P_{t+s}(j)}{P_{t+s}}-mc_{t+s}\Big)Y_{t+s}(j) -\frac{\psi_p}{2}\Big(\frac{P_{t+s}(j)}{\tilde{\pi}_{{t+s}-1}P_{{t+s}-1}(j)} -1\Big)^2{Y_{t+s}}\Big];
\end{equation*}
$mc_{t}$ is the real marginal cost with $mc_t:=\frac{{(W_t/P_t)}^{1-\alpha}{(R_t^K/P_t)}^{\alpha}}{\alpha^{\alpha}{1-\alpha}^{1-\alpha}}$. $W_t$ and $R_t^K$ denote the nominal wage and nominal rental rate of capital services respectively. $\frac{\Lambda_{t+s}}{\Lambda_{t}}$ is the stochastic discount factor from the optimization exercise of households.

\subsection{Households}
There exists a continuum of infinitely lived monopolistically competitive households indexed by $i \in [0,1]$ supplying differentiated labor service, $N_t(i)$. Household $i$ sells labor service $N_t(i)$ to a representative employment agency which combines the differentiated labor services across the spectrum of households into a composite $N_t$, which in turn is supplied to the intermediate goods producers for production in a perfectly competitive market. Differentiated labor services is aggregated via a Dixit-Stiglitz aggregator given by
\begin{equation*}
N_t=\Big(\int_{0}^{1}N_t(i)^{\frac{\epsilon_w-1}{\epsilon_w}}di\Big)^{\frac{\epsilon_w}{\epsilon_w-1}}.
\end{equation*}
The parameter $\epsilon_w$ is the constant elasticity of substitution across different varieties of labor services. Households maximize lifetime utility,
\begin{equation*}
E_0\sum_{t=0}^{\infty}\beta^t\Big[\Big(C_{t}(i)-\gamma C_{t-1}(i)\Big)-\psi_L\frac{N_{t}(i)^{1+\sigma_L}}{1+\sigma_L} -\psi_{t}^w(i)\Big],
\end{equation*}
with $C_{t+s}(i)$ consumption, $N_{t+s}(i)$ being labor services, and $\psi_{t}^w(i)$ denotes the loss in utility due to adjusting nominal wages $W_t(i)$. The utility cost of adjusting nominal wages is quadratic. The cost of adjusting nominal wages is 
\begin{equation*}
\psi_{t}^w(i)=\frac{\phi_w}{2}\Bigg(\frac{W_{t}(j)}{\tilde{\pi}_{t-1}^wW_{t-1}(j)} -1\Bigg)^2.
\end{equation*}
The parameter $\phi_w$ governs the size of wage adjustment costs. Wage contracts are indexed to productivity and inflation with
\begin{equation*}
\tilde{\pi}_{t}^w=G_z\overline{\pi}^{a_w}(\exp(\epsilon_z,t)\pi_{t-1})^{1-a_w}.
\end{equation*}
The parameter $\gamma$ governs the degree of habit formation in household utility.\footnote{Habits are internal to households.} Households also engage in savings. Households can save by  investing in  risk-free nominal bonds ($B_{t+s}(i)$) and owning physical capital stock $\overline{K}_t(i)$. Additionally, households can choose the extent of capital utilization $u_t(i)$. Utilization is combined with physical capital stock to transform physical capital to capital services $K_t(i)=u_t(i)\overline{K}_t(i)$. 
In the process of transforming physical capital to capital services via utilization $u_t(i)$, households incur costs of utilization
\begin{equation*}
a(u_t(i)):=\frac{r^k}{\sigma_a}\Big[\exp(\sigma_a(u_t(i)-1))  -1 \Big].
\end{equation*}
Households choose $C_t(i)$, $N_t(i)$, $B_{t+1}(i)$, investment $I_t(i)$, $\overline{K}_t(i)$ and $u_t(i)$ subject to the budget constraint
\begin{equation*}
C_t(i)+I(i)+\frac{B_{t+1}/{P_t}}{R_t\eta_t}+a(u_t(i))\overline{K}_t(i)=\frac{W_t(i)}{P_t}N_t(i)+\frac{R_t^K}{P_t}{K}_{t-1}(i)+\frac{B_t(i)}{P_t}+\frac{D_t}{P_t}-T_t, \text{ for any }  t.
\end{equation*}
The law of motion of capital
\begin{equation*}
\overline{K}_{t}(i)=(1-\delta)\overline{K}_{t-1}(i)+\mu_t(1-S_t(i))I_t(i).
\end{equation*}
Accumulation of capital is subject to the investment adjustment costs
\begin{equation*}
S_t(i)=\frac{\psi_I}{2}\Big[\frac{I_t(i)}{G_ZI_{t-1}(i)-1}\Big]^2.
\end{equation*}
Consistent with \citet{SW}, $\eta_t$ in the budget constraint and $\mu_t$ in the law of motion of capital represent shocks to demand for risk-free assets and an exogenous disturbance to the marginal efficiency of transforming final goods in $t$ to physical capital in $t+1$ with
\begin{equation*}
\ln(\eta_t)=\rho_{\eta}\ln(\eta_{t-1})+\epsilon_{\eta,t}, \epsilon_{\eta,t} \overset{iid}{\sim} N(0,\sigma_{\eta}^2)
\end{equation*}
and
\begin{equation*}
\ln(\mu_t)=\rho_{\mu}\ln(\mu_{t-1})+\epsilon_{\mu,t}, \epsilon_{\mu,t} \overset{iid}{\sim} N(0,{\sigma_{\mu}^2}).
\end{equation*}
\subsection{Monetary and fiscal policy}
The central bank sets the nominal interest rate $R_t^N$ according to a Taylor rule with
\begin{equation*}
R_t=\big(R_{t-1}\big)^{\rho_R}\Big[R\Big(\frac{\pi_t}{\overline{\pi}}\Big)^{\gamma_\pi}x_{g,t}^{\gamma_x}\Big(\frac{Y_t}{G_zY_{t-1}}\Big)^{\gamma_g}\Big]^{1-\rho_R}{\epsilon_{R,t}}.
\end{equation*}
The parameter $\rho_R \in (0,1)$ governs the extent of interest rate smoothing; $R$ denotes the interest rate in the non-stochastic steady state with $R=\beta^{-1}G_z\overline{\pi}$; $\overline{\pi}$ is the inflation target set by the central bank; $x_{g,t}$ in the Taylor rule is a proxy to quantify the output-gap in the economy with nominal rigidities relative to the flexible price benchmark. Consistent with \citet{gust2017empirical} we deviate from \citet{SW}  and specify the the output-gap $x_{g,t}$ with
\begin{equation*}
x_{g,t}=u_t^{\alpha}\Big(\frac{N_t}{N}\Big)^{1-\alpha},
\end{equation*}
where $u_t$ is the aggregate level of capital utilization (with non-stochastic steady state of 1) and $N$ is the non-stochastic steady state level of aggregate labor. We estimate both versions of the model where the output gap is proxied by $x_{g,t}$ as well as using the canonical definition of the output gap -- characterized as deviations of output in the model with nominal rigidities relative to the model without nominal rigidities and the Taylor rule being 
\begin{equation*}
R_t=\big(R_{t-1}\big)^{\rho_R}\Big[R\Big(\frac{\pi_t}{\overline{\pi}}\Big)^{\gamma_\pi}(x_t)^{\gamma_x}\Big(\frac{Y_t}{G_zY_{t-1}}\Big)^{\gamma_g}\Big]^{1-\rho_R}{\epsilon_{R,t}},
\end{equation*}
with $x_t=\frac{Y_t}{Y_t^N}$; $Y_t^N$ is the output in the flexible price model. The parameters $\gamma_{\pi} \geq 0$, $\gamma_g \geq 0$ and $\gamma_x \geq 0$ quantify the extent of the central bank's reaction to inflation, output-gap and output growth; ${\epsilon_{R,t}}$ is the exogenous disturbance to monetary policy with ${\epsilon_{R,t}} \overset{iid}{\sim} N(0,\sigma_R^2)$. 
Government spending evolves exogenously as a time-varying function of aggregate output $Y_t$ with 
\begin{equation*}
G_t=\big(1-\frac{1}{g_t}\big)Y_t,
\end{equation*} 
such that
\begin{equation*}
\ln(g_t)=(1-\rho_g)\ln(g)+\rho_g\ln(g_{t-1})+\epsilon_{g,t}; \text{ \space} {\epsilon_{g,t}} \overset{iid}{\sim} N(0,\sigma_g^2).
\end{equation*}
The government budget constraint satisfies $G_t=T_t$, for each period.

\subsection{Market clearing}
In a symmetric equilibrium, all intermediate goods firm choose the same price with $P_t(i)=P_t$, for all  $i$, and all households choose the same wage with $W_t(j)=W_t \text{ \space }, \forall j$. The aggregate production function is 
\begin{equation*}
Y_t=K_t^{\alpha}[Z_tN_t]^{1-\alpha}.
\end{equation*}
Market clearing implies
\begin{equation*}
Y_t=C_t+I_t+G_t+\frac{\phi_p}{2}\Bigg(\frac{P_{t}}{\tilde{\pi}_{t-1}P_{t-1}} -1\Bigg)^2{Y_t}+a(u_t)\overline{K}_t.
\end{equation*}

\subsection{First-order conditions}
The detailed first order conditions are summarized below:
\begin{equation}
[\frac{\pi_t}{\tilde{\pi}_{t-1}}-1]\frac{\pi_t}{\tilde{\pi}_{t-1}}=\beta E_t\Big[\frac{\Lambda_{t+1}}{\Lambda_{t}}\Big][\frac{\pi_{t+1}}{\tilde{\pi}_{t}}-1]\frac{\pi_{t+1}}{\tilde{\pi}_{t}}\frac{Y_{t+1}}{Y_t}+ \frac{\epsilon_p}{\psi_p}[mc_t-\frac{\epsilon_p-1}{\epsilon_p}]
\end{equation}
\begin{equation}
(1-\alpha)mc_t=\frac{W_tN_t}{P_tY_t}
\end{equation}
\begin{equation}
P_tr_t^K=\frac{\alpha}{1-\alpha}\frac{W_tN_t}{u_t\overline{K}_t}
\end{equation}
\begin{equation}
\tilde{\pi}_{t-1}=\overline{\pi}^{a}{{\pi}_{t-1}^{1-a}}
\end{equation}
\begin{equation}
Y_t=(u_t\overline{K}_t)^{\alpha}{\big(Z_tN_t\big)}^{1-\alpha}
\end{equation}
\begin{equation}
\Lambda_t=\big[C_t-\gamma C_{t-1}\big]^{-1}-\gamma\beta E_t\big[C_{t+1}-\gamma C_{t}\big]^{-1}
\end{equation}
\begin{equation}
1=\beta R_t \eta_tE_t \frac{\Lambda_{t+1}}{\Lambda_t}/\pi_{t+1}
\end{equation}
\begin{equation}
[\frac{\pi_{w,t}}{\tilde{\pi}_{w,t}}-1]\frac{\pi_{w,t}}{\tilde{\pi}_{w,t}}=\beta E_t\Big[\frac{\Lambda_{t+1}}{\Lambda_{t}}\Big][\frac{\pi_{w,t+1}}{\tilde{\pi}_{w,t+1}}-1]\frac{\pi_{w,t+1}}{\tilde{\pi}_{w,t+1}}+N_t\Lambda_t\frac{\epsilon_w}{\psi_w}\Big[\psi_L\frac{N_t^{\sigma_L}}{\Lambda_t}-\frac{\epsilon_w-1}{\epsilon_w}\frac{W_t}{P_t}\Big]
\end{equation}
\begin{equation}
\pi_{w,t}=\frac{W_t}{W_{t-1}}
\end{equation}
\begin{equation}
\tilde{\pi}_{w,t}=G_z\overline{\pi}^{a_w}\big(\exp(\epsilon_{z,t}) \pi_{t-1}\big)^{1-a_w}
\end{equation}
\begin{equation}
q_t=\beta E_t\Big[ \frac{\Lambda_{t+1}}{\Lambda_{t}}\big(r_{t+1}^K-a(u_{t+1}) +(1-\delta)q_{t+1}\big)\Big]
\end{equation}
\begin{multline}
1=q_t\mu_{t}\Big[1-\frac{\psi_I}{2}\big(\frac{I_t}{G_zI_{t-1}}-1\big)^2 -{\psi_I}\big(\frac{I_t}{G_zI_{t-1}}-1\big)\frac{I_t}{G_zI_{t-1}}\Big] \\+\beta \psi_IE_t\Big[q_{t+1}\frac{\Lambda_{t+1}}{\Lambda_{t}}\mu_{t+1}\big(\frac{I_{t+1}}{G_zI_{t}}-1\big) \big(\frac{I_{t+1}}{G_zI_{t}}\big)^2\Big]
\end{multline}
\begin{equation}
r_t^K=r^k{\exp(\sigma_a(u_t-1))}
\end{equation}
\begin{equation}
a(u_t)=\frac{r^k}{\sigma_a}\Big(\exp(\sigma_a(u_t-1))-1\Big)
\end{equation}
\begin{equation}
\overline{K}_{t+1}=(1-\delta)\overline{K}_{t}+\mu_t(1-\frac{\psi_I}{2}\Big[\frac{I_t}{G_ZI_{t-1}}-1\Big]^2)I_t
\end{equation}
\begin{equation}
R_t=\big(R_{t-1}\big)^{\rho_R}\Big[R\Big(\frac{\pi_t}{\overline{\pi}}\Big)^{\gamma_\pi}(x_{g,t})^{\gamma_x}\Big(\frac{Y_t}{G_zY_{t-1}}\Big)^{\gamma_g}\Big]^{1-\rho_R}{\epsilon_{R,t}}.
\end{equation}
When solving the model using the canonical definition of the output-gap, the Taylor rule is specified as follows where $Y_t^N$ is the output in the model with flexible prices and wages.
\begin{equation}
R_t=\big(R_{t-1}\big)^{\rho_R}\Big[R\Big(\frac{\pi_t}{\overline{\pi}}\Big)^{\gamma_\pi}(\frac{Y_t}{Y_t^N})^{\gamma_x}\Big(\frac{Y_t}{G_zY_{t-1}}\Big)^{\gamma_g}\Big]^{1-\rho_R}{\epsilon_{R,t}}
\end{equation}
\begin{equation}
Y_t=C_t+I_t+G_t+\frac{\phi_p}{2}\Bigg(\frac{P_{t}}{\tilde{\pi}_{t-1}P_{t-1}} -1\Bigg)^2{Y_t}+a(u_t)\overline{K}_t
\end{equation}

\begin{equation}
x_{g,t}=u_t^{\alpha}\Big(\frac{N_t}{N}\Big)^{1-\alpha}
\end{equation}

\begin{equation}
\ln(\eta_t)=\rho_{\eta}\ln(\eta_{t-1})+\epsilon_{\eta,t}; \text{ \space} \epsilon_{\eta,t} \overset{iid}{\sim} N(0,\sigma_{\eta}^2)
\end{equation}

\begin{equation}
\ln(\mu_t)=\rho_{\mu}\ln(\mu_{t-1})+\epsilon_{\mu,t}; \text{ \space} \epsilon_{\mu,t} \overset{iid}{\sim} N(0,{\sigma_{\mu}^2}).
\end{equation}

\begin{equation}
\ln(g_t)=(1-\rho_g)\ln(g)+\rho_g\ln(g_{t-1})+\epsilon_{g,t}; \text{ \space} {\epsilon_{g,t}} \overset{iid}{\sim} N(0,\sigma_g^2).
\end{equation}

\begin{equation}
Z_t=Z_{t-1}G_z\exp(\epsilon_{z,t}); \text{ \space } \epsilon_{z,t} \overset{iid}{\sim} N(0,\sigma_z^2).
\end{equation}

\subsection{De-trended equilibrium conditions}
Labor augmenting technological progress $Z_t$	evolves as
\begin{equation*}
Z_t=Z_{t-1}G_z\exp(\epsilon_{z,t}); \text{ \space } \epsilon_{z,t} \overset{iid}{\sim} N(0,\sigma_z^2).
\end{equation*} The deterministic component $G_z$ of labor augmenting technological process is the source of deterministic growth in the model. We summarize the de-trended stationary equilibrium conditions such that lower case letters $y_t:=\frac{Y_t}{Z_t},c_t:=\frac{C_t}{Z_t},i_t:=\frac{I_t}{Z_t},\overline{k}_t:=\frac{\overline{K}_t}{Z_{t-1}},w_t:=\frac{W_t}{P_tZ_t}, \tilde{g}_t:=(1-\frac{1}{g_t})\frac{Y_t}{Z_t} \text{and} \lambda_t:={Z_t}{\Lambda_t}, $ denote stationary transformations.

\begin{equation}\label{startdetrend}
[\frac{\pi_t}{\tilde{\pi}_{t-1}}-1]\frac{\pi_t}{\tilde{\pi}_{t-1}}=\beta E_t\Big[\frac{\lambda_{t+1}}{\lambda_{t}}\Big][\frac{\pi_{t+1}}{\tilde{\pi}_{t}}-1]\frac{\pi_{t+1}}{\tilde{\pi}_{t}}\frac{y_{t+1}}{y_t}+ \frac{\epsilon_p}{\psi_p}[mc_t-\frac{\epsilon_p-1}{\epsilon_p}]
\end{equation}
\begin{equation}
(1-\alpha)mc_t=\frac{w_tN_t}{P_ty_t}
\end{equation}
\begin{equation}
r_t^K=\frac{\alpha}{1-\alpha}\frac{w_tN_t}{u_t\overline{k}_t}
\end{equation}
\begin{equation}
\tilde{\pi}_{t-1}=\overline{\pi}^{a}{{\pi}_{t-1}^{1-a}}
\end{equation}
\begin{equation}
y_t={G_{z,t}}^{-\alpha}(u_t\overline{k}_t)^{\alpha}{N_t}^{1-\alpha}
\end{equation}
\begin{equation}
\lambda_t=\big[c_t-\gamma c_{t-1}\big]^{-1}-\gamma\beta E_t\big[c_{t+1}-\gamma c_{t}\big]^{-1}
\end{equation}
\begin{equation}
1=\beta R_t \eta_tE_t \frac{\lambda_{t+1}}{\lambda_t}/\big({G_{z,t+1}}\pi_{t+1}\big)
\end{equation}
\begin{equation}
[\frac{\pi_{w,t}}{\tilde{\pi}_{w,t}}-1]\frac{\pi_{w,t}}{\tilde{\pi}_{w,t}}=\beta E_t\Big[\frac{\lambda_{t+1}}{\lambda_{t}}\Big][\frac{\pi_{w,t+1}}{\tilde{\pi}_{w,t+1}}-1]\frac{\pi_{w,t+1}}{\tilde{\pi}_{w,t+1}}+N_t\lambda_t\frac{\epsilon_w}{\psi_w}\Big[\psi_L\frac{N_t^{\sigma_L}}{\lambda_t}-\frac{\epsilon_w-1}{\epsilon_w}w_t\Big]
\end{equation}
\begin{equation}
\pi_{w,t}=\frac{w_t}{w_{t-1}}{G_{z,t}}{\pi}_{t}
\end{equation}
\begin{equation}
\tilde{\pi}_{w,t}=G_z\overline{\pi}^{a_w}\big(\exp(\epsilon_{z,t}) \pi_{t-1}\big)^{1-a_w}
\end{equation}
\begin{equation}
q_t=\beta E_t\Big[ \frac{\lambda_{t+1}}{{G_{z,t+1}}\lambda_{t}}\big(r_{t+1}^K-a(u_{t+1}) +(1-\delta)q_{t+1}\big)\Big]
\end{equation}
\begin{multline}
1=q_t\mu_{t}\Big[1-\frac{\psi_I}{2}\big(\frac{i_t\epsilon_{z,t}}{i_{t-1}}-1\big)^2 -{\psi_I}\big(\frac{i_t\epsilon_{z,t}}{i_{t-1}}-1\big)\frac{i_t\epsilon_{z,t}}{i_{t-1}}\Big]\\+\beta \psi_I E_t\Big[q_{t+1}\frac{\lambda_{t+1}}{\lambda_{t}}\mu_{t+1}\big(\frac{i_{t+1}\epsilon_{z,t+1}}{i_{t}}-1\big)\big(\frac{i_{t+1}}{i_{t}}\big)^2\epsilon_{z,t+1}\Big]
\end{multline}
\begin{equation}
r_t^K=r^k{\exp(\sigma_a(u_t-1))}
\end{equation}
\begin{equation}
a(u_t)=\frac{r^k}{\sigma_a}\Big(\exp(\sigma_a(u_t-1))-1\Big)
\end{equation}
\begin{equation}
\overline{k}_{t+1}=(1-\delta){G_{z,t}^{-1}}\overline{k}_{t}+\mu_t(1-\frac{\psi_I}{2}\Big[\frac{i_t\epsilon_{z,t}}{i_{t-1}}-1\Big]^2)i_t
\end{equation}
\begin{equation}\label{TRbaseline}
R_t=\big(R_{t-1}\big)^{\rho_R}\Big[R\Big(\frac{\pi_t}{\overline{\pi}}\Big)^{\gamma_\pi}(x_{g,t})^{\gamma_x}\Big(\frac{y_t\epsilon_{z,t}}{y_{t-1}}\Big)^{\gamma_g}\Big]^{1-\rho_R}{\epsilon_{R,t}}
\end{equation}
\begin{equation}
y_t=c_t+i_t+\tilde{g}_t+\frac{\phi_p}{2}\Bigg(\frac{P_{t}}{\tilde{\pi}_{t-1}P_{t-1}} -1\Bigg)^2{y_t}+a(u_t){G_{z,t}^{-1}}\overline{k}_t
\end{equation}

\begin{equation}\label{ogbaseline}
x_{g,t}=u_t^{\alpha}\Big(\frac{N_t}{N}\Big)^{1-\alpha}
\end{equation}

\begin{equation}
\ln(\eta_t)=\rho_{\eta}\ln(\eta_{t-1})+\epsilon_{\eta,t}; \text{ \space} \epsilon_{\eta,t} \overset{iid}{\sim} N(0,\sigma_{\eta}^2)
\end{equation}

\begin{equation}
\ln(\mu_t)=\rho_{\mu}\ln(\mu_{t-1})+\epsilon_{\mu,t}; \text{ \space} \epsilon_{\mu,t} \overset{iid}{\sim} N(0,{\sigma_{\mu}^2})
\end{equation}

\begin{equation}\label{enddetrend}
\ln(g_t)=(1-\rho_g)\ln(g)+\rho_g\ln(g_{t-1})+\epsilon_{g,t}; \text{ \space} {\epsilon_{g,t}} \overset{iid}{\sim} N(0,\sigma_g^2).
\end{equation}

Equations \ref{startdetrend}-\ref{enddetrend} summarize the detrended equilibrium conditions of the model. We solve the model by taking a second-order approximation of Equations \ref{startdetrend}-\ref{enddetrend}.
To show that the solution algorithm can handle a higher dimensional state space, we solve a version of the model keeping all the conditions unchanged except for the Taylor Rule (Equation \ref{TRbaseline}) which approximates the output gap using Equation \ref{ogbaseline} with 

\begin{equation}\label{TRSW}
R_t=\big(R_{t-1}\big)^{\rho_R}\Big[R\Big(\frac{\pi_t}{\overline{\pi}}\Big)^{\gamma_\pi}(\frac{y_t}{y_t^N})^{\gamma_x}\Big(\frac{y_t\epsilon_{z,t}}{y_{t-1}}\Big)^{\gamma_g}\Big]^{1-\rho_R}{\epsilon_{R,t}}.
\end{equation}

The specification of the Taylor rule in Equation \ref{TRSW} follows the canonical definition of the output gap defined as the difference between output in the model with nominal rigidities ($y_t$) and output in the model with flexible prices and wages ($y_t^N$). The number of state variables in the DSGE model increases from 12 to 21 when using the definition of the output gap consistent with the canonical definition and used in \citet{SW}.

\bibliographystyle{apalike}
\bibliography{references_v1}
\end{document}